\documentclass[a4paper,titlepage,11pt]{article}
\pdfoutput=1
\usepackage[utf8]{inputenc}
\usepackage{geometry}
\usepackage[english]{babel}
\usepackage{amsmath, amsfonts, graphicx}
\usepackage{mathtools}
\usepackage{graphicx}
\usepackage{float}
\usepackage{mciteplus}
\usepackage{bbold}
\usepackage{pstricks}
\usepackage[labelfont=bf,font={small}]{caption}
\usepackage{pgfplots}
\usepackage{empheq}
\usepackage{tikz}
\usetikzlibrary{positioning, arrows}
\usetikzlibrary{external}
\usepackage{footmisc}
\usepackage{multirow}
\usepackage{url}
\usepackage[breakable, theorems, skins]{tcolorbox}
\usepackage{enumerate}
\usepackage{physics}
\usepackage{bm}
\usepackage{braket}
\usepackage{epstopdf}
\usepackage{hyperref}
\usepackage{cancel}

\definecolor{blue(pigment)}{rgb}{0.2, 0.2, 0.6}
\definecolor{darkblue}{rgb}{0.0, 0.0, 0.55}

\hypersetup{
    colorlinks,
    citecolor=blue,
    filecolor=blue,
    linkcolor=black,
    urlcolor=blue
}
\usepackage{xfrac}

\def\be{\begin{equation}}
\def\ee{\end{equation}}

\newtagform{normalsize}[\normalsize]{\normalsize(}{\normalsize)}

\def\bea{\begin{eqnarray}}
\def\eea{\end{eqnarray}}

\newcommand{\beq}{\begin{equation}}
\newcommand{\eeq}{\end{equation}}
 
\newcommand{\barr}{\!\begin{array}}
\newcommand{\earr}{\end{array}\!}

\def\be{\bea}
\def\ee{\eea}

	\numberwithin{equation}{section}

\begin{document}
\usetagform{normalsize}

\begin{titlepage}

\setcounter{page}{1} \baselineskip=15.5pt \thispagestyle{empty}

%\begin{flushright}
%hep-th/13mmnnn\\
%\end{flushright}
\vfil

${}$
\vspace{1cm}

\begin{center}

\def\thefootnote{\fnsymbol{footnote}}
\begin{changemargin}{0.05cm}{0.05cm} 
\begin{center}
{\Large \bf Gravitational wavefunctions in JT supergravity}
\end{center} 
\end{changemargin}

~\\[1cm]
{Andreas Belaey,\footnote{{\protect\path{andreas.belaey@ugent.be}}} Francesca Mariani,\footnote{{\protect\path{francesca.mariani@ugent.be}}} Thomas G. Mertens\footnote{{\protect\path{thomas.mertens@ugent.be}}}}
\\[0.3cm]
{\normalsize { \sl Department of Physics and Astronomy
\\[1.0mm]
Ghent University, Krijgslaan, 281-S9, 9000 Gent, Belgium}}\\[3mm]

\end{center}

%\vspace{2cm}
%\vspace{1cm}

%\hrule
 \vspace{0.2cm}
\begin{changemargin}{01cm}{1cm} 
{\small  \noindent 
\begin{center} 
\textbf{Abstract}
\end{center} 
We determine explicit expressions for the continuous two-sided gravitational wavefunctions in supersymmetric versions of JT gravity, focusing mainly on $\mathcal{N}=2$ JT supergravity. Our approach is based on representation theory of the associated supergroup, for which we determine the relevant mixed parabolic matrix elements that implement asymptotic AdS boundary conditions at the quantum level. We match our expressions with those found by solving the energy-eigenvalue equation  \cite{Lin:2022zxd}. We discuss gravitational applications by computing several amplitudes of interest, and address how our framework can be generalized further.
}
\end{changemargin}
 \vspace{0.3cm}
%\hrule
\vfil
\begin{flushleft}
\today
\end{flushleft}

\end{titlepage}
\newpage
 \tableofcontents

\newpage
\hypersetup{linkcolor=blue}

\setcounter{footnote}{0}
\section{Introduction}

The past years have seen considerable interest in formulating and solving models of lower-dimensional quantum gravity. This surge started with the JT gravity model \cite{Jackiw:1984je, Teitelboim:1983ux}, as studied extensively in e.g. \cite{Almheiri:2014cka, Jensen:2016pah, Maldacena:2016upp, Engelsoy:2016xyb, Cotler:2016fpe, Stanford:2017thb, Kitaev:2018wpr, Mertens:2017mtv, Mertens:2018fds, Lam:2018pvp, Harlow:2018tqv, Yang:2018gdb, Blommaert:2018oro, Blommaert:2018iqz, Iliesiu:2019xuh, Saad:2019lba,Saad:2019pqd,Blommaert:2019wfy,Okuyama:2019xbv,Blommaert:2020seb,Saad:2021rcu,Post:2022dfi,Altland:2022xqx,Jafferis:2022wez,Blommaert:2021fob,Griguolo:2023aem, Iliesiu:2024cnh} and reviewed in \cite{Mertens:2022irh}. This model is particularly interesting since it describes the  near-horizon physics of a large class of higher-dimensional nearly extremal black holes, and hence describes universality of black hole physics even in our universe, see e.g. \cite{Nayak:2018qej,Iliesiu:2020qvm,Castro:2021csm,Iliesiu:2022kny,Castro:2022cuo} for recent work along these lines.

Attention since then has shifted in two directions. Firstly, more general classes of dilaton gravity models can be considered (see \cite{Grumiller:2002nm} for a nice review), and there has been substantial progress in obtaining results there. This includes approaches based on deformations of the JT gravity datapoint
\cite{Maxfield:2020ale,Witten:2020wvy,Turiaci:2020fjj,Eberhardt:2023rzz,Kruthoff:2024gxc}, studies of the Liouville string as a particular model of dilaton gravity (based on a hyperbolic sine dilaton potential) \cite{Mertens:2020hbs,Fan:2021bwt,Blommaert:2023wad,Kyono:2017pxs,Collier:2023cyw}, and the double-scaled SYK (DSSYK) model, first solved in \cite{Berkooz:2018qkz,Berkooz:2018jqr,Berkooz:2020xne} and subsequently reinterpreted in \cite{Lin:2022rbf,Berkooz:2022mfk,Lin:2023trc,Blommaert:2023opb,Goel:2023svz,Narovlansky:2023lfz,Verlinde:2024zrh,Verlinde:2024znh,Almheiri:2024ayc,Almheiri:2024xtw} and phrased as a sine dilaton gravity model in some of our recent work \cite{Blommaert:2024ydx}.

In another direction, supersymmetric versions of JT gravity have been formulated \cite{Chamseddine:1991fg,Astorino:2002bj,Forste:2017kwy,Forste:2017apw} and subsequently solved as well (see e.g. \cite{Mertens:2017mtv,Fan:2021wsb,Lin:2022zxd, Lin:2022rzw} for the boundary correlators and \cite{Stanford:2019vob,Turiaci:2023jfa} for the higher genus and multiboundary amplitudes).
Investigating both of these types of generalized models is important as it allows us to understand the generality of physical lessons we might uncover. 
In this respect, the JT supergravity models are special to connect with higher-dimensional black holes, for which one needs precision holography to couple to the collection of existing theoretical solutions, the majority of which are found for supersymmetric black holes. Upon dimensionally reducing to the near-horizon JT model, one typically retains some supersymmetry (typically $\mathcal{N}=2$ or $\mathcal{N}=4$), see e.g. \cite{Boruch:2022tno,Heydeman:2020hhw} for recent progress in this direction.

Motivated by these considerations, in this work we will revisit $\mathcal{N}=2$ JT supergravity. The canonical disk partition function of $\mathcal{N}=2$ JT supergravity has been computed from either fermionic localization of the super-Schwarzian model \cite{Stanford:2017thb}, or from the Virasoro conformal bootstrap as \cite{Mertens:2017mtv}:
\begin{equation}
\label{eq:fullpf}
Z(\beta) = 4\sum_{\abs{q} < 1/2} \sin 2\pi \abs{q} + 2\sum_{q \in \mathbb{Z}/2N}\int_0^{+\infty} dk \frac{k \sinh 2\pi k}{\pi (k^2+q^2)}e^{-\beta (k^2+q^2)}
\end{equation}
The first term represents a discrete zero-energy sector, physically interpretable as the BPS sector of short multiplets which only appears starting with $\mathcal{N}=2$ supersymmetry. The charge quantum number $q$ is usually quantized as the notation above suggests. If it is fractionally quantized in units of $1/2N$ for an integer $N$, then there are BPS states whenever $N>1$.\footnote{One can also have half-integer quantization of $q \in (\mathbb{Z}+1/2)/2N$ which would result in a cos(.) function instead and is the choice of relevance for even N SYK and the higher-dimensional susy black hole in AdS$_5 \times S^5$ \cite{Heydeman:2022lse}. Our notation corresponds to odd N SYK instead. We emphasize that none of our results depend on this choice except for the precise range of $q$ in our expressions. Our notation of the quantum number $q$ is more closely related to the representation theory notation of e.g. \cite{Frappat:1996pb}, and is related to that of \cite{Lin:2022zxd} (denoted $j$ there) by a $1/2$ shift of $q$ (and factor of $2$ rescaling).} The second term is the continuous sector of states. Throughout this work, we will focus on the continuous sector of states only. Alternatively, we can set $N=1$ for which the discrete sector is absent altogether, and we will use this notation in what follows.

In this work, we will demonstrate how the two-sided gravitational wavefunctions of the $\mathcal{N}=2$ JT supergravitational bulk Hilbert space are constructed using the first-order BF gauge theoretic description in terms of the positive  orthosymplectic OSp$^+(2|2,\mathbb{R})$ semi-supergroup. These supergravitational wavefunctions have been obtained before as eigenfunctions of the $\mathcal{N}=2$ Liouville minisuperspace Hamiltonian in \cite{Lin:2022zxd}. On a technical level, we elaborate on our previous work on the representation theory of OSp$(2|2,\mathbb{R})$ in \cite{Belaey:2023jtr}, where we explicitly constructed the continuous principal series representations. Since principal series representations are irreducible, the corresponding representation matrix elements are solutions of the Casimir eigenvalue equation. We will demonstrate how certain constrained representation matrix elements match with the two-sided gravitational wavefunctions, satisfying asymptotic AdS$_{2|4}$ Brown-Henneaux gravitational boundary conditions at the holographic boundaries. Our main result is the precise formulation on how we can implement these gravitational boundary conditions at the level of the representation theory of higher rank superalgebras, made precise in e.g. equation \eqref{eq:fermleft} below. We formulate this in such a way that further generalization is in principle straightforward. Compared to directly solving the Liouville Schr\"odinger equation \cite{Lin:2022zxd}, our procedure straightforwardly fixes the overall and relative normalization of the different states in the fermionic multiplet, which leads in particular to an unambiguous bottom-up determination of the  (super)gravitational density of states. 

In the second part of this work, we illustrate our calculational scheme to derive several gravitational amplitudes of interest, focusing mainly on how the gauge/group theoretical language translates into the gravitational language. In particular, we demonstrate how the smooth connectivity of the two boundaries of the no-boundary wavefunction fixes uniquely the set of states propagating in the supergravitational bulk. 
Next to this, the wavefunction of a matter Wilson line is a solution of the Casimir eigenvalue problem, which we interpret as a highest weight discrete series matrix element. Following the analysis of \cite{Belaey:2023jtr}, we physically interpret this as the worldline path integral of a charged particle traveling along a trajectory connecting the two entangled boundaries. 
A similar interpretation can be made for amplitudes containing an end-of-the-world (EOW) brane. These ``open'' EOW brane correlators are the analogues of the ``closed'' circular EOW brane solutions considered from the BF gauge theory perspective in our previous work \cite{Belaey:2023jtr}. \\

The remainder of this work is organized as follows.
\textbf{Section \ref{s:sectmain}} contains our main technical result: using the positive semigroup OSp$^+(2|2,\mathbb{R})$ (which we define below), we explicitly construct its mixed parabolic matrix element in equation \eqref{eq:finalWhittakerfunction}, and match this to a solution of the $\mathcal{N}=2$ super-Liouville Schr\"odinger equation.

In \textbf{section \ref{s:applications}} we apply this result to obtain some concrete supergravitational amplitudes (the gravitational propagator, the TFD state, end-of-the-world brane amplitudes). 

We present some more speculative material in a concluding \textbf{section \ref{s:concl}}, including a starting point for similar  $\mathcal{N}=4$ JT supergravity amplitudes, for which such results are at the moment unknown using any technique.

The appendices contain some additional technical results that complement the points made in the main text. In particular \textbf{appendix \ref{appendix:representationtheory}} collects various details on the representation theory of OSp$(2|2,\mathbb{R})$, including its regular representations and Haar measure. \textbf{Appendix \ref{sec:hamiltonianreduction}} reports detailed expressions on the Hamiltonian reduction of $\mathfrak{osp}(2|2,\mathbb{R})$ to the $\mathcal{N}=(2,2)$ superconformal algebra, generalizing the approach of \cite{Bershadsky:1989mf,Bershadsky:1989tc} to the $\mathcal{N}=2$ case in one dimension less. In \textbf{appendix \ref{appendix:N=1}} we apply the methods of the main text to the simpler case of $\mathcal{N}=1$ JT supergravity.

\section{Supergroup structure of $\mathcal{N}=2$ JT supergravity}
\label{s:sectmain}

The classical rewriting of $\mathcal{N}=2$ JT supergravity in terms of an $\mathfrak{osp}(2|2,\mathbb{R})$ BF gauge theory (see \cite{Gomis:1991cc} and appendix C of \cite{Belaey:2023jtr}) is insufficient to describe the quantum gravitational amplitudes, since this requires information on the proper group exponentiation of this superalgebra. In particular, we know that actual (super)gravity \emph{cannot} be described by BF amplitudes of \emph{any} group $G$, which has to do with the fact that perfectly reasonable gauge connections can correspond to singular geometries in Euclidean signature, which are to be excluded in the Euclidean gravitational path integral. More technically, for the BF gauge theory only the hyperbolic so-called Teichm\"uller component of the moduli space of flat gauge connections corresponds to physical geometries. These hyperbolic holonomies can then in turn be singled out by restricting the underlying algebraic structure to the positive semigroup $G^+ \subset G$. 

In this section, we argue that the $\mathcal{N}=2$ supergravitational two-boundary wavefunctions are in fact described by the positive semi-supergroup OSp$^+(2|2,\mathbb{R})$, in analogy with the $\mathcal{N}=0,1$ cases \cite{Blommaert:2018oro, Blommaert:2018iqz, Fan:2021wsb}, which we first briefly review in the next subsection.

\subsection{Review of the bosonic story}\label{section:bosonicstory}
To study JT gravity in the first order form, one considers an $\mathfrak{sl}(2,\mathbb{R})$ BF gauge theory \cite{Fukuyama:1985gg, Isler:1989hq,  Chamseddine:1989yz}, with the appropriate boundary term \cite{Mertens:2018fds}: 
\begin{equation}
\label{C4:BFactionprop}
    I_{BF} = -\int_{\mathcal{M}} \text{Tr}(\mathbf{B}\mathbf{F})+\frac{1}{2} \oint_{\partial\mathcal{M}}d\tau\; \text{Tr}(\mathbf{B}\mathbf{A}_\tau),
\end{equation} 
in terms of an auxiliary field $\mathbf{B}=B_{-}E^-+B_{+}E^++B_HH$, and a one-form gauge field $\mathbf{A}_\mu=(A_\mu)_{-}E^-+(A_\mu)_{+}E^++(A_\mu)_HH$ containing the first-order frame fields and spin connection, in terms of the fundamental $\mathfrak{sl}(2,\mathbb{R})$ generators $E^-,E^+,H$.
The required boundary conditions defining JT gravity are the divergent bare boundary length $\ell_{\text{bare}}$, and a divergent dilaton profile $\Phi$ near the AdS$_2$ boundary regulated with $\epsilon\rightarrow0$ as:
\begin{equation}
\label{C4:secondorderboundary}  \ell_{\text{bare}}=\frac{\beta}{\epsilon}, \qquad \Phi\sim\frac{1}{\epsilon},
\end{equation}
with the renormalized boundary length $\beta$. 
In the first order BF formulation, we use the boundary condition $\mathbf{B}\rvert_\partial= \mathbf{A}_\tau\rvert_\partial$. 
One proceeds by path integrating out the auxiliary field $\mathbf{B}$ in the bulk action, which localizes the model to flat $\mathfrak{sl}(2,\mathbb{R})$ gauge connections.
Plugged into the boundary term, the BF action \eqref{C4:BFactionprop} reduces to the holographic particle-on-SL$(2,\mathbb{R})$ description. 

Additionally, we impose the asymptotically AdS$_2$ (Brown-Henneaux \cite{Brown:1986nw}) boundary conditions on the BF fields: 
\begin{equation}
\label{eq:gaugeboundaryconditions}
    B_{-}=\frac{1}{\epsilon},
\end{equation} 
which further constrains the dynamics of the model, reducing it to the boundary gravitational degrees of freedom only.

In the gauge theoretic description, only the hyperbolic component corresponds to non-singular geometries, which is naturally achieved by restraining to the semigroup of positive group elements SL$^+(2,\mathbb{R})$ \cite{Blommaert:2018iqz}. Crucially, the only unitary irreducible representations in the Plancherel decomposition on SL$^+(2,\mathbb{R})$ are the continuous (principal) series representations labeled in terms of a continuous momentum label $k$: 
\begin{align}
    L^2(\text{SL}^+(2,\mathbb{R}))=\int_{\oplus} d\mu(k)\;P_k\otimes P_k, \qquad d\mu(k)=k\sinh(2\pi k),
\end{align}
whose Plancherel measure $d\mu(k)$ matches the Schwarzian density of states. Hence the representation matrix elements evaluated in these irreducible representations describe a complete set of wavefunctions of the gravitational Hilbert space.
The representation matrix elements themselves solve the Casimir eigenvalue problem of the associated particle-on-SL$(2,\mathbb{R})$ system, which can be found as follows. 
Parametrizing the SL$(2,\mathbb{R})$ group manifold in terms of Gauss-Euler coordinates ($\beta,\phi,\gamma$) as:
\begin{align}
    g=e^{\beta E^-}e^{2\phi H}e^{\gamma E^+},
\end{align}
one can write down the two mutually commuting left- and right-regular generators as
\begin{alignat}{3}
    \hat{L}_{E^-}&=-\partial_\beta, \qquad &\hat{L}_H&=-\frac{1}{2}\partial_\phi+\beta\partial_\beta,\qquad &\hat{L}_{E^+}&=\beta^2\partial_\beta-\beta \partial_\phi-e^{-2\phi}\partial_\gamma,\label{eq:leftsl(2,R)}\\ 
    \hat{R}_{E^+}&=\partial_\gamma,\qquad &\hat{R}_H&=\frac{1}{2}\partial_\phi-\gamma\partial_\gamma,\qquad &\hat{R}_{E^-}&=\gamma\partial_\phi-\gamma^2\partial_\gamma+e^{-2\phi}\partial_\beta.\label{eq:rightSL(2,R)}
\end{alignat}
One can then evaluate the mutual Casimir in this realization as: 
\begin{align}
    \hat{\mathcal{C}}_2&\equiv(\hat{L}_H)^2+\frac{1}{2}\left(\hat{L}_{E^-}\hat{L}_{E^+}+\hat{L}_{E^-}\hat{L}_{E^+}\right)=(\hat{R}_H)^2+\frac{1}{2}\left(\hat{R}_{E^-}\hat{R}_{E^+}+\hat{R}_{E^-}\hat{R}_{E^+}\right)\nonumber\\
    \label{eq:casbos}
    &= \frac{1}{4}\partial_\phi^2+\frac{1}{2}\partial_\phi-e^{-2\phi}\hat{L
}_{E^-}\hat{R}_{E^+}.
\end{align} 
To label the representation matrices uniquely, we should consider a maximal set of commuting operators in the set $\{\mathcal{C}_2,\hat{L}_X,\hat{R}_Y\}$. Since $\hat{L}_X$ and $\hat{R}_Y$ commute for any of the generators $X,Y$, we can choose to specify the representation labels by the eigenvalue of the Casimir $j(j+1) = -(k^2 +1/4), \,\, k\in \mathbb{R}^+$, and the eigenvalues of the left-regular generator $\hat{L}_{E^-}=-\nu$ and the right-regular generator $\hat{R}_{E^+}=-\lambda$. 
This implies that there exist matrix elements of the principal series representations which solve the mutual Casimir eigenvalue equation:
\begin{align}
    \hat{\mathcal{C}}_2\; R_{k,\nu\lambda}(g) = -\left(k^2+\tfrac{1}{4}\right)\;  R_{k,\nu\lambda}(g), \qquad k \in \mathbb{R}^+.
\end{align} 
Since the left- and right-regular generators implementing these boundary conditions take the simple one-derivative forms \eqref{eq:leftsl(2,R)}, \eqref{eq:rightSL(2,R)}, the total matrix element is only dependent on a single function of the hyperbolic parameter $\phi$, multiplied by exponential prefactors: 
\begin{align}
    R_{k,\nu\lambda}(g)= e^{\nu\beta}e^{-\lambda\gamma}\; \Psi_{k,\nu\lambda}(\phi).
\end{align} 
Mathematically, we will refer to the non-trivial matrix element $\Psi_{k,\nu\lambda}(\phi)$ of the Cartan generator $H$ as the Whittaker function, given for the \emph{semi}group SL$^+(2,\mathbb{R})$ by the mixed parabolic matrix element $\Psi_{k,\nu\lambda}(\phi)\equiv\braket{\nu\mid e^{2\phi H}\mid \lambda}$:
\begin{align}
    \Psi_{k,\nu\lambda}(\phi)=\,\, \raisebox{-0.3\height}{\includegraphics[height=0.8cm]{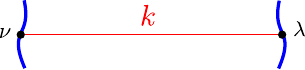}}\,\,=e^{-\phi}K_{2ik}\left(2\sqrt{\nu\lambda}e^{-\phi}\right),
\end{align} 
where we graphically draw the representation matrix element as an interval where each boundary endpoint carries one of the parabolic indices $\nu$ or $\lambda$. We will drop these subscripts $\nu\lambda$ in $\Psi_{k,\nu\lambda}$ from here on.

To go to gravity, we impose the Brown-Henneaux boundary conditions \eqref{eq:gaugeboundaryconditions} in terms of the left and right parabolic generators $L_{E^-}$ and $R_{E^+}$ as \cite{Blommaert:2018oro, Blommaert:2018iqz}:
\begin{align}
\label{eq:bosonicbrownbenneaux}
    \nu=\lambda\equiv \frac{1}{\epsilon},
\end{align} 
which reduces the Casimir \eqref{eq:casbos} to the Liouville Hamiltonian: 
\begin{align}
\label{eq:bosonicLiouville}
    \left(-\partial_\ell^2+\frac{1}{\epsilon^2}e^{-\ell}\right)\psi_k(\ell)=k^2\;\psi_k(\ell).
\end{align}
The corresponding gravitational solutions $\psi_k(\ell)=K_{2ik}(2e^{-\ell/2}/\epsilon)$ span the two-sided bulk Hilbert space of $1+1$d JT gravity \cite{Harlow:2018tqv} in terms of the bare geodesic length separating the two asymptotic boundaries $\ell=2\phi$  \eqref{eq:wilsonlinegeodesiclength}. We can absorb the UV-regulator in terms of the renormalized length variable connecting the two asymptotic AdS$_2$ boundaries
\begin{align}
    \ell_{\text{bare}} = \ell_{\text{ren}}-\ln\epsilon^2,
\end{align} 
leading to the renormalized gravitational wavefunctions $\psi_k(\ell)=K_{2ik}(2e^{-\ell/2})$, where we dropped the subscript on $\ell$ for clarity. In terms of the renormalized group parameters, this shift corresponds to $ \phi_{\text{bare}}=\phi_{\text{ren}}-\ln\epsilon$. 

This semigroup structure was shown in \cite{Fan:2021wsb} to generalize to the case of $\mathcal{N}=1$ JT supergravity, where the relevant structure describing the supergravitational amplitudes corresponds to the OSp$^+(1|2,\mathbb{R})$ semi-supergroup, which contains an SL$^+(2,\mathbb{R})$ bosonic subblock of positive entries. By restricting to this semigroup, all holonomies are automatically contained in the hyperbolic conjugacy class corresponding to non-singular geometries. Furthermore, this restriction again singles out the continuous principal series representations in the Plancherel decomposition, as the states in the supergravitational bulk Hilbert space. 

\subsection{Supergravity as a semigroup}
In this section, we will demonstrate how the underlying positive semigroup structure persists in $\mathcal{N}=2$ JT supergravity. In particular, we will demonstrate how the two-boundary supergravitational wavefunctions of the $\mathcal{N}=2$ JT supergravitational bulk Hilbert space are reproduced as constrained matrix elements of the semigroup OSp$^+(2|2,\mathbb{R})$. 

The positive semigroup OSp$^+(2|2,\mathbb{R})$ is defined by constraining the top-left SL$(2,\mathbb{R})$ bosonic subblock of the fundamental $2|2$-dimensional OSp$(2|2,\mathbb{R})$ group element to positive entries: 
\begin{align}
    \text{OSp}^+(2|2,\mathbb{R})\equiv \left\{\left[\begin{array}{c c | c c} 
	a & b & \alpha_1 & \beta_1\\ 
    c & d & \gamma_1 & \delta_1\\	
 \hline 
	\alpha_2 & \beta_2 & w & y\\
    \gamma_2 & \delta_2 & z & u\\
\end{array}\right]\in \text{OSp}(2|2,\mathbb{R})\; \Big\rvert\quad a,b,c,d>0\right\}. \label{eq:defsubsemigroup}
\end{align}
in terms of 8 bosonic supernumbers $a,b,c,d,w,y,z,u$ and 8 fermionic supernumbers \\$\alpha_{1,2},\beta_{1,2},\gamma_{1,2},\delta_{1,2}$ that satisfy several relations between them coming from preserving the orthosymplectic form (see e.g. \cite{Belaey:2023jtr} for details in our convention). We collect some information on the supergroup OSp$(2|2,\mathbb{R})$ in appendix \ref{appendix:representationtheory}. Note that in general, $a,b,c,d$ are supernumbers whose positivity is defined through the positivity of their bodies.\footnote{\label{footnote:positivity}We can expand any supernumber $z$ in a Grassmann expansion. We call the purely bosonic piece $z_0$ the body of the supernumber. The remainder is dubbed the soul, and can have both odd and even parity. Any supergroup number is positive iff its body is positive: $z>0\;\leftrightarrow\; z_0>0$.} In terms of the real Gauss-Euler decomposition \eqref{eq:GEreal}, this constrains only the parabolic coordinates $\gamma, \beta>0$, while $\phi$ can still take any value on the real number line. Conjugacy classes $M$ of OSp$(2|2,\mathbb{R})$ mirror those of the bosonic subgroup SL$(2,\mathbb{R}) \otimes  \text{U}(1)$, and are given in three inequivalent classes, distinguished by their trace in the $2\times 2$ bosonic upper diagonal block:
\begin{align}
\label{eq:hypclass}
&\text{hyperbolic:  }M =\text{diag}\left(\left[\begin{array}{c c} 
	\cosh \phi & \sinh \phi  \\ 
    \sinh \phi & \cosh \phi \end{array}\right],\left[\begin{array}{c c} 
	\cos \theta & -\sin \theta  \\ 
    \sin \theta & \cos \theta \end{array}\right]\right), \\
    &\text{parabolic: }M =\text{diag}\left(\left[\begin{array}{c c} 
	1 & \pm 1  \\ 
    0 & 1 \end{array}\right],\left[\begin{array}{c c} 
	\cos \theta & -\sin \theta  \\ 
    \sin \theta & \cos \theta \end{array}\right]\right), \\
    &\text{elliptic: }M =\text{diag}\left(\left[\begin{array}{c c} 
	\cos \phi & -\sin \phi  \\ 
    \sin \phi & \cos \phi \end{array}\right],\left[\begin{array}{c c} 
	\cos \theta & -\sin \theta  \\ 
    \sin \theta & \cos \theta \end{array}\right]\right).
\end{align}
Assuming now $a,b,c,d >0$, the (absolute value of the) trace of the $2 \times 2$ bosonic upper block is given by
\begin{equation}
\abs{a+d} = \abs{a + \frac{1}{a} + \frac{bc}{a} + \frac{\gamma_1\alpha_1 + \delta_1 \beta_1}{a}} > \abs{ a + \frac{1}{a}} \geq 2,
\end{equation}
where we used one of the relations $ad-bc - \gamma_1\alpha_1 - \delta_1 \beta_1 =1$ between the variables of an $\text{OSp}(2|2,\mathbb{R})$ matrix in \eqref{eq:defsubsemigroup}. We conclude that only matrices that are in the hyperbolic conjugacy class \eqref{eq:hypclass} are considered when restricting to the positive semigroup OSp$^+(2|2,\mathbb{R})$. This is physically required for Euclidean gravity since only these correspond geometrically to smooth surfaces.\footnote{Elliptic conjugacy classes correspond to conical defects, and parabolic conjugacy classes to cusp-like singularities. Both of these are to be excluded when restricting to smooth Euclidean geometries.}

To describe the gravitational Hilbert space, we are interested in the matrix elements in a specific set of representations. These are the continuous (principal) series representations, labeled by a continuous label $j$ and a discretized U$(1)$ charge quantum number $q$. We have mathematically constructed these representations for the full group $\text{OSp}(2|2,\mathbb{R})$ from first principles in \cite{Belaey:2023jtr} and refer to that work for technical details. Here we immediately present the result applied to the positive semigroup.
The natural carrier space of the principal series representations for the positive semigroup OSp$^+(2|2,\mathbb{R})$ is the set of square-integrable functions $f(x|\vartheta_1,\vartheta_2)\in L^2(\mathbb{R}^{+1|2})$ on the positive real superline 
\begin{align}
    \mathbb{R}^{+1|2} \equiv \{(x|\vartheta_1,\vartheta_2)\mid x>0\}.
\end{align}
We now define the principal series representation of $\text{OSp}^+(2|2,\mathbb{R})$ by the group action (with the group element parametrized as in \eqref{eq:defsubsemigroup}):
\begin{align}
\label{eq:semgroupact}
    f(x|&\vartheta_1,\vartheta_2) \to e^{2iq \psi} \, (bx+d+\beta_2\vartheta_1 + \delta_2 \vartheta_2 )^{2j} \\
&\times f\left(\frac{ax+c+\alpha_2 \vartheta_1 + \gamma_2 \vartheta_2}{bx+d+\beta_2\vartheta_1 + \delta_2 \vartheta_2}\Big\rvert-\frac{\alpha_1 x + \gamma_1 -w \vartheta_1 - z\vartheta_2}{bx+d+\beta_2\vartheta_1 + \delta_2 \vartheta_2}, -\frac{\beta_1x + \delta_1 - y \vartheta_1 - u \vartheta_2}{bx+d+\beta_2\vartheta_1 + \delta_2 \vartheta_2}\right). \nonumber
\end{align}  
The supernumber angle $\psi$ is defined in terms of the other variables as: 
\begin{align}
\label{eq:cos_psi}
\cos \psi &= -\beta_1 \vartheta_2 + u + \frac{\beta_1 x + \delta_1 - y \vartheta_1 - u\vartheta_2}{b x+d+\beta_2\vartheta_1 + \delta_2 \vartheta_2} (b\vartheta_2 + \delta_2), \\
 \label{eq:sin_psi}
 \sin \psi &= -\alpha_1 \vartheta_2 + z + \frac{\alpha_1  x + \gamma_1 - w \vartheta_1 - z\vartheta_2}{b x+d+\beta_2\vartheta_1 + \delta_2 \vartheta_2} (b\vartheta_2 + \delta_2).
\end{align}  
Some comments are in order. Firstly, compared to these irreps of the full group $\text{OSp}(2|2,\mathbb{R})$ \eqref{eq:gact}, we stripped the absolute value and sign factors which are always trivial for the positive semigroup. Secondly, unitarity of the principal series action restricts the continuous representation label to be purely imaginary $j=ik$, in terms of the momentum label $k\in \mathbb{R}^+$, and the charge label to be real $q\in\mathbb{R}$. The latter is further discretized $q\in \mathbb{Z}/2$ to correspond to the compact U$(1)$ R-symmetry group. Thirdly, the specifics of the parabolic induction procedure (as we used in \cite{Belaey:2023jtr}) identifies the auxiliary coordinates $(x\vert\vartheta_1,\vartheta_2)$ with the coordinates of the lower-triangular parabolic subgroup $\bar{N}$ in the decomposition of the group $G = \bar{N} MAN$, as $(x\vert\vartheta_1,\vartheta_2) \, \leftrightarrow \,(\beta \vert \theta_x^-,\theta_y^-)$, which indeed constrains $x>0$ since $\beta>0$ by definition. Finally, we note that the super-M\"obius transformation of an element $g\in \text{OSp}^+(2|2,\mathbb{R})$ on $x>0$ transforms it to the element $\frac{ax+c+\alpha_2 \vartheta_1 + \gamma_2 \vartheta_2}{bx+d+\beta_2\vartheta_1 + \delta_2 \vartheta_2}>0$,  which is still positive, since the positivity of a supernumber is determined entirely by its body (see footnote \ref{footnote:positivity}).

\subsection{Gravitational matrix elements}
\label{sec: Grav Matrix element}
We now consider the representation matrix elements in these representations, of a Gauss-Euler decomposed group element \eqref{eq:GEreal}:
\begin{align}\label{eq:GEmatrixelement}
     g&= e^{\sqrt{2}\theta^-_{y} F^-_y} e^{\sqrt{2}\theta^-_{x} F^-_x} e^{\beta E^-} e^{2\phi H} e^{2i \theta Z}e^{\gamma E^+} e^{\sqrt{2}\theta^+_{x} F^+_x} e^{\sqrt{2}\theta^+_{y} F^+_y},
\end{align}
which parametrizes the group manifold in terms of two real hyperbolic coordinates $\{\phi, \theta\}$, and six  parabolic coordinates $\{\beta, \gamma\mid\theta_x^-,\theta_y^-,\theta_x^+,\theta_y^+\}$. 

The generators here are taken in the principal series representation as \cite{Belaey:2023jtr}:
\begin{align}
2H&=-2j+2x\partial_x+\vartheta_2\partial_{\vartheta_2}+\vartheta_1\partial_{\vartheta_1}, \label{eq:generatorH}\\
    2iZ&=2iq-\vartheta_1\partial_{\vartheta_2}+\vartheta_2\partial_{\vartheta_1},\label{eq:generatorZ}\\
    E^+&=2jx-2iq\vartheta_1\vartheta_2-x\vartheta_2\partial_{\vartheta_2}-x\vartheta_1\partial_{\vartheta_1}-x^2\partial_x, \label{eq:generatorE+}\\
    E^-&=\partial_x, \label{eq:generatorE-}\\
    \sqrt{2}F_x^+&=2j \vartheta_1-2iq\vartheta_2-\vartheta_1\vartheta_2\partial_{\vartheta_2}-x\partial_{\vartheta_1}-x\vartheta_1\partial_{x},\label{eq:generatorFx+}\\
    \sqrt{2}F_x^-&=\partial_{\vartheta_1}+\vartheta_1\partial_x,\label{eq:generatorFx-}\\
    \sqrt{2}F_y^+&=-2j\vartheta_2 - 2iq\vartheta_1-\vartheta_1\vartheta_2\partial_{\vartheta_1}+x\partial_{\vartheta_2}+x\vartheta_2\partial_x,\label{eq:generatorFy+}\\
    \sqrt{2}F_y^-&=-\partial_{\vartheta_2}-\vartheta_2\partial_x. \label{eq:generatorFy-}
\end{align}
As differential operators, these generators are found to satisfy the opposite $\mathfrak{osp}(2|2,\mathbb{R})$ superalgebra, which differs by a sign in the anticommutation relations of the real matrix algebra \eqref{eq:fullrealalgebra}. 

The natural generalization of the Brown-Henneaux boundary conditions corresponding to a particle moving on AdS$_{2|4}$, instructs us to consider mixed parabolic type matrix elements. These matrix elements are evaluated between eigenstates which simultaneously diagonalize the respective outer parabolic factors to the left of the matrix element. Mathematically, one refers to these gravitational boundary eigenstates as Whittaker vectors, which we construct for the OSp$^+(2|2,\mathbb{R})$ semigroup in the next subsection.

\subsubsection{Whittaker vectors; or gravitational boundary eigenstates}
\label{eq:whitdefi}
We want to define the right Whittaker vector as diagonalizing the three right parabolic generators $E^+,F_x^+,F_y^+$ simultaneously while respecting their (opposite) algebra relations:
\begin{align} 
\label{eq:oppositealgebra}
          \{F_x^{+},F_x^{+}\}=\{F_y^{+},F_y^{+}\}=- E^{+},\qquad \{F_x^+,F_y^+\}=0.
\end{align}
The last relation in particular, $\{F_x^+,F_y^+\}=0$, prohibits us from diagonalizing directly all three generators in terms of bosonic eigenvalues. This was possible for the $\mathcal{N}=1$ superalgebra in \cite{Fan:2021wsb}, which only involves $E^+$ and $F^+$ with the relation $\{F^+,F^+\}=-E^+$, allowing a $\mathbb{Z}_2$ sign factor $\epsilon$ in the diagonalization of $F^+$. The crucial new ingredient that starts at $\mathcal{N}=2$, is the presence of multiple anticommuting fermionic generators.
Whereas we cannot diagonalize all generators simultaneously in terms of ordinary numbers, the strategy seems obvious to instead ``diagonalize'' the fermionic generators in terms of gamma matrices, for which the last expression of \eqref{eq:oppositealgebra} then simply becomes the Dirac/Clifford algebra.

Given these considerations, we define the right Whittaker vector in coordinate space with the notation $\braket{x,\vartheta_1,\vartheta_2\mid \lambda, \psi_3, \psi_4}$ by the following three relations:
\begin{align}
      E^+ \braket{x,\vartheta_1, \vartheta_2\mid \lambda, \psi_3,\psi_4}&= -\lambda \braket{x,\vartheta_1, \vartheta_2\mid\lambda, \psi_3,\psi_4}, \nonumber \\
      \label{eq:fermleft}
    \sqrt{2}F_x^+\;\braket{x,\vartheta_1,\vartheta_2\mid \lambda, \psi_3, \psi_4}&=-i\psi_3 \sqrt{\lambda}\;\braket{x,\vartheta_1,\vartheta_2\mid \lambda, \psi_3, \psi_4}, \nonumber \\
    \sqrt{2}F_y^+\;\braket{x,\vartheta_1,\vartheta_2\mid \lambda, \psi_3, \psi_4}&= -i\psi_4\sqrt{\lambda}\;\braket{x,\vartheta_1,\vartheta_2\mid \lambda, \psi_3, \psi_4}, 
\end{align}
with the real parameter $\lambda$ playing the role of the eigenvalue of $E^+$, reminiscent of the bosonic case. Additionally, we have introduced two real Majorana fermions $\psi_{3}$, $\psi_4$ satisfying the Dirac algebra: 
\begin{align}
\label{eq:cliffordalgebra}
    \{\psi_i,\psi_j\}=2\delta_{ij},
\end{align}
and which are taken to anticommute with all other Grassmann variables appearing in the group element and the carrier space later on. Note that we choose factors of $i$ in the fermionic eigenvalues here to diagonalize to the left of the eigenvector. This choice will lead in the end to the correct gravitational wavefunctions, and can be viewed as part of the definition of our construction. 

The above relations can be directly solved by the ket eigenfunction:
\begin{align}
    &\braket{x,\vartheta_1,\vartheta_2\mid\lambda, \psi_3,\psi_4}= \frac{1}{\sqrt{2\pi}}e^{-\lambda/x}x^{2j} e^{-i\psi_3\sqrt{\lambda} \vartheta_1/x}e^{i\psi_4\sqrt{\lambda} \vartheta_2/x}e^{-2iq \vartheta_1\vartheta_2/x}\nonumber\\ &\quad= \frac{1}{\sqrt{2\pi}}e^{-\lambda/x}x^{2j} \left(1-i\psi_3 \sqrt{\lambda} \frac{\vartheta_1}{x}+i\psi_4 \sqrt{\lambda} \frac{\vartheta_2}{x}-\psi_3\psi_4 \lambda\frac{\vartheta_1\vartheta_2}{x^2}-2iq \frac{\vartheta_1\vartheta_2}{x}\right).\label{eq:rightWhittakervector}
\end{align} 
The first line is written as a suggestive generalization of the bosonic answer, whereas in the second line we have expanded the expression in the Grassmann numbers.

One checks immediately that this expression is indeed consistent with the algebra relations \eqref{eq:oppositealgebra}: 
\begin{align}
(F_x^+)^2\braket{x,\vartheta_1,\vartheta_2\mid \hspace{-0.7mm}\lambda,\psi_3,\psi_4}&=(F_y^+)^2\braket{x,\vartheta_1,\vartheta_2\mid \hspace{-0.7mm}\lambda,\psi_3,\psi_4}= \tfrac{1}{2}\lambda \braket{x,\vartheta_1,\vartheta_2\mid \hspace{-0.7mm}\lambda,\psi_3,\psi_4}, \nonumber\\
F_x^+F_y^+\hspace{-0.7mm}\braket{x,\vartheta_1,\vartheta_2 \hspace{-0.7mm}\mid \hspace{-0.7mm}\lambda,\psi_3,\psi_4}&=-F_y^+F_x^+ \hspace{-0.7mm}\braket{x,\vartheta_1,\vartheta_2\mid \hspace{-0.7mm}\lambda,\psi_3,\psi_4}=\tfrac{1}{2}\psi_4\psi_3\lambda\braket{x,\vartheta_1,\vartheta_2\hspace{-0.7mm}\mid \hspace{-0.7mm}\lambda,\psi_3,\psi_4},
\end{align} 
where we have used that the successive action of the Grassmann-valued generators anticommute with the $\psi$'s.

We can analogously consider the left parabolic eigenvector that simultaneously diagonalizes the right adjoint actions of the bosonic left parabolic generator $E^-$ \eqref{eq:generatorE-} and both anticommuting fermionic left parabolic generators $F_{x}^-$ \eqref{eq:generatorFx-}, $F_y^-$ \eqref{eq:generatorFy-}, which satisfy the opposite algebra relations\begin{align}
    \{F_x^{-},F_x^{-}\}=\{F_y^{-},F_y^{-}\}= E^{-},\qquad \{F_x^-,F_y^-\}=0.
\end{align} 
The left Whittaker bra eigenfunction satisfying these conditions, is given by: 
\begin{align}
    \braket{\nu, \psi_1,\psi_2\mid x,\vartheta_1,\vartheta_2}&= \frac{1}{\sqrt{2\pi}}e^{-\nu x}e^{-\psi_1 \sqrt{\nu}\vartheta_1}e^{\psi_2\sqrt{\nu} \vartheta_2}\label{eq:leftWhittaker}\\&= \frac{1}{\sqrt{2\pi}}e^{-\nu x} \left(1-\psi_1 \sqrt{\nu}\vartheta_1+\psi_2 \sqrt{\nu}\vartheta_2+\psi_1\psi_2 \nu \vartheta_1\vartheta_2\right), \nonumber
\end{align}
which trivially diagonalizes the right adjoint action of $E^-=\partial_x$: 
\begin{align}
    \braket{\nu, \psi_1,\psi_2\mid x,\vartheta_1,\vartheta_2} E^-&= \braket{\nu, \psi_1,\psi_2\mid x,\vartheta_1,\vartheta_2} (-\overleftarrow{\partial_x})=\nu \braket{\nu, \psi_1,\psi_2\mid x,\vartheta_1,\vartheta_2},
\end{align} 
in terms of the real parameter $\nu$. We have also introduced two additional independent real Majorana fermions on the left boundary $\psi_1$, $\psi_2$ satisfying the Dirac algebra \eqref{eq:cliffordalgebra} while anticommuting with all other Grassmann numbers. In particular, they anticommute with the already introduced Majorana fermions $\psi_{3,4}$. One now readily checks that this eigenvector indeed simultaneously diagonalizes the right adjoint action\footnote{The adjoint action of the fermionic generators on general functions can be deduced from the product rule of Grassmann derivatives in the partial integration.} 
 of both fermionic generators to the left of the eigenvector:\footnote{\label{footnote:leftwhittakersign} Contrary to the right Whittaker vector, we do not introduce a factor $i$ in the fermionic eigenvalues, since this would diagonalize the fermionic generators to the right of the eigenvectors.}
\begin{align}
    \braket{\nu, \psi_1,\psi_2 \hspace{-0.7mm}\mid \hspace{-0.7mm} x,\vartheta_1,\vartheta_2}\hspace{-0.7mm}\sqrt{2} F_x^- \hspace{-0.7mm} &= \hspace{-0.7mm}\braket{\nu, \psi_1,\psi_2 \hspace{-0.7mm}\mid \hspace{-0.7mm} x,\vartheta_1,\vartheta_2} \hspace{-0.7mm}(\hspace{-0.7mm}\overleftarrow{\partial}\hspace{-0.7mm}_{\vartheta_1}\hspace{-1.4mm}-\hspace{-0.7mm}\vartheta_1 \hspace{-0.7mm}\overleftarrow{\partial}\hspace{-0.7mm}_x)\label{eq:adjointrightaction}\hspace{-0.7mm}=\hspace{-0.7mm} -\psi_1 \sqrt{\nu} \braket{\nu, \psi_1,\psi_2 \hspace{-0.7mm}\mid \hspace{-0.7mm} x,\vartheta_1,\vartheta_2}\nonumber\\
     \braket{\nu, \psi_1,\psi_2 \hspace{-0.7mm} \mid \hspace{-0.7mm} x,\vartheta_1,\vartheta_2}\hspace{-0.7mm}\sqrt{2}F_y^- \hspace{-0.7mm}&= \hspace{-0.7mm}\braket{\nu, \psi_1,\psi_2\hspace{-0.7mm}\mid \hspace{-0.7mm} x,\vartheta_1,\vartheta_2} \hspace{-0.7mm}(\hspace{-0.7mm}-\hspace{-0.7mm}\overleftarrow{\partial}\hspace{-0.7mm}_{\vartheta_2}\hspace{-1.4mm}+\hspace{-0.7mm}\vartheta_2 \hspace{-0.7mm}\overleftarrow{\partial}\hspace{-0.7mm}_x)  
     \hspace{-0.7mm}= \hspace{-0.7mm}-\psi_2 \sqrt{\nu}\braket{\nu, \psi_1,\hspace{-0.7mm}\psi_2\hspace{-0.7mm}\mid \hspace{-0.7mm}x,\vartheta_1,\hspace{-0.7mm}\vartheta_2}.
\end{align} 
This is again consistent with the algebra relations: 
\begin{align}
     \braket{\nu, \psi_1,\psi_2\hspace{-0.7mm}\mid\hspace{-0.7mm} x,\vartheta_1,\vartheta_2} (F_x^-)^2 \hspace{-0.7mm}&=
      \braket{\nu, \psi_1,\psi_2\hspace{-0.7mm}\mid \hspace{-0.7mm}x,\vartheta_1,\vartheta_2} (F_y^-)^2=  \tfrac{1}{2}\nu \braket{\nu, \psi_1,\psi_2\hspace{-0.7mm}\mid\hspace{-0.7mm} x,\vartheta_1,\vartheta_2} \nonumber \\
       \braket{\nu, \psi_1,\psi_2\hspace{-0.7mm}\mid \hspace{-0.7mm}x,\vartheta_1,\vartheta_2} F_x^-F_y^- \hspace{-0.7mm}&=  \braket{\nu, \psi_1,\psi_2\hspace{-0.7mm}\mid \hspace{-0.7mm}x,\vartheta_1,\vartheta_2}(-F_y^- F_x^-) \hspace{-0.7mm}=  \tfrac{1}{2}\psi_1\psi_2\nu  \braket{\nu, \psi_1,\psi_2\hspace{-0.7mm}\mid\hspace{-0.7mm} x,\vartheta_1,\vartheta_2}.
\end{align}
The left and right Whittaker vectors are related to each other. In particular, the right Whittaker vector is transformed to a left Whittaker vector by a principal series  transformation \eqref{eq:gact} of the group element $\omega$ given by\footnote{Since $\omega$ is not part of the semigroup, we use the action on the right Whittaker vector of the full group OSp$(2|2,\mathbb{R})$, which contains an additional absolute value in $|x|^{2j}$. 
For this, it is important that we work with a \emph{sub}semigroup.}
\begin{align} \label{eq:omega}
    \omega=\begin{bmatrix}
        0&1&0&0\\-1&0&0&0\\0&0&1&0\\0&0&0&1
    \end{bmatrix}.
\end{align} 
From \eqref{eq:sin_psi} we can deduce the associated supernumber angle of this transformation as $\psi=-\frac{\vartheta_1\vartheta_2}{x}$. Acting on the right Whittaker vector of the full group yields: \begin{align}
    & \omega\;\cdot\; e^{-\lambda/x}|x|^{2j} \left(1-i\psi_3 \sqrt{\lambda} \frac{\vartheta_1}{x}+i\psi_4 \sqrt{\lambda} \frac{\vartheta_2}{x}-\psi_3\psi_4 \lambda\frac{\vartheta_1\vartheta_2}{x^2}-2iq \frac{\vartheta_1\vartheta_2}{x}\right)\nonumber\\ &= e^{\lambda x}\left( 1+i\psi_3\sqrt{\lambda}\vartheta_1-i\psi_4\sqrt{\lambda}\vartheta_2-\psi_3\psi_4 \lambda\vartheta_1\vartheta_2\right).
\end{align}
This is identified with the left Whittaker vector \eqref{eq:leftWhittaker}:\begin{align}
    \braket{-\lambda,-\psi_3,-\psi_4\mid x,\vartheta_1,\vartheta_2}=e^{\lambda x}\left( 1+\psi_3\sqrt{-\lambda}\vartheta_1-\psi_4\sqrt{-\lambda}\vartheta_2-\psi_3\psi_4 \lambda\vartheta_1\vartheta_2\right).
\end{align}
In the fundamental Gauss-Euler parametrization \eqref{eq:GEexpli}, the element $\omega$ corresponds to $\phi\rightarrow-\infty$, $\gamma=-\beta=e^{-\phi}$,  $\theta=0$, $\theta_x^+=\theta_y^+=\theta_x^-=\theta_y^-=0$. Since this group element transforms a right Whitakker vector into a left Whittaker vector, which each simultaneously diagonalize \emph{different} parabolic generators on both boundaries, it determines the analogue of the unit group element in the mixed-parabolic set-up. This will turn out to be the relevant group element describing the no-boundary condition of the Hartle-Hawking state in section \ref{sec:tfd}. 

\subsubsection{Whittaker function; or two-boundary gravitational wavefunction}
Since both the left and right Whittaker vectors diagonalize the individual parabolic generators to the left as in \eqref{eq:adjointrightaction} and \eqref{eq:fermleft}, the mixed parabolic matrix element \eqref{eq:GEmatrixelement} reduces to the Whittaker function of the remaining $H$ and $Z$ Cartan generators, multiplied by the exponential parabolic prefactors to the left:  
\begin{align}
    \braket{\nu, \psi_1,\psi_2\mid g\mid \lambda, \psi_3,\psi_4}=e^{-\theta_x^- \psi_1\sqrt{\nu}} e^{-\theta_y^- \psi_2 \sqrt{\nu}}e^{-i\theta_x^+\psi_3\sqrt{\lambda}}e^{-i\theta_y^+\psi_4\sqrt{\lambda}}e^{\beta\nu}e^{-\lambda \gamma}\; \Psi_{j,q}(\phi,\theta), \label{eq:parabolicfactorization}
\end{align} 
where the overlap in the Whittaker function $\Psi_{j,q}(\phi,\theta)\equiv\braket{\nu,\psi_1,\psi_2\mid e^{2\phi H}e^{2i\theta Z}\mid \lambda,\psi_3,\psi_4}$ is deduced by inserting the Whittaker vectors \eqref{eq:rightWhittakervector} and \eqref{eq:leftWhittaker} and integrating over the carrier space coordinates:
\begin{align}
    &\Psi_{j,q}(\phi,\theta)=\frac{1}{2\pi} \int_0^{\infty}dx \int d\vartheta_1d\vartheta_2 \;e^{-\nu x}(1-\psi_1 \sqrt{\nu}\vartheta_1+\psi_2 \sqrt{\nu}\vartheta_2+\psi_1\psi_2\nu\vartheta_1\vartheta_2)\nonumber\\ &\times e^{2\phi H}e^{2i\theta Z}\;\cdot\; e^{-\lambda/x}x^{2j}\Big(1-i\psi_3\sqrt{\lambda}\frac{\vartheta_1}{x}+i\psi_4\sqrt{\lambda}\frac{\vartheta_2}{x}-\psi_3\psi_4 \lambda\frac{\vartheta_1\vartheta_2}{x^2}-\frac{2iq}{x}\vartheta_1\vartheta_2\Big).
\end{align}
Crucially, the integration range is over the positive half-line $x>0$ only, in the restriction to the positive semigroup OSp$^+(2|2,\mathbb{R})$. Integrating over $\vartheta_1$ and $\vartheta_2$, and using the principal series action of $e^{2\phi H}$ and $e^{2i\theta Z}$ given by \eqref{eq:exponentiatedgroupactions}, finally leads to: 
 \begin{align}
    &\Psi_{j,q}(\phi,\theta) = \frac{e^{2iq\theta}}{\pi}\Big[-\psi_3\psi_4 \frac{\lambda^{j+1/2}}{\nu^{j-1/2}}e^{-\phi}K_{2j-1}(2\sqrt{\nu\lambda}e^{-\phi})-2iq \left(\frac{\lambda}{\nu}\right)^jK_{2j}(2\sqrt{\nu\lambda}e^{-\phi})\nonumber\\& +\frac{\lambda^{j+1/2}}{\nu^{j-1/2}}e^{-\phi}K_{2j}(2\sqrt{\nu\lambda}e^{-\phi})\Big(-i\psi_1\psi_3\sin\theta+i\psi_1\psi_4\cos\theta-i\psi_2\psi_3\cos\theta-i\psi_2\psi_4 \sin\theta\big)\nonumber\\ &+\psi_1\psi_2 \frac{\lambda^{j+1/2}}{\nu^{j-1/2}}e^{-\phi}K_{2j+1}(2\sqrt{\nu\lambda}e^{-\phi}) \Big],\label{generalWhittakerfunction}
\end{align} 
using the integral representation of the modified Bessel functions of the second kind \begin{align}\label{eq:integralBessel}
    \int_0^\infty dx\; e^{-\nu x}e^{-\lambda/x}x^{2j-1}=2\left(\frac{\lambda}{\nu}\right)^jK_{2j}(2\sqrt{\nu\lambda}).
\end{align} 
Specializing to gravity, we implement the asymptotic AdS$_{2|4}$ boundary conditions corresponding to the gravitational $\mathfrak{sl}(2,\mathbb{R})$ subblock \eqref{eq:bosonicbrownbenneaux} and set:
\begin{align}\label{eq:asymototicbrownhenneaux}
    \lambda=\nu=\frac{1}{\epsilon},\qquad \epsilon\rightarrow0,
\end{align} 
in terms of the near-boundary UV-parameter $\epsilon$. This leads to the final gravitational wavefunction in terms of the renormalized hyperbolic parameter $\phi_{\text{bare}}\rightarrow\phi_{\text{ren}}-\ln\epsilon$:
\begin{align}
    &\Psi_{j,q}(\phi,\theta)=\raisebox{-0.3\height}{\includegraphics[height=0.8cm]{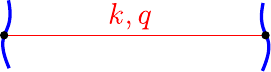}}\nonumber\\
    &=\frac{e^{2iq\theta}}{\pi}\Big[-\psi_3\psi_4 e^{-\phi}K_{2j-1}(2e^{-\phi})-2iq K_{2j}(2e^{-\phi})+\psi_1\psi_2 e^{-\phi}K_{2j+1}(2e^{-\phi})\nonumber\\& +e^{-\phi}K_{2j}(2e^{-\phi})\Big(-i\psi_1\psi_3\sin\theta+i\psi_1\psi_4\cos\theta-i\psi_2\psi_3\cos\theta-i\psi_2\psi_4 \sin\theta\Big) \Big]. \label{eq:finalWhittakerfunction}
\end{align} 
This expression is our main result. We demonstrate in the next subsection that these gravitational wavefunctions indeed diagonalize the two-sided $\mathcal{N}=2$ Liouville Hamiltonian with energy $E(k,q)=k^2+q^2$, matching our result with that found in \cite{Lin:2022zxd}.

\subsection{Hamiltonian reduction and Liouville quantum mechanics} \label{sec:hamreduction}
The gravitational Whittaker functions implementing the Brown-Henneaux boundary conditions are solutions to the $\mathcal{N}=2$ Liouville minisuperspace Hamiltonian \cite{Lin:2022zxd}:
\begin{align}
\label{eq:N=2Liouville2}
    \Big(-\partial_\ell^2-\frac{1}{4}\partial_\theta^2+e^{-\ell} +ie^{-\ell/2-i\theta}\;\overline{\psi}_-\psi_++ie^{-\ell/2+i\theta}\;\psi_-\overline{\psi}_+\Big)\Psi_{k,q}(\ell,\theta)=E(k,q)\; \Psi_{k,q}(\ell,\theta),
\end{align}
 in terms of the renormalized geodesic length $\ell=2\phi$, U$(1)$-holonomy variable $\theta$ and energy variable $E(k,q)=k^2+q^2$. This equation descends from the harmonic analysis of the (left- or right-) regular Casimir eigenvalue equation: \begin{align}\label{eq:casimireigenvalueequation}
    \hat{\mathcal{C}}_2\; \braket{\nu, \psi_1,\psi_2\mid g\mid \lambda, \psi_3,\psi_4}=(j^2-q^2)\braket{\nu, \psi_1,\psi_2\mid g\mid \lambda, \psi_3,\psi_4}.
\end{align} 
Since the principal series representations are irreducible, the action of the quadratic Casimir is diagonalized as a consequence of Schur's lemma \eqref{eq:quadcas}: $\mathcal{C}_2=j^2-q^2$.
Since $j=ik$ is purely imaginary as a consequence of unitarity, this is identified with the energy propagating inside the bulk: $E=|\mathcal{C}_2|=k^2+q^2$.

Choosing mixed parabolic boundary conditions and the asymptotic value of the AdS$_{2|4}$ Brown-Henneaux boundary conditions \eqref{eq:asymototicbrownhenneaux}, the Casimir eigenvalue equation \eqref{eq:casimireigenvalueequation} naturally reduces to the Liouville Hamiltonian acting on the Whittaker function, if we choose to diagonalize the parabolic generators to the left of the matrix element, as in \eqref{eq:parabolicfactorization}. We leave a more technical discussion to appendix \ref{sec:hamiltonianreduction}. Additionally, we work out the Hamiltonian reduction of the $\mathfrak{osp}(2|2,\mathbb{R})$ superconformal algebra to an emergent $\mathcal{N}=(2,2)$ supersymmetric quantum mechanical algebra, after implementing these boundary conditions.

\subsection{$\mathcal{N}=2$ multiplet representation and relation with earlier work}
To connect with earlier work, we can construct two pairs of conjugate lightcone fermions $\psi_-$, $\overline{\psi}_-$, and $\psi_+$, $\overline{\psi}_+$ through
\begin{equation}
\label{eq:fermiontransf}
    \begin{aligned}
        \psi_-&\equiv\tfrac{1}{2}\left(\psi_1+i\psi_2\right), \qquad \overline{\psi}_-\equiv\tfrac{1}{2}\left(\psi_1-i\psi_2\right), \\ \psi_+&\equiv\tfrac{1}{2}\left(\psi_3+i\psi_4\right), \qquad \overline{\psi}_+\equiv\tfrac{1}{2}\left(\psi_3-i\psi_4\right).
    \end{aligned}
\end{equation}
 These lightcone combinations correspond with the four respective fermionic lightcone generators $F^-$, $\overline{F}^-$, $F^+$,$\overline{F}^+$ defined from \eqref{eq:generatorstransf}. Using the real Dirac algebra \eqref{eq:cliffordalgebra}, the lightcone combinations satisfy
 \begin{align}
    \{\psi_-,\overline{\psi}_-\}=1, \qquad \{\psi_+,\overline{\psi}_+\}=1.
    \label{eq:diracAlgebra}
\end{align} 
These fermions make up a four-dimensional $\mathcal{N}=2$ fermionic multiplet by acting with all possible raising combinations on the vacuum state $\ket{0}$: \begin{align}\label{eq:multiplet}
    \ket{0}, \qquad \overline{\psi}_-\ket{0}, \qquad \overline{\psi}_+\ket{0}, \qquad \overline{\psi}_+\overline{\psi}_- \ket{0}.
\end{align} 
This vacuum state is by definition annihilated by the lowering operators on each side: \begin{align}
    \psi_-\ket{0}= \psi_+\ket{0}=0.
\end{align} 

We can minimally realize the four lightcone fermions spanning the four-dimensional multiplet in a $4\times 4$ matrix representation, using the Dirac algebra \eqref{eq:diracAlgebra}:
\begin{alignat}{2}
\label{eq:lightconefermions}
    \overline{\psi}_-=\begin{bmatrix}
    0&0&0&0\\1&0&0&0\\0&0&0&0\\0&0&-1&0
    \end{bmatrix}, \, \psi_-=\begin{bmatrix}
        0&1&0&0\\ 0&0&0&0\\ 0&0&0&-1\\0&0&0&0
    \end{bmatrix},  \, \overline{\psi}_+=\begin{bmatrix}
        0&0&0&0\\ 0&0&0&0\\ 1&0&0&0\\ 0&1&0&0
    \end{bmatrix}, \, \psi_+= \begin{bmatrix}
        0&0&1&0\\ 0&0&0&1\\ 0&0&0&0\\0&0&0&0
    \end{bmatrix}
\end{alignat} 
where the rows and columns are ordered as:
\begin{align}
    \ket{0}=\begin{bmatrix}
        1\\0\\0\\0
    \end{bmatrix}, \qquad \overline{\psi}_-\ket{0}=\begin{bmatrix}
        0\\1\\0\\0
    \end{bmatrix}, \qquad \overline{\psi}_+\ket{0}=\begin{bmatrix}
        0\\0\\1\\0
    \end{bmatrix}, \qquad \overline{\psi}_+\overline{\psi}_-\ket{0}=\begin{bmatrix}
        0\\0\\0\\1
    \end{bmatrix}.
\end{align} 
Transforming back to the real basis of $\psi_{1,2,3,4}$ in terms of four-dimensional gamma matrices, we can write the wavefunction \eqref{eq:finalWhittakerfunction} in the matrix multiplet representation as
\begin{align}\label{eq:generalmatrixexpression}
    \Psi_{j,q}(\phi,\theta)=\raisebox{-0.3\height}{\includegraphics[height=0.8cm]{gravprop4.pdf}}=\frac{e^{2iq\theta}}{\pi}\begin{bmatrix}
        f_1&0&0&0\\ 0&g_1&h_1&0\\0&g_2&h_2&0\\0&0&0&f_2
    \end{bmatrix},
\end{align} 
with the matrix entries
\begin{align}
    f_1&=-2iqK_{2j}(2e^{-\phi})+i e^{-\phi}(K_{2j+1}(2e^{-\phi})-K_{2j-1}(2e^{-\phi})) \label{eq:entryf1}, \\
    g_1&=-2iqK_{2j}(2e^{-\phi})-ie^{-\phi}(K_{2j+1}(2e^{-\phi})+K_{2j-1}(2e^{-\phi})),\\
    g_2&= 2e^{-\phi}e^{i\theta}K_{2j}(2e^{-\phi}),\\
    h_1&= 2e^{-\phi}e^{-i\theta}K_{2j}(2e^{-\phi}),\\
    h_2&= -2iq  K_{2j}(2e^{-\phi}) +ie^{-\phi}(K_{2j+1}(2e^{-\phi})+K_{2j-1}(2e^{-\phi})),\\
    f_2&= -2iq K_{2j}(2e^{-\phi})+ie^{-\phi}(K_{2j-1}(2e^{-\phi})-K_{2j+1}(2e^{-\phi})). \label{eq:entryf2}
\end{align} 
Using this matrix representation, we can identify each column of \eqref{eq:generalmatrixexpression} as an independent eigenstate of the Liouville eigenvalue problem, by multiplying with the respective states of the multiplet on both sides.
Using the recurrence relation of the Bessel functions:
\begin{equation}
\label{besselrecursion}
    e^{-\phi}\left(K_{2j+1}(2e^{-\phi})-K_{2j-1}(2e^{-\phi})\right)=2j K_{2j}(2e^{-\phi}),
\end{equation}
we can identify the four linearly independent states:
\begin{align}
    &\ket{F^+}=2i\frac{e^{2iq\theta}}{\pi}\left(j-q\right)K_{2j}(2e^{-\phi})\;\ket{0}     \label{eq:states1}, \nonumber \\
    &\ket{H}=\frac{e^{2iq\theta}}{\sqrt{\pi}}\Big[\hspace{-1.4mm}-\hspace{-0.7mm}i\hspace{-0.7mm}\left(2e^{-\phi}K_{2j-1}(2e^{-\phi})\hspace{-0.7mm}+\hspace{-0.7mm}(2q\hspace{-0.7mm}+\hspace{-0.7mm}2j) K_{2j}(2e^{-\phi})\right)\overline{\psi}_- \hspace{-0.7mm}+2 e^{-\phi+i\theta} K_{2j}(2e^{-\phi})\overline{\psi}_+\Big]\hspace{-1mm}\ket{0}, \nonumber \\ 
    &\ket{L}= \frac{e^{2iq\theta}}{\sqrt{\pi}}\Big[i\hspace{-0.7mm}\left(2e^{-\phi} K_{2j-1}(2e^{-\phi})\hspace{-0.7mm}+\hspace{-0.7mm}(2j\hspace{-0.7mm}-\hspace{-0.7mm}2q) K_{2j}(2e^{-\phi})\right)\overline{\psi}_+ \hspace{-0.7mm}+\hspace{-0.7mm}2 e^{-\phi-i\theta}K_{2j}(2e^{-\phi}) \overline{\psi}_-\Big]\hspace{-0.7mm}\ket{0}, \nonumber \\
    &\ket{F^-}=2i\frac{e^{2iq\theta}}{\pi}\left(-j-q\right)K_{2j}(2e^{-\phi})\;\overline{\psi}_+\overline{\psi}_-\ket{0}.
\end{align} 
 Due to the fact that $(\pm j-q)/\sqrt{E}$ (with $E=E(k,q)=k^2+q^2$) is a phase factor, the bottom $\ket{F^+}$ and top $\ket{F^-}$ states in the multiplet are equivalent to: 
\begin{align}
    \ket{F^+}\simeq 2i\frac{e^{2iq\theta}}{\pi}\sqrt{E}\;K_{2j}(2e^{-\phi})\;\ket{0}, \qquad \ket{F^-}\simeq2i\frac{e^{2iq\theta}}{\pi}\sqrt{E}\;K_{2j}(2e^{-\phi})\;\overline{\psi}_+\overline{\psi}_-
    \;\ket{0}.
\end{align} 
Up to an overall representation-independent prefactor, these states reproduce the continuous energy eigenstates of the $\mathcal{N}=2$ Liouville minisuperspace problem, as determined in \cite{Lin:2022zxd}.

It is worthwhile at this stage to state the benefit of our representation theoretic approach compared to other approaches. 
Our main benefit is that, comparing to directly diagonalizing the Liouville Hamiltonian, the overall \emph{and} relative prefactors of the different states in the multiplet are \emph{fixed} by the normalization of the left and right Whittaker vectors. This leads to an unambiguous normalization that allows a bottom-up determination of the gravitational density of states (Plancherel measure), as we will write down in the next section, that matches earlier work but crucially without invoking that earlier work to fix the precise normalization of the different states. This issue of the relative normalization only appears when supersymmetry is included. A second benefit is that by phrasing the problem fully in terms of the representation theory of the underlying supergroup, our techniques in principle also apply for more general superconformal algebras that contain gravity. We will come back to this in the concluding section \ref{s:concl}.

As mentioned earlier, the structure of a non-trivial Dirac algebra for the fermionic generators only arises starting with $\mathcal{N}=2$. However, we can also formulate the $\mathcal{N}=1$ case in this language. In appendix \ref{appendix:N=1} we rederive the gravitational matrix elements according to the conventions of the fermionic multiplet of this paper. As an addition to \cite{Fan:2021wsb}, we discuss the different parity sectors and collect formulas for the reduction of the Casimir operator of OSp$(1|2,\mathbb{R})$ to the $\mathcal{N}=1$ Liouville minisuperspace Hamiltonian and the associated supercharges formulated in \cite{Douglas:2003up}. This serves as a simpler example of the techniques presented in this work.

\section{Supergravity amplitudes and applications}
\label{s:applications}
In this section, we will apply our mixed parabolic representation matrix element from the previous section to analyze concrete (super)gravitational amplitudes in the $\mathcal{N}=2$ JT supergravity model. Our aim is to develop the bulk understanding of the representation theoretic objects, as has been done in the past for the $\mathcal{N}=0$ and $\mathcal{N}=1$ cases in \cite{Blommaert:2018oro, Blommaert:2018iqz, Fan:2021wsb}. 

\subsection{Gravitational density of states}\label{section:densityofstates}
Given the gravitational Whittaker function \eqref{eq:generalmatrixexpression} in the matrix multiplet representation with that precise normalization factor, the gravitational density of states $\rho(k,q)$, or Plancherel measure in the representation theory language, is obtained by calculating the overlap over the group coordinates
\begin{equation}
    \int_{0}^{2\pi}d\theta\int_{-\infty}^{+\infty}d\phi ~\Psi_{j',q'}^\dag(\phi,\theta)\;\Psi_{j,q}(\phi,\theta)\equiv \frac{\delta_{q,q'}\delta(k-k')}{\rho(k,q)}\times\mathbb{1}_{4\times4},
    \label{eq:integralPsi}
\end{equation}
and reading off the factor $\rho(k,q)$ on the RHS. In this expression, we have used the appropriate flat Haar measure factor on the supergroup manifold, as we explicitly derived in Appendix \ref{subsection:HaarMeasure}. Note that the renormalized $\phi$ variable can be negative and ranges over the entire real axis in the restriction to the semigroup.
Making use of the orthogonality property between modified Bessel functions
\begin{equation}
    \int_{-\infty}^{+\infty} d\phi K_{2ik}(2e^{-\phi})K_{2ik'}(2e^{-\phi})=\frac{\pi^2\delta(k-k')}{8k \sinh(2\pi k)},
    \label{eq:ortBesselK}
\end{equation}
and the general relation
\begin{equation}
\int_{-\infty}^{+\infty} d\phi\; e^{-\alpha\phi}K_{2ik}(2e^{-\phi})K_{2ik'}(2e^{-\phi})=\frac{1}{8\Gamma(\alpha)}\Gamma\left(\frac{\alpha\pm 2ik\pm 2ik'}{2}\right),
    \label{eq:generic_alpha}
\end{equation}
it is straightforward to verify that
\begin{equation}
 \int_{0}^{2\pi}d\theta\int_{-\infty}^{+\infty}d\phi ~\Psi_{j',q'}^\dag(\phi,\theta)\Psi_{j,q}(\phi,\theta)=\frac{\pi(k^2+q^2)}{k\sinh2\pi k}\delta_{q,q'}\delta(k-k')\times\mathbb{1}_{4\times4}.
\end{equation}
When taking the product between Whittaker functions in \eqref{eq:integralPsi}, only the diagonal terms survive after integrating over $\theta$ and $\phi$. These expressions are identified with the overlap between the states $\braket{F^{+}\mid F^{+}}$, $\braket{H\mid H}$, $\braket{L\mid L}$ and $\braket{F^{-}\mid F^{-}}$ respectively, using the ``lightcone'' algebra relations \eqref{eq:diracAlgebra}. All the off-diagonal terms vanish, as a consequence of the orthogonality between the $\ket{L}$ and $\ket{H}$ states $\braket{L\mid H}=\braket{H\mid L}=0$. 
With the energy of the system given by $ E(k,q)=k^2+q^2$, 
the gravitational density of states reads
\begin{equation}
\label{eq:desnityofstates}
\rho(k,q)=\frac{k\sinh 2\pi k}{\pi E(k,q)}.
\end{equation}
This is indeed the known continuous part of the $\mathcal{N}=2$ super-Schwarzian density of states \cite{Stanford:2017thb,Mertens:2017mtv}.
\subsection{Gravitational two-boundary propagator}\label{sec:propagator}
We start by considering the gravitational propagator of two-boundary states. 

The two-sided gravitational wavefunctions can be used to write down propagation amplitudes from an initial state characterized by $\phi_i,\theta_i$, to a final state characterized by $\phi_f,\theta_f$, over a Euclidean time span $\tau$ as:
\begin{align}
    \raisebox{-0.45\height}{\includegraphics[height=2.5cm]{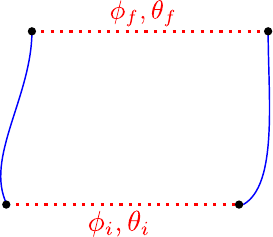}}=\braket{\phi_f,\theta_f \hspace{-0.7mm}\mid \hspace{-0.7mm} e^{-\tau H } \hspace{-0.7mm} \mid \hspace{-0.7mm}\phi_i,\theta_i}=\hspace{-1.4mm}\sum_{q \in \mathbb{Z}/2}\int_0^{+\infty} \hspace{-2.8mm} dk \braket{\phi_f,\theta_f \hspace{-0.7mm}\mid \hspace{-0.7mm} k,q} \braket{k,q \hspace{-0.7mm}\mid \hspace{-0.7mm}\phi_i,\theta_i} e^{-\tau E(k,q)},
\end{align}
where according to the Peter-Weyl theorem, the (normalized) representation matrices that propagate inside the bulk furnish (the continuous part of) the two-sided basis of the square-integrable functions on OSp$(2|2,\mathbb{R})$, which can be represented in the matrix multiplet representation as \eqref{eq:generalmatrixexpression}:
\begin{align}
    \braket{\phi, \theta\mid k,q}\equiv\sqrt{\rho(k,q)}\;\Psi_{k,q}(\phi,\theta)= \sqrt{\rho(k,q)}\;\frac{e^{2iq\theta}}{\pi}\begin{bmatrix}
        f_1&0&0&0\\ 0&g_1&h_1&0\\0&g_2&h_2&0\\0&0&0&f_2
    \end{bmatrix},
\end{align}  
together with the Hermitian conjugate $\braket{k,q\mid \phi,\theta}=\braket{\phi,\theta\mid k,q}^\dag$.

Initial and final states are characterized by the parameters $\phi$ and $\theta$ corresponding to $H$ and $Z$ respectively. The physical interpretation of these parameters is in terms of a Wilson line connecting both boundaries along a curve $\mathcal{C}$ as 
\begin{align}
    \mathcal{W}(\mathcal{C})=\exp\left(-\int_{\mathcal{C}} dx^M \mathbf{A}_M \right)\simeq e^{2\phi H}e^{2i\theta Z},
\end{align}
illustrated for e.g. the initial state as
\begin{align}
    \raisebox{-0.45\height}{\includegraphics[height=2.5cm]{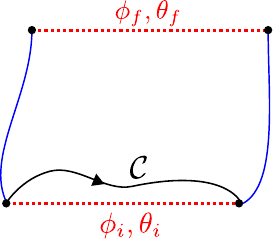}}
\end{align}
The parameters $\theta_{i,f}$ characterize a U(1) phase difference between both boundaries at the respective initial and final time. The physical interpretation for the parameter $\phi$ is well-known and corresponds to the geodesic distance between both boundary endpoints in the bulk AdS$_2$ geometry. 

It is illuminating to see this gravitational interpretation of the group theory variables directly as follows. We first go back to the bosonic $\mathcal{N}=0$ JT model and come back to the supersymmetric versions below. We consider the boundary-to-boundary holonomy variable
\begin{equation}
\int_{\mathcal{C}}dx^\mu \mathbf{A}_\mu 
\end{equation}
with the usual dictionary $\mathbf{A}_\mu = (e^1_\mu, e^2_\mu, \omega_\mu)$. We now perform the following steps. Firstly, we deform the Wilson line to be along the geodesic that connects these boundary points. For any curve, the extrinsic curvature trace $K$ can be written as:\footnote{See e.g. appendix D of \cite{Ebert:2022cle} for a nice derivation.}
\begin{equation}
K = \partial_a n^a + e^\mu_a \omega^a_{\mu b}n^b,
\end{equation}
where $n$ is a unit-normalized ($n_a n^a =1$) vector field perpendicular to the curve $C$. $K$ is invariant under local Lorentz transformations. Such local Lorentz transformation acts on the unit vector field $n_{a}(x)$ as $n_{a}\rightarrow O^{ab}(x)n_b(x)$ where $O^{ab}(x) \in \text{SO}(2)$.  There exists a choice of local Lorentz frame for which the resulting term $\partial_a n^a$ vanishes. 
This is precisely when the frame field is aligned with the curve $\mathcal{C}$ with one axis parallel to the curve (say $a=1$) and one perpendicular (say $a=2$) to the curve. Indeed, for such a choice of frame we have $\partial_a n^a = \partial_{2} n^2 = 0$ where we used that the normal component of the normal vector is maximal on the curve and then has to decrease on either side since its norm is fixed to 1. With this choice of frame, and with $\omega^{ab} \equiv \epsilon^{ab} \omega$, the extrinsic trace becomes:
\begin{equation}
K = e^\mu_1 n^2 \omega.
\end{equation}
If the curve $C$ is now a geodesic, we have $\omega \sim K =0$, and we reduce the Wilson line to 
\begin{equation}
-\int_{\mathcal{C}}dx^\mu \mathbf{A}_\mu  = \int_{\mathcal{C}}dx^\mu e^1_\mu  H,
\end{equation}
since $e^2_\mu dx^\mu =0$ along the curve, and where $H$ is the Cartan generator of $\mathfrak{sl}(2)$. The pull-back metric to the curve is given by
\begin{equation}
ds^2  = e^1_\mu \dot{x}^\mu e^1_\nu \dot{x}^\nu dt^2.
\end{equation}
We hence have $\sqrt{h} = e^1_\mu \dot{x}^\mu $ and the Wilson line just becomes the geodesic length computed between both boundary endpoints.

We can thus write the Wilson line evaluated in the representation $R$ along a curve $\mathcal{C}$ connecting two points on the boundary as: 
\begin{align}
    \mathcal{W}_R(\mathcal{C})=\exp\left(-\int_{\mathcal{C}}dx^\mu\mathbf{A}_\mu \right)=e^{\ell H}, \label{eq:wilsonlinegeodesiclength}
\end{align} 
identifying $\ell \equiv 2 \phi$ as the geodesic length between the two endpoints in the Gauss-Euler parametrization. If the two endpoints are located on the AdS boundary, this quantity is the \emph{bare} geodesic length and diverges as usual due to the AdS boundary asymptotics. 

This construction generalizes to the supergravity case of interest here. In particular, we have the analogous first-order superspace expression:
\begin{equation}
K = \partial^A n_A + n^B E^{\;M}_A \Omega^A_{M B}.
\end{equation}
Let us focus on a Wilson line whose endpoint coordinates are bosonic (this is the bottom component of the Wilson line). Then the connecting geodesic is purely in the bosonic subspace along the entire trajectory. This means that $ dx^Mf^\pm_M= 0$ and only the bosonic components of $A_M$ are turned on. Performing the same SO$(2)$ rotation to align the bosonic frame fields with the curve $C$, we obtain exactly the same result as before, supplemented with the Wilson line of the R-sector. E.g. for $\mathcal{N} =2$ we have:
\begin{equation}
\int_{\mathcal{C}} dx^M \mathbf{A}_M = -\int_{\mathcal{C}} dx^M e^{\;\;1}_M H + \int_{\mathcal{C}}dx^M \sigma_M J^{\text{R}},
\end{equation}
where the field $\sigma_M$ is the U$(1)$ gauge field. 

\subsection{Thermofield double state}\label{sec:tfd}

In this section, we will demonstrate how the requirement of the connectivity of the two boundaries selects one particular physical linear combination of the four representation-theoretic states in the thermofield double state.

The Hartle-Hawking state is obtained by propagating from the unit group element, characterizing the geometrically smooth boundary conditions of the no-boundary wavefunction, onto the group element characterized by $\phi,\theta$: \begin{align}
    \raisebox{-0.45\height}{\includegraphics[height=2.35cm]{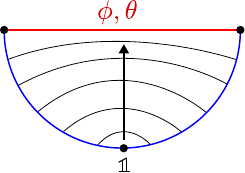}}=\braket{\phi,\theta\mid e^{-\beta H/2 }\mid \mathbb{1}}=\sum_{q \in \mathbb{Z}/2}\int_0^{+\infty} dk \braket{\phi,\theta\mid k,q} \braket{k,q\mid \mathbb{1}} e^{-\beta E(k,q)/2}.
\end{align}  
The unit group element in a mixed parabolic basis is by definition the matrix element $\braket{\nu,\psi_1,\psi_2\mid \omega\mid \lambda, \psi_3,\psi_4}$, with $\omega$ \eqref{eq:omega} transforming the right Whittaker vector into a left Whittaker vector. In the Gauss-Euler parametrization, this is described by the locus 
\begin{equation}
\phi\rightarrow-\infty, \qquad \theta\rightarrow 0,
\end{equation}
and all fermions turned to zero. Physically, this boundary condition implies that the \emph{bare} length between the two boundaries is zero and the two boundaries come together in the initial state of the Euclidean propagation, as illustrated above. Furthermore, the holonomy of the U$(1)$ gauge field vanishes between the two boundaries if the gauge field is continuous and no charged defect is present on the bottom semicircle.

All Bessel functions strongly decay to zero independently of the representation label $\alpha$ as $z\rightarrow\infty$, using the asymptotics $K_\alpha(z)\sim e^{-z}\sqrt{\frac{\pi}{2z}}$.
With $z=e^{-\phi}$, this is a double exponential decay for $\phi\rightarrow-\infty$. Looking at the entries of the representation matrix \eqref{eq:entryf1}-\eqref{eq:entryf2}, the dominant terms in the asymptotic regime $\phi\rightarrow-\infty$ are proceeded by an exponential counterterm $e^{-\phi}$:
\begin{alignat}{3}
    f_1&\rightarrow 0, \qquad     &g_1&\rightarrow -2ie^{-\phi}K_{2j}(2e^{-\phi}), \qquad     &g_2&\rightarrow 2e^{-\phi}e^{i\theta}K_{2j}(2e^{-\phi}),\\
    f_2&\rightarrow 0 , \qquad &h_1&\rightarrow 2e^{-\phi}e^{-i\theta}K_{2j}(2e^{-\phi}), \qquad     &h_2&\rightarrow  2ie^{-\phi}K_{2j}(2e^{-\phi})\, .
\end{alignat}
For $\theta=0$, we can regularize the unit (normalized) representation matrix element 
\begin{align}
    \braket{\mathbb{1}\mid k,q}=\sqrt{\rho(k,q)}\;\Psi_{k,q}(\omega)
\end{align} 
in terms of an overall infinitesimal constant $\epsilon=\lim_{\phi\rightarrow-\infty}\;2e^{-\phi/2}e^{-e^{-\phi}}$:\footnote{This unit representation matrix element satisfies the postulated defining conditions of \cite{Lin:2022zxd} for the infinite temperature TFD state, as annihilating both \begin{align}
    \left(\psi_-+i\psi_+\right)\braket{\mathbb{1}\mid k,q}=\left(\overline{\psi}_-+i\overline{\psi}_+\right)\braket{\mathbb{1}\mid k,q}=0,
\end{align}which can be checked explicitly in the matrix multiplet representation (\ref{eq:lightconefermions}).}  \begin{align}
    \braket{\mathbb{1}\mid k,q}&\rightarrow \sqrt{\rho(k,q)}\begin{bmatrix}
        0&0&0&0\\0&-i&1&0\\0&1&i&0\\0&0&0&0
    \end{bmatrix}.
\end{align}
Multiplying the representation matrices with the (conjugate) unit matrix element restricts the bulk representation matrices to the inner subspace corresponding to the $\ket{H}$ and $\ket{L}$ states.
Note that the $\ket{F^+}$ and $\ket{F^-}$ states are automatically removed in the physical Hartle-Hawking state for any remaining $\theta$ holonomy if the two sides are connected. This stems from the physical restriction that the opposite boundaries should have opposite $R$-charges in the TFD state. Using the $R$-charges, as derived in \eqref{eq:reducedRcharges}, we have 
\begin{align}
    \hat{J}_- &\Psi_{k,q}(\phi,\theta)=
    \frac{e^{2iq\theta}}{\pi}\begin{bmatrix}
        (-2q+1/2)f_1&0&0&0\\ 0&(-2q-1/2)g_1&(-2q+1/2)h_1&0\\0&(-2q-1/2)g_2&(-2q+1/2)h_2&0\\0&0&0&(-2q-1/2)f_2
    \end{bmatrix}, \nonumber \\
    \hat{J}_+ &\Psi_{k,q}(\phi,\theta)= \frac{e^{2iq\theta}}{\pi}\begin{bmatrix}
        (2q+1/2)f_1&0&0&0\\ 0&(2q+1/2)g_1&(2q-1/2)h_1&0\\0&(2q+1/2)g_2&(2q-1/2)h_2&0\\0&0&0&(2q-1/2)f_2
    \end{bmatrix}.
\end{align} 
such that only the inner diagonal block satisfies $(\hat{J}_-+\hat{J}_+)\;\Psi_{k,q}(\phi,\theta)=0$.

In the expression for the Hartle-Hawking state, we thus encounter the matrix product with the conjugate $\braket{k,q\mid \mathbb{1}}= \braket{\mathbb{1}\mid k,q}^\dag$, leading to: 
\begin{center}
\begin{equation}
\raisebox{-0.45\height}{\includegraphics[height=2cm]{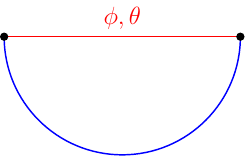}}=
     \sum_{q \in \mathbb{Z}/2}\int_0^{+\infty} dk\; \rho(k,q) e^{-\beta E(k,q)/2}\;\frac{e^{2iq\theta}}{\pi}\begin{bmatrix}
       0&0&0&0\\0&ig_1+h_1&g_1-ih_1&0\\ 0&ig_2+h_2&g_2-ih_2 &0\\0&0&0&0
    \end{bmatrix}
    \label{eq:fullHartleHawking}
\end{equation} 
\end{center}
Multiplying the Hartle-Hawking state with the basis states of the fermionic multiplet, we encounter one linearly independent thermofield double state:\footnote{The second option (third column) has an irrelevant multiplicative phase factor of $-i$.}
\begin{align}
    \ket{\text{TFD}}=\sum_{q \in \mathbb{Z}/2}\int_0^{+\infty} dk  \;\rho(k,q) e^{-\beta E(k,q)/2} \Big(i\ket{H}+\ket{L}\Big).\label{eq:TFD}
\end{align}
Thus, the unit group element selects one physical linear combination of the four representation-theoretic states in the thermofield double.\footnote{This state is to be identified with the one determined in \cite{Lin:2022zxd}, where they argue in favor of this specific physical combination by consistency with the behavior of a boundary correlator at small distances. Comparing to the conventions of \cite{Lin:2022zxd}, we have $\ket{H}_{\text{here}}= -\ket{H}_{\text{there}}$, $\ket{L}_{\text{here}}=-i\ket{L}_{\text{there}}$, such that up to an overall phase factor, (\ref{eq:TFD}) matches with the continuous part of the thermofield double state of \cite{Lin:2022zxd}.}
Taking the overlap between two thermofield double states with flat measure yields the finite-temperature disk partition function: 
 \begin{align}
    \braket{\text{TFD}\mid \text{TFD}}= \sum_{q \in \mathbb{Z}/2}\int_0^{+\infty} dk \; 2\rho(k,q) e^{-\beta E(k,q)},
\end{align}
which coincides with the continuous part of the Schwarzian disk partition function \cite{Mertens:2017mtv}. 

As an interesting generalization, if we were to introduce a non-zero holonomy in the U(1) sector $\Theta$, the wavefunction for the initial state is instead given by
\begin{align}
    \braket{\mathbb{1}, \Theta\mid k,q}\,\, \rightarrow \,\, e^{2iq\Theta}\begin{bmatrix}
        0&0&0&0\\ 0&-i&e^{-i\Theta}&0\\
        0&e^{i\Theta}&i&0\\ 0&0&0&0
    \end{bmatrix},
\end{align} 
which gives the final state:
\begin{align}
\raisebox{-0.45\height}{\includegraphics[height=2cm]{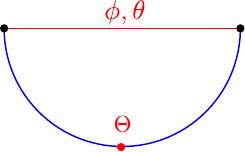}} \hspace{-0.7mm}=\hspace{-1.4mm} \sum_{q \in \mathbb{Z}/2}\int_0^{+\infty} \hspace{-3.5mm} dk\rho(k,q) e^{-\frac{\beta}{2} E(k,q)}\;\frac{e^{2iq(\theta-\Theta)}}{\pi}\hspace{-2mm}
\begin{bmatrix}
0&\hspace{-2mm}0&0&\hspace{-2mm}0\\
0&\hspace{-2mm}ig_1\hspace{-0.7mm}+\hspace{-0.7mm}e^{i\Theta}h_1&e^{-i\Theta}g_1\hspace{-0.7mm}-\hspace{-0.7mm}ih_1&\hspace{-2mm}0\\0& \hspace{-2mm}ig_2\hspace{-0.7mm}+\hspace{-0.7mm}e^{i\Theta}h_2&e^{-i\Theta}g_2\hspace{-0.7mm}-\hspace{-0.7mm}ih_2 &\hspace{-2mm}0\\0&\hspace{-2mm}0&0&\hspace{-2mm}0
    \end{bmatrix}
\end{align}
Considering any single column as an elementary wavefunction, and decomposing that state in its different $\mathcal{N}=2$ fermionic multiplet components, there is only a single independent state of the form:
\begin{align}
    \ket{\text{TFD}}&=\sum_{q \in \mathbb{Z}/2}\int_0^{+\infty} dk \;\rho(k,q)e^{-\beta E(k,q)/2}e^{-2iq\Theta}\Big(i\ket{H}+e^{i\Theta}\ket{L}\Big),
\end{align} 
which exhibits a relative phase factor in the physical combination of the states $\ket{H}$ and $\ket{L}$ in the fermionic multiplet.
This phase factor $e^{i\Theta}$ is a priori difficult to determine from the perspective of diagonalizing the Casimir/Liouville operator \eqref{eq:N=2Liouville2}. Here we see the physical interpretation of it as corresponding to a U(1) phase between both boundaries, or alternatively as a U(1) gauge theory defect on the connecting boundary.

\subsection{Matter Wilson lines as discrete series matrix elements}\label{section:discreteseries}
A Wilson line between the two asymptotic boundaries in the bulk BF language can be identified as the hyperbolic group element of the geodesic length between the two anchoring points, as demonstrated in section \eqref{sec:propagator}: 
\begin{align}
    \mathcal{W}(\mathcal{C})=\exp\left(-\int_{\mathcal{C}} dx^M \mathbf{A}_M \right)= e^{2\phi H}e^{2i\theta Z}.
\end{align}
The matrix elements of the Wilson lines then act as operator insertions in the resulting quantum mechanical model. We choose discrete lowest and highest weight states as the bra and ket of this matrix element respectively with mixed parabolic labels $\nu=\lambda=0$, and restrict to the Cartan generators.

Analogous to the case for SL$(2,\mathbb{R})$ and OSp$(1|2,\mathbb{R})$, it is possible to realize the discrete highest and lowest weight irreps in the principal series representation by a monomial basis on the superline $\mathbb{R}^{1 \vert 2}$ \cite{Belaey:2023jtr}. 
A highest weight state is annihilated by both $F_x^+$ and $F_y^+$ (and hence automatically $E^+$). This is the right Whittaker vector \eqref{eq:rightWhittakervector} with $\lambda=0$, which removes the fermionic structures $\psi_{3,4}$:
\begin{equation}
\braket{ x,\vartheta_1,\vartheta_2\mid j,Q} =  x^{2j}e^{-2i Q \frac{\vartheta_1\vartheta_2}{x}}=x^{2j}-2iQx^{2j-1}\vartheta_1\vartheta_2.
\end{equation} In addition, this eigenstate simultaneously diagonalizes the action of the Cartan generators $H,Z$ since the brackets with the right parabolic generators vanish: \begin{align}
    H=j,\qquad Z=Q.
\end{align}
If $2j \in - \mathbb{N}$,\footnote{Just as in the simpler cases of SL$(2,\mathbb{R})$ and OSp$(1\vert 2,\mathbb{R})$, this restriction to half integers is not visible at the level of the monomial representation, which only probes the universal cover of OSp$(2\vert 2,\mathbb{R})$.} the representation is unbounded from below, which generates the discrete highest weight irrep by acting with the lowering operators. The discrete series representation label $j$ is related to the conformal scaling dimension $h>0$ of the dual bilocal operator as $j\equiv-h$. 

The (conjugate) lowest weight state with $H=-j$ and $Z=Q$ is determined from the analytically continued (conjugate) lowest weight state of the finite dimensional representation as: \begin{align}
    \braket{-j,Q\mid x,\vartheta_1,\vartheta_2}=\delta(x|\vartheta_1,\vartheta_2).
\end{align}

The Whittaker function evaluated in a mixed parabolic basis of a (conjugate) lowest weight and a highest weight state is then
\begin{align}
    \braket{-j,Q\mid e^{2\phi H}e^{2i\theta Z}\mid j,Q} = \, e^{2j \ln \epsilon} e^{2\phi j}e^{2iQ\theta} ,
\end{align} 
where we regulated the infinity at $x=0$ by letting $x = \epsilon \to 0^+$. Using the identification $\ell_{\text{bare}}=2\phi_{\text{bare}}$ between the bare lengths, we see that the answer is automatically a function of the renormalized length $\ell_{\text{ren}}\equiv \ell_{\text{bare}} + \ln \epsilon^2$, and hence we can write the answer as $e^{2\phi_{\text{ren}} j}e^{2iQ\theta}$ with $\phi_{\text{ren}}=\phi_{\text{bare}}+\ln \epsilon$. This matches with the bulk wavefunctions between both boundaries also being determined in terms of renormalized variables.

The highest weight discrete series matrix element is thus diagonal in the $\mathcal{N}=2$ multiplet: \begin{align}\label{eq:highestweightwhittakerfunction}
    \braket{-j,Q\mid e^{2\phi H}e^{2i\theta Z}\mid j,Q} = \begin{bmatrix}
         e^{2\phi j}e^{2iQ\theta}&0&0&0\\0& e^{2\phi j}e^{2iQ\theta}&0&0\\ 0&0& e^{2\phi j}e^{2iQ\theta}&0\\0&0&0& e^{2\phi j}e^{2iQ\theta}
    \end{bmatrix}.
\end{align}  
For BPS operators with $j=Q$ (for which the energy $E=Q^2-j^2$ vanishes), we have $e^{j(2\phi+2i\theta)}$. Inserting this BPS operator in the disk partition function leads to the two-point correlators considered in \cite{Lin:2022zxd}, together with neutral correlators for which $Q=0$.

On the other hand, in the geodesic approximation, we can identify $-j=h\approx \mu\gg 1$ with $\mu$ the mass of the worldline particle. The highest weight matrix element of the Wilson line then coincides with the classical saddle of the ``open" charged worldline path integral $e^{-\mu \ell}e^{2i\theta Q}$, with $\ell$ the geodesic  length.\footnote{This should be compared to our result for the ``closed" charged worldline path integral in \cite{Belaey:2023jtr}, which yields an additional one-loop correction to this saddle from the Weyl denominator, leading to a non-trivial UV behavior at $\ell\rightarrow0$.}

We determine more general discrete series matrix elements evaluated in a mixed parabolic basis  from the reduced harmonic analysis in appendix \ref{section:discreteseriesharmonicanalysis}, which have a smooth $\nu\lambda\rightarrow0$ limit to the highest weight solution determined above \eqref{eq:highestweightwhittakerfunction}.

\subsection{End-of-the-world brane amplitudes}
Given the gravitational density of states \eqref{eq:desnityofstates} and the Hartle-Hawking state amplitude \eqref{eq:TFD}, we can determine the path integral on a disk topology with one end-of-the-world brane boundary. For $\mathcal{N}=2$, this brane describes a massive particle of mass $\mu$ and charge $Q$ whose worldline demarcates where space ends. 
Amplitudes involving open EOW brane boundaries have been computed for instance in \cite{Penington:2019kki} for the bosonic case and \cite{Boruch:2023trc} for the supersymmetric case. From the perspective of our work, it is sufficient to note that brane worldlines can be identified with Wilson lines in the BF gauge theoretical description, both in bosonic JT gravity \cite{Iliesiu:2019xuh} and its supersymmetric versions \cite{Belaey:2023jtr}. Hence amplitudes with EOW branes can be directly constructed using the techniques of this work, as we illustrate next.

Using the functional form of a Wilson line insertion $\mathcal{W}=e^{2j\phi+2iQ\theta}\mathbb{1}_{4 \times 4}$ evaluated in a representation labeled by $j,Q$, we can compute the EOW brane amplitude by gluing to the Euclidean preparation of the thermofield double along the hyperbolic group parameters $\phi,\theta$:
\begin{align}
 &\sum_{q \in \mathbb{Z}/2}\int_0^\infty dk\; \rho(k,q) e^{-\beta E(k,q)/2}\;\int_0^{2\pi} d\theta\int_{-\infty}^{+\infty} d\phi~ 
  e^{2j\phi+2iQ\theta}
 \frac{e^{2iq\theta}}{\pi}\begin{bmatrix}
      0&0&0&0\\0& ig_1+h_1&g_1-ih_1&0\\ 0&ig_2+h_2&g_2-ih_2  &0\\0&0&0&0
    \end{bmatrix}.
    \label{eq:EOWcalculation}
\end{align}
Performing the integral over $\phi$ and $\theta$ using $\int_{-\infty}^{+\infty} d\phi\; e^{-\alpha\phi} K_{2ik}(2e^{-\phi})=\frac{|\Gamma(\alpha+ik)|^2}{4}$, this yields
\begin{align}
\label{eq:halfmoon}
\raisebox{-0.45\height}{\includegraphics[height=2.2cm]{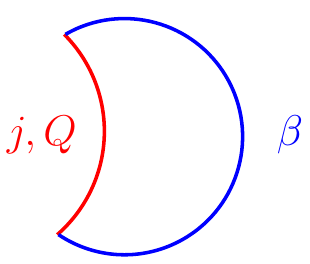}} = \sum_{q\in \mathbb{Z}/2}\int_0^\infty  dk\; \rho(k,q) e^{-\beta E(k,q)/2}\; \begin{bmatrix}
   0&0&0&0\\0&  w_1&y_1&0\\0& w_2&y_2&0\\0&0&0&0
 \end{bmatrix},
\end{align} 
with
\small
\begin{align}
    w_1&= \hspace{-0.7mm} q|\Gamma(ik\hspace{-0.7mm}-\hspace{-0.7mm}2j)|^2\delta_{q,-Q}+\hspace{-0.7mm}\frac{1}{2}\hspace{-0.7mm}\left(\hspace{-0.7mm}|\Gamma(\frac{3}{2}\hspace{-0.7mm}+\hspace{-0.7mm}ik\hspace{-0.7mm}-\hspace{-0.7mm}2j)|^2\hspace{-0.7mm}+|\Gamma(\frac{1}{2}\hspace{-0.7mm}+\hspace{-0.7mm}ik\hspace{-0.7mm}-\hspace{-0.7mm}2j)|^2\hspace{-0.7mm}\right)\hspace{-0.7mm}\delta_{q,-Q}\hspace{-0.7mm}+\hspace{-0.6mm}|\Gamma(1\hspace{-0.7mm}+\hspace{-0.7mm}ik\hspace{-0.7mm}-\hspace{-0.7mm}2j)|^2\delta_{q,\frac{1}{2}-Q}, \nonumber \\
    y_1&=\hspace{-0.7mm}-i q|\Gamma(ik\hspace{-0.7mm}-\hspace{-0.7mm}2j)|^2\delta_{q,-Q}-\hspace{-0.7mm}\frac{i}{2}\hspace{-0.7mm}\left(\hspace{-0.7mm}|\Gamma(\frac{3}{2}\hspace{-0.7mm}\hspace{-0.7mm}+\hspace{-0.7mm}ik\hspace{-0.7mm}-\hspace{-0.7mm}2j)|^2\hspace{-0.7mm}+|\Gamma(\frac{1}{2}\hspace{-0.7mm}+\hspace{-0.7mm}ik\hspace{-0.7mm}-\hspace{-0.7mm}2j)|^2\hspace{-0.7mm}\right)\hspace{-0.7mm}\delta_{q,-Q}\hspace{-0.7mm}-\hspace{-0.7mm}i|\Gamma(1\hspace{-0.7mm}+\hspace{-0.7mm}ik\hspace{-0.7mm}-\hspace{-0.7mm}2j)|^2\delta_{q,\frac{1}{2}-Q} , \nonumber \\
    w_2&= \hspace{-0.7mm} -i q|\Gamma(ik\hspace{-0.7mm}-\hspace{-0.7mm}2j)|^2\delta_{q,-Q}+\hspace{-0.7mm}\frac{i}{2}\hspace{-0.7mm}\left(\hspace{-0.7mm}|\Gamma(\frac{3}{2}\hspace{-0.7mm}+\hspace{-0.7mm}ik\hspace{-0.7mm}-\hspace{-0.7mm}2j)|^2\hspace{-0.7mm}+|\Gamma(\frac{1}{2}\hspace{-0.7mm}+\hspace{-0.7mm}ik\hspace{-0.7mm}-\hspace{-0.7mm}2j)|^2\hspace{-0.7mm}\right)\hspace{-0.7mm}\delta_{q,-Q}\hspace{-0.7mm}+\hspace{-0.7mm}i|\Gamma(1\hspace{-0.7mm}+\hspace{-0.7mm}ik\hspace{-0.7mm}-\hspace{-0.7mm}2j)|^2\delta_{q,\frac{1}{2}-Q}, \nonumber \\
    y_2&=\hspace{-0.7mm}- q|\Gamma(ik\hspace{-0.7mm}-\hspace{-0.7mm}2j)|^2\delta_{q,-Q}+\hspace{-0.7mm}\frac{1}{2}\hspace{-0.7mm}\left(\hspace{-0.7mm}|\Gamma(\frac{3}{2}\hspace{-0.7mm}+\hspace{-0.7mm}ik\hspace{-0.7mm}-\hspace{-0.7mm}2j)|^2\hspace{-0.7mm}+|\Gamma(\frac{1}{2}\hspace{-0.7mm}+\hspace{-0.7mm}ik\hspace{-0.7mm}-\hspace{-0.7mm}2j)|^2\hspace{-0.7mm}\right)\hspace{-0.7mm}\delta_{q,-Q}\hspace{-0.7mm}+\hspace{-0.7mm}|\Gamma(1\hspace{-0.7mm}+\hspace{-0.7mm}ik\hspace{-0.7mm}-\hspace{-0.7mm}2j)|^2\delta_{q,\frac{1}{2}-Q}.
\end{align}
\normalsize
Reading off the columns of \eqref{eq:halfmoon}, one finds only a single (linearly) independent wavefunction:\footnote{The other option again has an irrelevant multiplicative factor of $-i$.} \begin{align}
    \ket{\text{EOW}}=\sum_{q\in \mathbb{Z}/2}\int_0^\infty dk\;\rho(k,q)e^{-\beta E(k,q)/2}\left[w_1\overline{\psi}_-+w_2\overline{\psi}_+\right]\ket{0}.
\end{align}
This constitutes the ``open" counterpart to the closed circular EOW brane wavefunctions  considered in \cite{Belaey:2023jtr}.
In the geodesic approximation, we furthermore have the identification $-j\approx \mu$ between the conformal scaling dimension $-j=h$ and  the mass $\mu$ of the EOW  particle.

\section{Concluding remarks}
\label{s:concl}
In this work, we have used the supergroup structure to describe the continuous two-sided gravitational wavefunctions of  $\mathcal{N}=2$ JT supergravity as principal series matrix elements of the positive semigroup OSp$^+(2|2,\mathbb{R})$. These matrix elements are evaluated between mixed parabolic eigenstates satisfying e.g. \eqref{eq:fermleft}, which implement the asymptotically AdS$_{2|4}$ Brown-Henneaux boundary conditions. With these constraints, the corresponding Casimir eigenvalue problem, by consistency, reduces to the $\mathcal{N}=2$ Liouville minisuperspace description of JT gravity. We find that the gravitational matrix elements of the positive semigroup OSp$^+(2|2,\mathbb{R})$ yield the correct gravitational density of states, establishing this algebraic structure as the correct one describing gravity, echoing previous results for $\mathcal{N}=0,1$ JT (super)gravity \cite{Blommaert:2018oro, Fan:2021wsb} and as part of the classical limit of Liouville (super)gravity theories \cite{Fan:2021bwt}.

We end with some partially open problems for which a full solution is left to the future.

\subsubsection*{BPS states}
From the diagonalization of the Liouville problem \eqref{eq:N=2Liouville2}, it is known that next to the continuous series representations, the gravitational Hilbert space also contains normalizable discrete BPS states (the first term in \eqref{eq:fullpf}), as bound states of the exponential potentials in the Schr\"odinger problem. These were explicitly constructed in this fashion in \cite{Lin:2022zxd}, and can also be seen from the Schwarzian limit of the $\mathcal{N}=2$ super-Virasoro vacuum character \cite{Mertens:2017mtv}. From the perspective of representation theory of OSp$(2|2,\mathbb{R})$, these would be atypical representations with shorter multiplets. However, they seem quite special in the following sense. The Plancherel measure on these representations formally looks like an analytic continuation of the principal series representations to zero energy ($j = \pm q$) and hence imaginary $k$, violating the unitarity constraint of the principal series representations. We are unaware of any first principles construction of an atypical representation with such properties in the literature. It would hence be interesting to construct these explicitly from the underlying representation theory.

\subsubsection*{Higher rank superconformal algebras}
It is interesting to try to develop a general picture now on how these gravitational states arise in the most generic supergravity model. As a concrete first example of how our techniques could be generalized, we consider here the case of $\mathcal{N}=4$ JT supergravity. The relevant $\mathcal{N}=4$ supersymmetric Hamiltonian can be extracted from the $\mathcal{N}=4$ Liouville CFT \cite{Creutzig:2011qm,Hikida:2007sz}, by taking the minisuperspace limit:
\begin{align}
\label{eq:HN4}
\hat{H} =& -\hspace{-0.7mm}\partial_\ell^2 + \hspace{-0.7mm}H_{\text{QM}}^{\text{SU}(2)}[g] + \frac{1}{4}e^{-\frac{\ell}{2}} (\psi^3_a \hspace{-0.9mm}-\hspace{-0.7mm}i\psi^4_a)g^{-1}_{ab}(\psi^1_{b}\hspace{-0.9mm}+\hspace{-0.7mm}i\psi^2_{b}) + \frac{1}{4}e^{-\frac{\ell}{2}}(\psi^1_a\hspace{-0.9mm}-\hspace{-0.7mm}i\psi^2_a) g_{ab}(\psi^3_b\hspace{-0.9mm}+\hspace{-0.7mm}i\psi^4_b)+ e^{-\ell} \nonumber \\
=& -\hspace{-0.7mm}\partial_\ell^2 + \hspace{-0.7mm} H_{\text{QM}}^{\text{SU}(2)}[g]  + e^{-\ell/2}\; \overline{\psi}^+_ag^{-1}_{ab}\psi_b^- + e^{-\ell/2}\;\overline{\psi}_a^- g_{ab}\psi_b^+ + e^{-\ell},
\end{align}
in terms of the compact group Hamiltonian of the particle on the SU$(2)$ group manifold  $H_{\text{QM}}^{\text{SU}(2)}[g] \equiv \text{Tr} (g^{-1} \dot{g})^2$. The index $a=1,2$ runs through the SU(2) matrix elements in the fundamental representation $g_{ab}$. We introduced a set of 8 hermitian gamma matrices $\psi_{a}^{1,2,3,4}$ that satisfy the Euclidean Clifford algebra $\{\psi^{i}_a,\psi^{j}_{b}\} = 2 \delta^{ij}\delta_{ab}$. Such a Dirac algebra can be realized minimally with $16 \times 16$ matrices, leading to 16 different states in the supermultiplet. These can be found as usual \eqref{eq:fermiontransf}: we group the 8 gamma matrices in two sets: 4 raising operators and 4 lowering operators. Starting with an (assumed unique) state annihilated by all lowering operators $\left\vert\frac{1}{2}\frac{1}{2}\frac{1}{2}\frac{1}{2}\right\rangle$,\footnote{We denoted this vacuum state as $\ket{0}$ in previous sections.} we can then apply raising operators to flip one of the spins consecutively:
\begin{equation}
\left\vert\frac{1}{2}\frac{1}{2}\frac{1}{2}\frac{1}{2}\right\rangle \to 4\text{-vector} \to \text{antisym rank-2 tensor} \to 4\text{-vector} \to \left\vert-\frac{1}{2}-\frac{1}{2}-\frac{1}{2}-\frac{1}{2}\right\rangle,
\end{equation}
or in terms of dimensions $1-4-6-4-1$ with 8 bosonic and 8 fermionic states. The Hamiltonian $H$ \eqref{eq:HN4} is easily seen to be hermitian when using that $g_{ab}$ is a unitary matrix.

As a specific example, the simplest eigenstate of \eqref{eq:HN4} is the bottom state, on which the terms with $\psi$'s vanish, satisfying
\begin{equation}
\left(-\partial_\ell^2 + H_{\text{QM}}^{\text{SU}(2)}[g] + e^{-\ell}\right) \psi_{m,n}^{k,j} = E(k,j)\; \psi_{m,n}^{k,j}.
\end{equation}
This is readily solved by separation of variables into:
\begin{equation}
\label{eq:gs}
\psi_{m,n}^{k,j}(\phi,g) = D_{mn}^{j}(g) K_{2ik}(2e^{-\ell/2})\;
, \qquad m,n=-j\hdots +j,
\end{equation}
in terms of the SU(2) representation matrices $D_{mn}^{j}(.)$ (i.e. the Wigner $D$-functions), and the modified Bessel function of the second kind $K_{2ik}(.)$. Here $E= k^2 + j(j+1)$ is the total energy. Gravitationally, this eigenfunction has an interpretation as a 2-boundary wavefunction in $\mathcal{N}=4$ JT supergravity, schematically as
\begin{align}
   \psi_{m,n}^{k,j}(\phi,g) = \,\, \raisebox{-0.3\height}{\includegraphics[height=0.8cm]{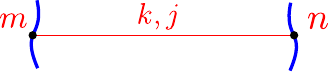}}.
\end{align}
The wavefunction describes a state with gravitational energy $k^2$ and $R$-charge energy $j(j+1)$. The SU(2) labels $m$ and $n$ can be thought of as living on the two different asymptotic boundaries. Unlike the eigenvalues of the bosonic parabolic generators $E^\pm$, these are free indices to be summed over in gravitational amplitudes. Next to the ground state, we are currently in the process of finding all eigenstates of this Hamiltonian \cite{toap} and hope to report on this soon.

The main point that we want to make here is that the structure of these eigenfunctions \eqref{eq:gs} synergizes nicely with our group-theoretical perspective.
Indeed, the analogue of the Gauss-Euler decomposition of the group element $g \in \text{PSU}(1,1\vert 2)$ (connected to the identity) has the form:
\begin{align}
\label{eq:GEcomplex}
g = \underline{e^{\bar{\theta}^-_\alpha \bar{F}^-_\alpha} e^{\theta^-_\alpha F^-_\alpha} e^{\beta E^-} {\color{blue}e^{i \varphi \frac{\sigma_3}{2}}}} e^{2\phi H} {\color{blue}e^{i\theta \frac{\sigma_1}{2}}} \underline{{\color{blue}e^{i \psi \frac{\sigma_3}{2}}} e^{\gamma E^+} e^{\theta^+_\alpha F^+_\alpha} e^{\bar{\theta}^+_\alpha \bar{F}^+_\alpha}}.
\end{align}
We have colored in blue the Euler angle decomposition (with angles $(\varphi,\theta, \psi)$) of the SU(2) R-charge sector. We have also underlined the parts of the decomposition that we want to simultaneously diagonalize for the bra and ket vector. For the bottom state, where we essentially ignore the fermionic degrees of freedom, the resulting matrix element with mixed parabolic boundary conditions is then indeed equation \eqref{eq:gs}.

This is part of a generic story as follows. Any superconformal algebra has as bosonic subalgebra $\mathfrak{sl}(2,\mathbb{R}) \oplus \mathfrak{g}_R$ with $\mathfrak{g}_R$ the maximally compact real form of $\mathfrak{g}_{R}^{\mathbb{C}}$. For the associated compact R-charge subgroup $G_R \simeq \text{exp}(\mathfrak{g}_R)$, we use a generalized Euler angle decomposition of the group element: this is the Cartan or $KAK$ decomposition, see e.g. \cite{Vilenkin2}. For the full group element $g$ then, the bra and ket states are chosen as simultaneous eigenvector of all fermionic generators (in the Dirac algebra sense), of $E^\pm$ for the gravitational part, and of a maximally set of commuting generators (+ Casimirs) in the subgroup $K \subset G_R$.\footnote{This construction has a known analogue when constructing Whittaker functions for higher spin gravity based on the $\mathfrak{sl}(N,\mathbb{R})$ algebra. In this case, we need to pick a set of positive simple roots in $\mathfrak{sl}(N,\mathbb{R})$, whereas the non-simple root generators need to be chosen to have eigenvalue zero by consistency. See \cite{Chervov} for the mathematical construction of these matrix elements, and \cite{Blommaert:2018oro} for the physical implementation in the current context.}

\subsection*{Supersymmetric quantum deformations}
Our techniques to compute the gravitational matrix elements from the bottom up, by constructing Whittaker vectors that implement the Brown-Henneaux gravitational boundary conditions at the left and right boundaries, can in principle be applied to quantum deformed supergroups as well, which describe Liouville supergravity models and supersymmetric DSSYK models. In particular, the $\mathcal{N}=1$ and $\mathcal{N}=2$ supersymmetric models have an underlying group theoretical structure that is governed respectively by suitable real forms of the OSp$_q(1|2)$ and OSp$_q(2|2)$ quantum supergroups, where $q$ is the deformation parameter. In both cases, one should first write down the analogous Gauss-Euler decomposition implementing the Hopf duality between the Hopf coordinate algebra and the quantum algebra $U_q(\mathfrak{osp}(1|2))$ and $U_q(\mathfrak{osp}(2|2))$. A natural guess for OSp$_q(1|2)$, of relevance for supersymmetric DSSYK, is then \cite{Blommaert:2023opb}:
\begin{equation}
    g=e^{2\theta^-F^-}e^{\gamma E^-}_{q^{-2}}e^{2\phi H} e^{\beta E^+}_{q^2}e^{2\theta^+F^+}, \quad 0<q<1, 
     \label{eq:gaussEulerDSSYK1}
\end{equation}
where the exponentials involving the bosonic generators $E^-$ and $E^+$ are $q$-deformed exponentials, while the fermionic generators $F^-$ and $F^+$ are inserted in ordinary exponentials involving the Grassmann parameters $\theta^-$ and $\theta^+$. Similarly, for OSp$_q(2|2,\mathbb{R})$ we can conjecture
\begin{equation}
    g=e^{\bar{\theta}^-\bar{F}^-}e^{\theta^-F^-}e^{\gamma E^-}_{q^{-2}}e^{2\phi H}e^{2i\theta Z} e^{\beta E^+}_{q^2}e^{\theta^+F^+}e^{\bar{\theta}^+\bar{F}^+}, \quad 0<q<1.
    \label{eq:gaussEulerDSSYK2}
\end{equation}
The objects in \eqref{eq:gaussEulerDSSYK1} and \eqref{eq:gaussEulerDSSYK2} go back to the undeformed Gauss-Euler decompositions \eqref{eq:GaussEulerN1} and \eqref{eq:GEcomplex} in the $q\rightarrow 1$ limit.
Equations \eqref{eq:gaussEulerDSSYK1} and \eqref{eq:gaussEulerDSSYK2} represent the starting point to construct the matrix elements and to implement gravitational boundary conditions, in an analogous formalism as the one we have developed in section \ref{sec: Grav Matrix element} and Appendix \ref{appendix:N=1} for $\mathcal{N}=2$ and $\mathcal{N}=1$ JT supergravity respectively. One new feature that will appear here is that when diagonalizing $E^\pm$, it is known that for quantum groups one has the option to diagonalize these up to the action of the Cartan generator in the form $q^{\alpha H}$ for some coefficient $\alpha$ \cite{Kharchev:2001rs}. To describe both Liouville gravity and DSSYK, one indeed has to choose a non-zero value of $\alpha$ to match with gravity \cite{Fan:2021bwt}. This was shown in  \cite{Fan:2021bwt} to be true for supersymmetric Liouville gravity as well in the ``diagonalization'' of the fermionic operators $F^\pm$. It would be interesting to combine these results with our current work and reach the most general statement on the role of these representation matrices as gravitational wavefunctions.

\section*{Acknowledgments}
We thank J. Papalini, T. Tappeiner and Q. Wu for discussions. AB acknowledges the UGent Special Research Fund (BOF) for financial support. FM is supported by the Research Foundation - Flanders (FWO) doctoral fellowship 11P9Z24N. TM acknowledges financial support from the European Research Council (grant BHHQG-101040024). Funded by the European Union. Views and opinions expressed are however those of the author(s) only and do not necessarily reflect those of the European Union or the European Research Council. Neither the European Union nor the granting authority can be held responsible for them.

\appendix

\section{OSp$(2|2,\mathbb{R})$ representation theory}
\label{appendix:representationtheory}
In this appendix, we give an overview of some relevant aspects of the representation theory of OSp$(2|2,\mathbb{R})$, using the same conventions as in \cite{Belaey:2023jtr}. In addition, we collect detailed expressions for the left- and right-regular representations and the Haar measure on the group manifold. 
\subsection{Conventions}

\iffalse The supergroup OSp$\left(2|2,\mathbb{R}\right)$ is defined as the subgroup of GL$\left(2|2,\mathbb{R}\right)$ matrices 
\begin{equation}
g=\left[\begin{array}{c c | c c} 
	a & b & \alpha_1 & \beta_1\\ 
    c & d & \gamma_1 & \delta_1\\	
 \hline 
	\alpha_2 & \beta_2 & w & y\\
    \gamma_2 & \delta_2 & z & u\\
\end{array}\right],
\label{eq:OSp2matrix}
\end{equation}
with 8 bosonic supernumbers $a,b,c,d,w,y,z,u$ and 8 fermionic supernumbers \\$\alpha_{1,2},\beta_{1,2},\gamma_{1,2},\delta_{1,2}$,
that preserve the orthosymplectic form $\Omega$: $g\Omega g^{{st}^3}= \Omega$, with\footnote{The supertranspose operation consists in flipping the sign of one block of fermionic variables so that the property $(g_1g_2)^{st}=g_2^{st}g_1^{st}$ is preserved,
\begin{equation}\label{eq:generalgroupelement}
    \left[\begin{array}{c | c} 
	 A & B\\ 	
 \hline 
	 C & D\\
    
\end{array}\right]^{st}=\left[\begin{array}{c | c} 
	 A^T & -C^T\\ 	
 \hline 
	 B^T & D^T\\
    
\end{array}\right].
\end{equation} } 
\begin{equation}
   \Omega=\left[\begin{array}{c c | c c} 
	0 &-1 & 0 & 0\\ 
    1 & 0 & 0 & 0\\	
 \hline 
	0 & 0 & 1 & 0\\
    0 & 0 & 0 & 1\\
    
\end{array}\right].
\end{equation}
In particular, the conformal OSp$(2,\mathbb{R})$ supergroup contains an SL$(2,\mathbb{R})$ subgroup in the relations of the top-left bosonic subblock described by $a,b,c,d$. 
\fi
The $\mathfrak{osp}(2|2)$ Lie superalgebra is a $4 \vert 4$-dimensional vector space, with bosonic generators $H,E^\pm, Z$ and fermionic generators $F^\pm$, $\bar{F}^\pm$. In the Cartan-Weyl basis, these generators satisfy the superalgebra relations (see e.g. \cite{Frappat:1996pb}):
\begin{alignat}{3}
\label{eq:superalgebra}
[H,E^\pm] &= \pm E^\pm, \qquad &[H,F^\pm] &= \pm \frac{1}{2} F^\pm, \quad &[H, \bar{F}^\pm ] &= \pm \frac{1}{2} \bar{F}^\pm, \nonumber \\
[Z,H] &= [Z,E^\pm] =0, \quad &[Z,F^\pm] &= \frac{1}{2} F^\pm, \quad &[Z,\bar{F}^\pm] &= - \frac{1}{2}\bar{F}^\pm, \nonumber \\
[E^\pm,F^\pm] &= [E^\pm,\bar{F}^\pm] =0, \quad &[E^\pm,F^\mp] &=-F^\pm, \quad &[E^\pm,\bar{F}^\mp] &= \bar{F}^\pm, \nonumber \\
\{F^\pm,F^\pm\} &= \{\bar{F}^\pm,\bar{F}^\pm\} = 0, \qquad &\{F^\pm,F^\mp\} &= \{\bar{F}^\pm,\bar{F}^\mp\} =0, \qquad &\{F^\pm,\bar{F}^\pm\} &= E^\pm, \nonumber \\
[E^+,E^-] &= 2H, \quad &\{F^\pm,\bar{F}^\mp\} &= Z\mp H. \quad &
\end{alignat}
The Cartan subalgebra is spanned by the $H$ and $Z$ generators, whose eigenvalues can be raised and lowered by half a unit by acting with the different $F^\pm$, $\bar{F}^\pm$, and the eigenvalue of $H$ by a full unit by acting with $E^\pm$. 
 
Exponentiating these generators yields a connected piece of the OSp$(2|2,\mathbb{R})$ supergroup manifold.\footnote{The full supermanifold has two connected pieces, coming from the two pieces of the O(2) R-symmetry subgroup. The second piece does not seem to play a role for supergravitational purposes.} One convenient parametrization connected to the identity is the Gauss-Euler decomposition:
\begin{align}
\label{eq:GEcomplex}
g = e^{\bar{\theta}^- \bar{F}^-} e^{\theta^- F^-} e^{\beta E^-} e^{2\phi H} e^{2i \theta Z}e^{\gamma E^+} e^{\theta^+ F^+} e^{\bar{\theta}^+ \bar{F}^+}.
\end{align}
Here, the supergroup manifold is parametrized in terms of two real coordinates $\{\phi, \theta\}$ corresponding to the Cartan subalgebra, and six coordinates $\{\beta, \gamma\mid \overline{\theta}^\pm, \theta^\pm\}$, where $\beta, \gamma$ are real numbers and the Grassmann numbers $\theta_\pm$ and $\bar{\theta}_\pm$ are complex conjugates.

It is often convenient to transfer to new real fermionic generators defined as:
\begin{equation}\label{eq:generatorstransf}
F^+_x \equiv \frac{F^++\bar{F}^+}{\sqrt{2}}, \quad F^+_y \equiv \frac{F^+-\bar{F}^+}{\sqrt{2}i}, \quad F^-_x \equiv \frac{F^--\bar{F}^-}{\sqrt{2}}, \quad F^-_y \equiv \frac{F^-+\bar{F}^-}{\sqrt{2}i},
\end{equation} 
which satisfy the $\mathfrak{osp}(2|2)$ matrix algebra in the form:
\begin{alignat}{3}
[H,E^\pm] &= \pm E^\pm, \qquad &[E^+,E^-]&=2H, \qquad &[Z,H]&=0,\nonumber\\ 
\{F_x^\pm, F_x^\pm\}&=\pm E^+, \qquad &\{F_y^\pm,F_y^\pm\}&=\pm E^+, \qquad &\{F_x^\pm,F_y^\pm\}&=0,\nonumber \\\{F_x^+,F_x^-\}&=H, \qquad &\{F_y^+,F_y^-\}&=H, \qquad &\{F_x^\pm,F_y^\mp\}&=\mp iZ,\nonumber\\
 [Z,F_y^\pm]&= \frac{1}{2i}F_x^\pm, \qquad &[Z,F_x^\pm]&=-\frac{1}{2i}F_y^\pm, \qquad &[H, F_x^\pm]&=\pm\frac{1}{2}F_x^\pm,\nonumber\\ \qquad [H, F_y^\pm]&=\pm\frac{1}{2}F_y^\pm,\qquad  &[E^{\pm},F_x^{\mp}]&=-F_x^{\pm}, \qquad &[E^{\pm}, F_y^{\mp}]&=-F_y^{\pm},\nonumber \\  [E^{\pm}, F_x^{\pm}]&=0, \qquad &[E^{\pm}, F_y^{\pm}]&=0, \qquad &[E^\pm, Z]&=0.\label{eq:fullrealalgebra}
\end{alignat}
Transforming to real fermionic coordinates defined by\begin{align}\label{eq:coordiantetransf}
    \theta^+ \equiv \theta_x^+ - i \theta_y^+, \quad \bar{\theta}^+ \equiv \theta_x^+ + i \theta_y^+,\quad \theta^- \equiv \theta_x^- - i \theta_y^-, \quad \bar{\theta}^- \equiv -\theta_x^- - i \theta_y^-,
\end{align} 
the Gauss-Euler decomposition \eqref{eq:GEcomplex} is written up to a shift in $\gamma$ and $\beta$ as: \begin{align}\label{eq:GEreal}
    g = e^{\sqrt{2}\theta^-_{y} F^-_y} e^{\sqrt{2}\theta^-_{x} F^-_x} e^{\beta E^-} e^{2\phi H} e^{2i \theta Z}e^{\gamma E^+} e^{\sqrt{2}\theta^+_{x} F^+_x} e^{\sqrt{2}\theta^+_{y} F^+_y}.
\end{align}
In the fundamental $2|2$-dimensional matrix representation, this exponentiated group element reads: 
\tiny
\begin{align}
\label{eq:GEexpli}
\left[\begin{array}{c c | c c} 
\hspace{-0.1cm} e^\phi & e^\phi \gamma & e^\phi \theta^+_x & - e^{\phi} \theta^+_{y} \\
\hspace{-0.1cm} e^\phi \beta & \hspace{-0.1cm} e^{-\phi} \hspace{-0.1cm} + \hspace{-0.05cm}\beta \gamma e^{\phi} \hspace{-0.1cm} - \hspace{-0.05cm}(\theta^-_x \hspace{-0.05cm}\cos\theta \hspace{-0.05cm}- \hspace{-0.05cm}\theta^-_y \hspace{-0.05cm} \sin\theta) \theta^+_x \hspace{-0.1cm}- (\theta^-_x \hspace{-0.05cm} \sin\theta + \theta^-_y \hspace{-0.05cm} \cos\theta) \theta^+_y \hspace{-0.1cm} & \hspace{-0.05cm} e^\phi \beta \theta^+_x \hspace{-0.05cm} -\hspace{-0.05cm}\theta^-_x \hspace{-0.05cm}\cos\theta \hspace{-0.05cm}+\hspace{-0.05cm}\theta^-_y \hspace{-0.05cm}\sin\theta & -e^\phi \beta \theta^+_y \hspace{-0.05cm} + \hspace{-0.05cm} \theta^-_x \hspace{-0.05cm}\sin\theta \hspace{-0.05cm}+\hspace{-0.05cm}\theta^-_y \hspace{-0.05cm}\cos\theta \hspace{-0.1cm}\\
\hline
\hspace{-0.1cm} e^\phi \theta^-_x & e^\phi \gamma \theta^-_x + \theta^+_x \cos\theta + \theta^+_y \sin \theta & e^\phi \theta^-_x\theta^+_x + \cos \theta & -e^{\phi} \theta^-_x \theta^+_y - \sin\theta \\
\hspace{-0.1cm} -e^\phi \theta^-_y & -e^\phi \gamma \theta^-_y + \theta^+_x \sin\theta - \theta^+_y \cos \theta & -e^{\phi} \theta^-_y \theta^+_x + \sin\theta & e^\phi \theta^-_y\theta^+_y + \cos \theta 
\end{array} \right]
\end{align}
\normalsize

\subsection{Principal series representations}
The principal series representations of the supergroup OSp$(2|2,\mathbb{R})$ were constructed in Appendix B of \cite{Belaey:2023jtr} using the method of parabolic induction. The representations are labeled by a discrete U$(1)$ representation label $q\in \mathbb{Z}/2$, and an imaginary continuous representation label $j=ik$, $k\in \mathbb{R}^+$ as a consequence of unitarity. They act on the space of square-integrable functions on the real superline $f(x| \vartheta_1,\vartheta_2)\in L^2(\mathbb{R}^{1|2})$ as:
\begin{align}
\label{eq:gact}
f(x|&\vartheta_1,\vartheta_2) \to e^{2iq \psi} \, \text{sgn}(bx+d+\beta_2\vartheta_1 + \delta_2 \vartheta_2)^{\epsilon} \vert bx+d+\beta_2\vartheta_1 + \delta_2 \vartheta_2 \vert^{2j} \\
&\times f\left(\frac{ax+c+\alpha_2 \vartheta_1 + \gamma_2 \vartheta_2}{bx+d+\beta_2\vartheta_1 + \delta_2 \vartheta_2}\Big\rvert -\frac{\alpha_1 x + \gamma_1 -w \vartheta_1 - z\vartheta_2}{bx+d+\beta_2\vartheta_1 + \delta_2 \vartheta_2}, -\frac{\beta_1x + \delta_1 - y \vartheta_1 - u \vartheta_2}{bx+d+\beta_2\vartheta_1 + \delta_2 \vartheta_2}\right). \nonumber
\end{align}in terms of the supernumbers corresponding to the fundamental matrix representation of $g\in \text{OSp}(2|2,\mathbb{R})$, the supernumber angle $\psi$ defined by \eqref{eq:cos_psi}-\eqref{eq:sin_psi}, and a $\mathbb{Z}_2$ sign label $\epsilon$.

For this representation, we collect the action of the various one-parameter subgroups on $f(x, \vartheta_1,\vartheta_2)$ as:
\begin{align} 
   \left(e^{2\phi H} \circ f \right)( x,\vartheta_1, \vartheta_2) &=e^{-2\phi j}f(e^{2\phi} x,e^{\phi}\vartheta_1,e^{\phi}\vartheta_2), \nonumber \\
\left(e^{\gamma E_+} \circ f \right)( x,\vartheta_1, \vartheta_2) & =|1+\gamma  x|^{2j}\left(1-\frac{i\gamma\vartheta_1\vartheta_2}{1+\gamma  x}\right)^{2q} f\left( \frac{ x}{1+\gamma  x},\frac{\vartheta_1}{1+\gamma  x},\frac{\vartheta_2}{1+\gamma  x}\right),\nonumber \\
   \left(e^{\beta E_-} \circ f \right)( x,\vartheta_1, \vartheta_2)&=f( x+\beta,\vartheta_1,\vartheta_2),\nonumber \\ 
   \left(e^{2i\theta Z} \circ f \right)( x,\vartheta_1, \vartheta_2) & = (\cos\theta+i\sin\theta)^{2q}f( x,\vartheta_1\cos \theta+\vartheta_2\sin \theta,\vartheta_2\cos\theta-\vartheta_1\sin\theta),\nonumber \\
   \left(e^{\sqrt{2}\theta_x^{+} F_{x}^{+}} \circ f \right)( x,\vartheta_1, \vartheta_2)&=(1\hspace{-0.05cm}-\hspace{-0.05cm}i\theta_x^{+}\vartheta_2)^{2q}|1\hspace{-0.05cm}+\hspace{-0.05cm}\theta_x^{+}\vartheta_1|^{2j}f\hspace{-0.05cm}\left(\hspace{-0.05cm}\frac{ x}{1\hspace{-0.05cm}+\hspace{-0.05cm}\theta_x^{+}\vartheta_1},\frac{\vartheta_1\hspace{-0.05cm}-\hspace{-0.05cm} x\theta_x^{+}}{1\hspace{-0.05cm}+\hspace{-0.05cm}\theta_x^{+}\vartheta_1},\frac{\vartheta_2}{1\hspace{-0.05cm}+\hspace{-0.05cm}\theta_x^{+}\vartheta_1}\hspace{-0.05cm}\right),\nonumber \\
   \left(e^{\sqrt{2}\theta_x^{-} F_{x}^{-}} \circ f \right)( x,\vartheta_1, \vartheta_2)&=f( x+\theta_x^-\vartheta_1,\vartheta_1+\theta_x^-,\vartheta_2),\nonumber \\
    \left(e^{\sqrt{2}\theta_y^{+} F_{y}^{+}} \circ f \right)( x,\vartheta_1, \vartheta_2)  &=\left( 1\hspace{-0.05cm}-\hspace{-0.05cm}i\theta_y^{+}\vartheta_1\right)^{2q}\hspace{-0.05cm}|1\hspace{-0.05cm}-\hspace{-0.05cm}\theta_y^{+}\vartheta_2|^{2j}f\hspace{-0.05cm}\left(\hspace{-0.05cm}\frac{ x}{1\hspace{-0.05cm}-\hspace{-0.05cm}\theta_y^{+}\vartheta_2},\frac{\vartheta_1}{1\hspace{-0.05cm}-\hspace{-0.05cm}\theta_y^{+}\vartheta_2},\frac{\vartheta_2\hspace{-0.05cm}+\hspace{-0.05cm} x\theta_y^{+}}{1\hspace{-0.05cm}-\hspace{-0.05cm}\theta_y^{+}\vartheta_2}\hspace{-0.05cm}\right),\nonumber \\
     \left(e^{\sqrt{2}\theta_y^{-} F_{y}^{-}}\circ f\right)( x,\vartheta_1, \vartheta_2)&=f( x-\theta_y^-\vartheta_2,\vartheta_1,\vartheta_2-\theta_y^-).\label{eq:exponentiatedgroupactions}
\end{align}
These are the exponentiated versions of the differential generators defined in the main text \eqref{eq:generatorH}-\eqref{eq:generatorFy-}. These Grassmann-valued generators satisfy the \textit{opposite} $\mathfrak{osp}(2|2)$ superalgebra, which differs by a sign in the anticommutators of \eqref{eq:fullrealalgebra}. 

The equations of the quadratic and cubic Casimirs for the opposite $\mathfrak{osp}(2|2)$ Lie superalgebra take the form:
\begin{align}
\label{eq:Casimir2}
     \mathcal{C}_2 &=H^2-Z^2+E^-E^+-(F^-\bar{F}^+-\bar{F}^-F^+), \\
    \label{eq:cubicCasimir2}
    \mathcal{C}_3 &=(H^2-Z^{2})Z+E^-E^+(Z-\tfrac{1}{2})+\tfrac{1}{2}F^-\bar{F}^+(H-3Z+1)\\
    &\hspace{1cm}+\tfrac{1}{2}\bar{F}^-F^+(H+3Z+1)-\tfrac{1}{2}E^-\bar{F}^+F^+-\tfrac{1}{2}\bar{F}^-F^-E^+.\nonumber
\end{align} 
One can show that both Casimirs in this representation are proportional to the identity operator and commute with all generators\begin{align}
\label{eq:quadcas}
  \mathcal{C}_2 =j^2-q^2 = - k^2 - q^2,  \qquad     \mathcal{C}_3 =q(j^2-q^2) =  -q(k^2 + q^2).
\end{align} 
This is a consequence of Schur's lemma for the \emph{irreducible} principal series representations.

\subsection{Left- and right-regular realizations of $\mathfrak{osp}(2|2,\mathbb{R})$} \label{appendix:regularrepresentations}
The left- and right-regular generators of $\mathfrak{osp}(2|2,\mathbb{R})$ are realized as first-order differential operators satisfying respectively
\begin{equation}
    \hat{L}_ig=-X_ig, \qquad 
    \label{eq:left}
    \hat{R}_ig=g X_i.
\end{equation}
For the applications of solving the harmonic analysis, we take the Gauss-Euler parametrization \eqref{eq:GEreal} in the principal series representation
\begin{align}
    g = e^{\sqrt{2}\theta^-_{y} F^-_y} e^{\sqrt{2}\theta^-_{x} F^-_x} e^{\beta E^-} e^{2\phi H} e^{2i \theta Z}e^{\gamma E^+} e^{\sqrt{2}\theta^+_{x} F^+_x} e^{\sqrt{2}\theta^+_{y} F^+_y},
\end{align} 
where the exponentiated generators are taken to be the generators of the principal series representation \eqref{eq:generatorH}-\eqref{eq:generatorFy-}, satisfying the opposite superalgebra relations.

Using the exponentiated superalgebra relations, we can deduce the differential operators of the left-regular generators acting on group elements of the principal series representation:
\begin{align}
\sqrt{2}\hat{L}_{F^-_{x}} &= -\left(\partial_{\theta_x^-} - \theta_x^- \partial_\beta \right),\qquad \sqrt{2}\hat{L}_{F^-_{y}} = -\left(\partial_{\theta_y^-} - \theta_y^- \partial_\beta \right),\qquad \hat{L}_{E^-} = -\partial_\beta, \nonumber\\
2\hat{L}_{H} &= -\left(\partial_{\phi} - 2\beta \partial_\beta -\theta_x^- \partial_{\theta_x^-}-\theta_y^-\partial_{\theta_y^-}\right),\qquad 2i\hat{L}_{Z} = - \left(\partial_\theta + \theta_y^- \partial_{\theta_x^-} - \theta_x^- \partial_{\theta_y^-}\right), \nonumber\\
\hat{L}_{E^+} &= -\Big(e^{-2\phi} \partial_\gamma + \beta \partial_\phi - \beta^2 \partial_\beta - \beta \theta_x^- \partial_{\theta_x^-} - e^{-\phi}(\theta_x^-\cos\theta - \theta_y^- \sin \theta ) (\partial_{\theta_x^+}+\theta_x^+ \partial_\gamma)\nonumber\\& - e^{-\phi}(\theta_y^-\cos\theta + \theta_x^- \sin \theta ) (\partial_{\theta_y^+}+\theta_y^+ \partial_\gamma) - \theta_y^- \theta_x^- \partial_\theta-\beta \theta_y^- \partial_{\theta_y^-}\Big)\nonumber,\\
\sqrt{2}\hat{L}_{F^+_{x}} &= -\Big(\hspace{-0.06cm}e^{-\phi} \cos \theta (\partial_{\theta_x^+} \hspace{-0.05cm} + \hspace{-0.05cm} \theta_x^+ \partial_\gamma)\hspace{-0.05cm} + \hspace{-0.05cm} e^{-\phi} \sin \theta (\partial_{\theta_y^+} \hspace{-0.05cm} + \hspace{-0.05cm} \theta_y^+ \partial_\gamma) \hspace{-0.05cm} + \hspace{-0.05cm} \beta (\partial_{\theta_x^-}\hspace{-0.05cm} - \hspace{-0.05cm} \theta_x^- \partial_\beta) \hspace{-0.05cm}\nonumber\\&\qquad+\theta_x^- \partial_\phi\hspace{-0.05cm} - \hspace{-0.05cm}\theta_x^-\theta_y^- \partial_{\theta_y^-} \hspace{-0.05cm}- \hspace{-0.05cm}\theta_y^- \partial_\theta \hspace{-0.06cm} \Big), \nonumber\\
\sqrt{2}\hat{L}_{F^+_{y}} &= -\Big(\hspace{-0.06cm}e^{-\phi} \cos \theta (\partial_{\theta_y^+} \hspace{-0.05cm} + \hspace{-0.05cm} \theta_y^+ \partial_\gamma) \hspace{-0.05cm}- \hspace{-0.05cm}e^{-\phi} \sin \theta (\partial_{\theta_x^+} \hspace{-0.05cm}+ \hspace{-0.05cm}\theta_x^+ \partial_\gamma) \hspace{-0.05cm}+ \hspace{-0.05cm}\beta (\partial_{\theta_y^-}\hspace{-0.05cm}-\hspace{-0.05cm}\theta_y^- \partial_\beta) \hspace{-0.05cm}+\hspace{-0.05cm}\theta_y^- \partial_\phi\hspace{-0.05cm}\nonumber\\ &\qquad-\hspace{-0.05cm} \theta_y^-\theta_x^- \partial_{\theta_x^-} \hspace{-0.05cm}+\hspace{-0.05cm} \theta_x^- \partial_\theta\hspace{-0.06cm} \Big).
\end{align}
These generators themselves satisfy the opposite Lie superalgebra, which can be checked explicitly. Alternatively, since the Grassmann-valued fermionic generators of the left-regular realization anticommute with the Grassmann-valued fermionic principal series generators, the former satisfy \begin{align}
\label{eq:intermediatealgebra}
    \{\hat{L}_{X_i},\hat{L}_{X_j}\}\;g
    =-\{X_i,X_j\}\;g
\end{align}
from the definition of the left-regular action \eqref{eq:left}, which leads to the same structure constants of the principal series generators $X_i$. 

Analogously, we can deduce the right-regular generators as:
\begin{align}
    \sqrt{2}\hat{R}_{F_y^+}&=\partial_{\theta_y^+}-\theta_y^+\partial_\gamma,\qquad\qquad
    \sqrt{2}R_{F_x^+}=\partial_{\theta_x^+}-\theta_x^+\partial_\gamma,\qquad\qquad
    \hat{R}_{E^+}=\partial_\gamma,\nonumber\\
    2i\hat{R}_Z&=\partial_\theta+\theta_x^+\partial_{\theta_y^+}-\theta_y^+\partial_{\theta_x^+},\qquad\qquad
    2\hat{R}_H=\partial_\phi -2\gamma\partial_\gamma-\theta_x^+\partial_{\theta_x^+}-\theta_y^+\partial_{\theta_y^+},\nonumber\\
    \hat{R}_{E^-}&=e^{-2\phi }\partial_\beta+\gamma\partial_\phi-\gamma^2\partial_\gamma-\theta_x^+\theta_y^+\partial_\theta-\gamma( \theta_x^+\partial_{\theta_x^+}+\theta_y^+\partial_{\theta_y^+})\nonumber\\&+e^{-\phi}\left(\theta_x^+\cos\theta +\theta_y^+\sin\theta\right)(\partial_{\theta_x^-}+\theta_x^-\partial_\beta)
    + e^{-\phi}\left(\theta_y^+\cos\theta-\theta_x^+\sin\theta\right)(\partial_{\theta_y^-}+\theta_y^-\partial_\beta),\nonumber\\
    \sqrt{2}\hat{R}_{F_x^-}&=e^{-\phi}\cos\theta \hspace{-0.05cm}\left(\hspace{-0.05cm}\partial_{\theta_x^-}\hspace{-0.05cm}+\hspace{-0.05cm}\theta_x^-\partial_\beta\hspace{-0.05cm}\right)\hspace{-0.05cm}-\hspace{-0.05cm}e^{-\phi}\sin\theta\hspace{-0.05cm} \left(\hspace{-0.05cm}\partial_{\theta_y^-}\hspace{-0.05cm}+\hspace{-0.05cm}\theta_y^-\partial_\beta\hspace{-0.05cm}\right)\hspace{-0.05cm}-\gamma \hspace{-0.05cm}\left(\partial_{\theta_x^+} \hspace{-0.05cm} - \hspace{-0.05cm} \theta_x^+\partial_\gamma\right)\nonumber\\&\quad-\theta_x^+\partial_\phi-\theta_y^+ \partial_\theta -\theta_y^+\theta_x^+\partial_{\theta_y^+},\nonumber\\
    \sqrt{2}\hat{R}_{F_y^-}&=e^{-\phi}\sin\theta\hspace{-0.02cm}\left(\hspace{-0.05cm}\partial_{\theta_x^-}\hspace{-0.05cm}+\hspace{-0.05cm}\theta_x^-\partial_\beta\hspace{-0.05cm}\right)\hspace{-0.05cm}+\hspace{-0.05cm}e^{-\phi}\cos\theta \hspace{-0.05cm}\left(\hspace{-0.05cm}\partial_{\theta_y^-}\hspace{-0.05cm} + \hspace{-0.05cm} \theta_y^-\partial_\beta\hspace{-0.05cm}\right)\hspace{-0.05cm}-\hspace{-0.05cm}\gamma\hspace{-0.05cm}\left(\hspace{-0.05cm}\partial_{\theta_y^+}-\theta_y^+\partial_\gamma\right)\hspace{-0.05cm}\nonumber\\&\quad+\hspace{-0.05cm}\theta_x^+\partial_\theta\hspace{-0.05cm}-\hspace{-0.05cm}\theta_y^+\partial_\phi\hspace{-0.05cm} +\hspace{-0.05cm}\theta_y^+\theta_x^+\partial_{\theta_x^+}.
    \label{eq:rightgen}
\end{align} 
These generators also satisfy the opposite superalgebra, which can again be checked explicitly, or using $\{\hat{R}_{X_i},\hat{R}_{X_j}\}\;g=
    g\;\{X_i,X_j\}$ from the definition of the right-regular action \eqref{eq:left}, leading again to the same structure constants of the principal series generators $X_i$.
    
We can streamline the notation by introducing the lightcone fermionic variables
\begin{equation}
    \begin{aligned}
        \theta^+& = \theta^+_x-i\theta^+_y, \qquad \bar{\theta}^+ = \theta^+_x + i \theta^+_y,\\
\theta^-& = \theta^-_x-i\theta^-_y, \qquad \bar{\theta}^- = -\theta^-_x - i \theta^-_y,
\label{eq:thetapm}
    \end{aligned}
\end{equation}
and the lightcone generators \eqref{eq:generatorstransf}: 
\begin{alignat}{2}
     F^+&=\frac{1}{\sqrt{2}}(F_x^++iF_y^+), \qquad &\overline{F}^+&=\frac{1}{\sqrt{2}}(F_x^+-iF_y^+),\\ F^-&=\frac{1}{\sqrt{2}}(F_x^-+iF_y^-), \qquad &\overline{F}^-&=-\frac{1}{\sqrt{2}}(F_x^--iF_y^-).
\end{alignat}
The generators can then be rewritten in this lightcone basis as:
\begin{align}
    \hat{L}_{F^-}&=-\Big(\partial_{\theta^-}+\frac{1}{2} \overline{\theta}^-\partial_\beta\Big),\qquad 
    \hat{L}_{\overline{F}^-}=-\Big(\partial_{\overline{\theta}^-}+\frac{1}{2}\theta^-\partial_\beta\Big),\qquad
    \hat{L}_{E^-}=-\partial_\beta,\nonumber  \\
   \hat{L}_H&=-\frac{1}{2}\left(\partial_\phi-2\beta \partial_\beta-\theta^-\partial_{\theta^-}-\overline{\theta}^-\partial_{\overline{\theta}^-}\right) ,\qquad
    \hat{L}_{Z}= \frac{1}{2}\left(i\partial_\theta -\theta^-\partial_{\theta^-}+\overline{\theta}^-\partial_{\overline{\theta}^-}\right),\nonumber\\
    \hat{L}_{E^+}&= -e^{-2\phi}\partial_\gamma-\beta\partial_\phi+\beta^2\partial_\beta+\beta\theta^-\partial_{\theta^-}+\beta\overline{\theta}^-\partial_{\overline{\theta}^-}\nonumber\\&+e^{-\phi-i\theta}\theta^-\Big(\partial_{\theta^+}+\frac{\overline{\theta}^+}{2}\partial_\gamma\Big)-e^{-\phi+i\theta}\overline{\theta}^-\Big(\partial_{\overline{\theta}^+}+\frac{\theta^+}{2}\partial_\gamma\Big)-\frac{i}{2}\theta^-\overline{\theta}^-\partial_\theta\nonumber,\\
    \hat{L}_{F^+}&=-e^{-\phi-i\theta}\Big(\partial_{\theta^+}+ \frac{\overline{\theta}^+}{2}\partial_\gamma\Big)-\beta\Big(\partial_{\theta^-}+\frac{\overline{\theta}^-}{2}\partial_\beta\Big)+\frac{\overline{\theta}^-}{2}\partial_\phi+i\frac{\overline{\theta}^-}{2}\partial_\theta+\frac{1}{2}\theta^-\overline{\theta}^-\partial_{\theta^-}~,\nonumber\\
    \hat{L}_{\overline{F}^+}&= -e^{-\phi+i\theta}\Big(\partial_{\overline{\theta}^+}+\frac{\theta^+}{2}\partial_\gamma \Big)+\beta \Big(\partial_{\overline{\theta}^-}+\frac{1}{2}\theta^-\partial_\beta\Big)-\frac{\theta^-}{2}\partial_\phi+i\frac{\theta^-}{2}\partial_\theta+\frac{1}{2}\theta^-\overline{\theta}^-\partial_{\overline{\theta}^-}~.
    \label{eq:leftreg}
\end{align}
and
\begin{align}\label{eq:oppositealgebrarightgeneratorslightcone}
    \hat{R}_{F^+}&=\partial_{\theta^+}-\frac{\overline{\theta}^+}{2}\partial_\gamma,\qquad
    \hat{R}_{\overline{F}^+}= \partial_{\overline{\theta}^+}-\frac{\theta^+}{2}\partial_\gamma,\qquad \hat{R}_{E^+}=\partial_\gamma,\nonumber\\
    \hat{R}_Z&=-\frac{i}{2}\partial_\theta+\frac{1}{2}\overline{\theta}^+\partial_{\overline{\theta}^+}-\frac{1}{2}\theta^+\partial_{\theta^+},\qquad
    \hat{R}_H=\frac{\partial_\phi}{2}-\gamma\partial_\gamma-\frac{\theta^+}{2}\partial_{\theta^+}-\frac{\overline{\theta}^+}{2}\partial_{\overline{\theta}^+},\nonumber\\
    \hat{R}_{E^-}&=e^{-\phi+i\theta}\theta^+\Big(\partial_{\theta^-}-\frac{\overline{\theta}^-}{2}\partial_\beta\Big)-e^{-\phi-i\theta}\overline{\theta}^+\Big(\partial_{\overline{\theta}^-}-\frac{\theta^-}{2}\partial_\beta\Big)+e^{-2\phi}\partial_\beta\nonumber\\
    &+\gamma\partial_\phi-\gamma^2\partial_\gamma-\gamma(\theta^+\partial_{\theta^+}+\overline{\theta}^+\partial_{\overline{\theta}^+})+\frac{i}{2}\theta^+\overline{\theta}^+\partial_\theta,  \nonumber\\
    \hat{R}_{F^-}&=e^{-\phi+i\theta}\Big(\partial_{\theta^-}-\frac{\overline{\theta}^-}{2}\partial_\beta\Big)-\gamma \Big(\partial_{\theta^+}-\frac{\overline{\theta}^+}{2}\partial_\gamma\Big)-\frac{\overline{\theta}^+}{2}\partial_\phi+\frac{i}{2}\overline{\theta}^+\partial_{\theta}-\frac{1}{2}\theta^+\overline{\theta}^+\partial_{\theta^+},\nonumber\\
    \hat{R}_{\overline{F}^-}&=e^{-\phi-i\theta}\Big(\partial_{\overline{\theta}^-}-\frac{\theta^-}{2}\partial_\beta\Big)+\gamma\Big(\partial_{\overline{\theta}^+}-\frac{\theta^+}{2}\partial_\gamma\Big)+\frac{\theta^+}{2}\partial_\phi+\frac{i}{2}\theta^+\partial_\theta-\frac{1}{2}\theta^+\overline{\theta}^+\partial_{\overline{\theta}^+}.
\end{align}

\subsection{Haar measure on the OSp$(2|2,\mathbb{R})$ group manifold}
\label{subsection:HaarMeasure}
Using the decomposition \eqref{eq:GEreal} in the fundamental representation \eqref{eq:GEexpli}, we can work out the Cartan-Maurer one-form $g^{-1}dg$ in terms of these generators into the lengthy expression (which we include here for reference):

\vspace{-0.4cm}
{\tiny
\begin{align}
        &g^{-1}dg=\;{\color{blue}2H} \Big[d\phi -e^{2\phi}\gamma d\beta-(e^{2\phi}\gamma \theta_x^-+e^\phi \theta_x^+\cos\theta+e^\phi \theta_y^+\sin\theta)d\theta_x^- -(e^{2\phi}\gamma \theta_y^--e^\phi \theta_x^+\sin\theta+\theta_y^+\cos\theta e^\phi)d\theta_y^-\Big] \nonumber\\
        &+{\color{blue}E^-}\Big[d\beta e^{2\phi}+e^{2\phi}\theta_x^-d\theta_x^-+e^{2\phi}\theta_y^-d\theta_y^-\Big] \nonumber\\
        &+{\color{blue}E^+}\Big[2\gamma d\phi+d\gamma-\gamma^2e^{2\phi}d\beta-(2e^\phi\gamma\theta_x^+\cos\theta +2e^\phi\gamma\theta_y^+\sin\theta +e^{2\phi}\gamma^2\theta_x^-)d\theta_x^-\nonumber\\&\qquad -\theta_x^+d\theta_x^++(2e^\phi \gamma\sin\theta\theta_x^+-2e^\phi\gamma \cos\theta\theta_y^+-e^{2\phi} \gamma^2\theta_y^-)d\theta_y^--\theta_y^+d\theta_y^++2\theta_y^+\theta_x^+d\theta\Big]\nonumber\\&
        +{\color{blue}2iZ}\Big[e^{2\phi}\theta_y^+\theta_x^+d\beta+\left(-e^\phi \theta_y^+\cos\theta+e^\phi\theta_x^+\sin\theta+e^{2\phi}\theta_x^-\theta_y^+\theta_x^+\right)d\theta_x^-+\left(e^\phi\theta_y^+\sin\theta+e^\phi \theta_x^+\cos\theta+e^{2\phi}\theta_y^-\theta_y^+\theta_x^+\right)d\theta_y^-+d\theta\Big]\nonumber\\
        &+{\color{blue}\sqrt{2}F_x^+}\Big[\theta_x^+d\phi-e^{2\phi}\gamma \theta_x^+d\beta+d\theta_x^++(\cos\theta e^\phi \gamma+\sin\theta e^\phi \theta_y^+\theta_x^++e^{2\phi}\gamma \theta_x^-\theta_x^+)d\theta_x^-\nonumber\\&\qquad+(-\sin\theta e^\phi \gamma +\cos\theta e^\phi \theta_y^+\theta_x^++e^{2\phi}\gamma \theta_y^-\theta_x^+)d\theta_y^-+\theta_y^+d\theta\Big]\nonumber\\
        &+{\color{blue}\sqrt{2}F_y^+}\Big[\theta_y^+d\phi-e^{2\phi}\gamma \theta_y^+d\beta +d\theta_y^++(e^\phi\gamma \sin\theta+e^{2\phi}\gamma \theta_x^-\theta_y^++e^\phi \cos\theta\theta_x^+\theta_y^+)d\theta_x^-\nonumber\\&\qquad+(e^\phi \gamma \cos\theta+e^{2\phi}\gamma \theta_y^-\theta_y^+-e^\phi \theta_x^+\theta_y^+\sin\theta)d\theta_y^--\theta_x^+d\theta\Big]\nonumber\\
        &+{\color{blue}\sqrt{2}F_x^-}\left[-e^{2\phi}\theta_x^+d\beta+(e^\phi \cos\theta+e^{2\phi}\theta_x^-\theta_x^+)d\theta_x^-+(e^{2\phi}\theta_y^-\theta_x^+-e^\phi \sin\theta)d\theta_y^-\right]\nonumber\\ 
        &+{\color{blue}\sqrt{2}F_y^-}\Big[-e^{2\phi} \theta_y^+d\beta+(e^\phi\sin\theta+e^{2\phi}\theta_x^-\theta_y^+)d\theta_x^-+(e^\phi\cos\theta+e^{2\phi}\theta_y^-\theta_y^+)d\theta_y^-\Big].
\end{align}}
\normalsize
We write this one-form in terms of a matrix $I^i_{\;\;j}$ as\begin{equation}
    g^{-1}dg=\sum_{ij}X_i I^i_{\;\;j} dx^j,
\end{equation}
with $X_i$ the fundamental algebra generators, and 
\begin{equation}
    I^i_{\;\;j}=\left[\begin{array}{c | c} 
A & B  \\
\hline 
C& D   \\
\end{array}\right]
\end{equation} 
a square $4|4$-dimensional matrix whose rows are ordered as $\{H,Z,E^+,E^-, F_x^+, F_x^-, F_y^+, F_y^-\}$ and the columns are ordered as $\{\phi, \beta, \gamma, \theta, \theta_x^+, \theta_x^-, \theta_y^+,\theta_y^-\}$. We collect the entries here:
\begin{equation}
   A= \begin{bmatrix}
      2&-2e^{2\phi}\gamma &0&0\\ 0&2i e^{2\phi}\theta_y^+\theta_x^+&0&2i\\ 2\gamma &-\gamma^2e^{2\phi}&1&2\theta_y^+\theta_x^+ \\ 0&e^{2\phi}&0&0  
    \end{bmatrix},
\end{equation}
\begin{align}
    &D\hspace{-0.05cm}-\hspace{-0.05cm}CA^{-1}B\hspace{-0.05cm}=\hspace{-0.1cm}\begin{bmatrix}
        \sqrt{2}&\hspace{-0.1cm}\sqrt{2}\cos\theta e^\phi\gamma\hspace{-0.05cm}-\hspace{-0.05cm}\sqrt{2}\sin\theta e^\phi \theta_y^+\theta_x^+&\hspace{-0.1cm}0\hspace{-0.1cm}&\hspace{-0.1cm}-\sqrt{2}\sin\theta e^\phi \gamma \hspace{-0.05cm}-\hspace{-0.05cm}\sqrt{2}\cos\theta e^\phi \theta_y^+\theta_x^+\\ 0 &\hspace{-0.1cm}\sqrt{2}e^\phi \cos\theta &\hspace{-0.1cm}0\hspace{-0.1cm}&\hspace{-0.1cm}-\sqrt{2}e^\phi \sin\theta\\ 0&\hspace{-0.1cm}\sqrt{2}e^\phi \gamma\sin\theta\hspace{-0.05cm}-\hspace{-0.05cm}\sqrt{2}e^\phi\cos\theta\theta_x^+\theta_y^+&\hspace{-0.1cm}\sqrt{2}\hspace{-0.1cm}&\hspace{-0.1cm}\sqrt{2}e^\phi\gamma \cos\theta\hspace{-0.05cm}+\hspace{-0.05cm}\sqrt{2}e^\phi \theta_x^+\theta_y^+\sin\theta\\ 0&\hspace{-0.1cm}\sqrt{2}e^\phi \sin \theta&\hspace{-0.1cm}0\hspace{-0.1cm}&\sqrt{2}e^\phi \cos\theta
    \end{bmatrix}\hspace{-0.1cm}.
\end{align}
Taking the determinant of $A$ and $D-CA^{-1}B$ leads to the Berezinian of $I$:
\begin{equation}
    \text{Ber}(I)\equiv\frac{\det(A)}{\det(D-CA^{-1}B)}=-i.
\end{equation}
Thus, the Haar measure of OSp$(2|2,\mathbb{R})$ in Gauss-Euler coordinates is an overall constant factor, independent of $\phi$ and $\theta$.

\subsection{Relation between OSp$(2\vert 2,\mathbb{R})$, SU$(1,1 \vert 1)$ and SL$(2 \vert 1, \mathbb{R})$}
For completeness, we review the relation between three distinct (but related) matrix supergroups: OSp$(2\vert 2,\mathbb{R})$ vs SU$(1,1 \vert 1)$ vs SL$(2 \vert 1, \mathbb{R})$. The first two are related by a 2:1 homomorphism as
\begin{equation}
    \text{OSp}(2|2,\mathbb{R})/\mathbb{Z}_2\, \simeq \, \text{SU}(1,1|1),
\end{equation}
where we mod out by the OSp$(2|2,\mathbb{R})$ supermatrix diag$(1,1 \vert 1,-1)$, which is an element in the second component of the supermanifold, coming from the difference of O(2) on the LHS with the SO$(2)\simeq \text{U}(1)$ bosonic substructure on the RHS.

The last supergroup SL$(2 \vert 1, \mathbb{R})$ is a different real form of the same complexified matrix supergroup. This can be seen explicitly as follows. 
The fundamental representation of $\mathfrak{sl}(2\vert 1)$ is constructed using the (super)matrix generators \cite{Frappat:1996pb}:
\begin{align}
\label{eq:sl21gen}
H &= \left[\begin{array}{ccc} \frac{1}{2} & 0 & 0 \\ 0 & -\frac{1}{2} & 0 \\ 0 & 0 & 0  \end{array} \right], Z = \left[\begin{array}{ccc} \frac{1}{2} & 0 & 0 \\ 0 & \frac{1}{2} & 0 \\ 0 & 0 & 1  \end{array} \right], E^+ =\left[\begin{array}{ccc} 0 & 1 & 0 \\ 0 & 0 & 0 \\ 0 & 0 & 0  \end{array} \right], E^- = \left[\begin{array}{ccc} 0 & 0 & 0 \\ 1 & 0 & 0 \\ 0 & 0 & 0  \end{array} \right], \nonumber \\
F^+ &= \left[\begin{array}{ccc} 0 & 0 & 0 \\ 0 & 0 & 0 \\ 0 & 1 & 0  \end{array} \right], \bar{F}^+ = \left[\begin{array}{ccc} 0 & 0 & 1 \\ 0 & 0 & 0 \\ 0 & 0 & 0  \end{array}\right], F^- = \left[\begin{array}{ccc} 0 & 0 & 0 \\ 0 & 0 & 0 \\ 1 & 0 & 0  \end{array} \right], \bar{F}^- = \left[\begin{array}{ccc} 0 & 0 & 0 \\ 0 & 0 & 1 \\ 0 & 0 & 0  \end{array} \right],
\end{align}
satisfying \eqref{eq:superalgebra}. These parametrize all $(2 \vert 1)$-dimensional real matrices whose supertrace vanishes. Exponentiating gives the real supergroup SL$(2 \vert 1, \mathbb{R})$, with maximal bosonic subgroup SL$(2,\mathbb{R}) \otimes \, \mathbb{R}$. One can flip sign $Z \to iZ$ to reach the real form whose maximal bosonic subgroup is SL$(2,\mathbb{R}) \otimes \, \text{U}(1)$. This real form is not equal to the real forms OSp$(2\vert 2,\mathbb{R})$ or $\text{SU}(1,1|1)$ of relevance for supergravity.
To transfer between them, in analogy with how the isomorphism $\text{SL}(2,\mathbb{R}) \simeq \text{SU}(1,1)$ works in the bosonic case, we can conjugate the $4 \vert 4$ generators \eqref{eq:sl21gen} with the SL$(2\vert 1, \mathbb{C})$ matrix: 
\begin{equation}
t = \frac{1}{\sqrt{2}} \left[\begin{array}{ccc} 1 & i & 0 \\ i & 1 & 0 \\ 0 & 0 & \sqrt{2}  \end{array} \right],
\end{equation}
and obtain:
\begin{align}
\label{eq:sugener}
H &= \left[\begin{array}{ccc} 0 & \frac{i}{2} & 0 \\ -\frac{i}{2} & 0 & 0 \\ 0 & 0 & 0  \end{array} \right]\hspace{-1.4mm}, Z = \left[\begin{array}{ccc} \frac{1}{2} & 0 & 0 \\ 0 & \frac{1}{2} & 0 \\ 0 & 0 & 1  \end{array} \right]\hspace{-1.4mm}, E^+ =\left[\begin{array}{ccc} \frac{i}{2} & \frac{1}{2} & 0 \\ \frac{1}{2} & -\frac{i}{2} & 0 \\ 0 & 0 & 0  \end{array} \right]\hspace{-1.4mm}, E^- = \left[\begin{array}{ccc} -\frac{i}{2} & \frac{1}{2} & 0 \\ \frac{1}{2} & \frac{i}{2} & 0 \\ 0 & 0 & 0  \end{array} \right]\hspace{-1.4mm}, \nonumber \\
F^+ &= \left[\begin{array}{ccc} 0 & 0 & \hspace{-1.4mm}0 \\ 0 & 0 & \hspace{-1.4mm}0 \\ \hspace{-1.4mm}-\frac{i}{\sqrt{2}} & \hspace{-1.4mm}\frac{1}{\sqrt{2}} & \hspace{-1.4mm} 0  \end{array} \right]\hspace{-1.4mm}, \bar{F}^+ = \left[\begin{array}{ccc} 0 & 0 & \hspace{-1.4mm}\frac{1}{\sqrt{2}} \\ 0 & 0 & \hspace{-1.4mm}\frac{i}{\sqrt{2}} \\ 0 & 0 & 0  \end{array}\right]\hspace{-1.4mm}, F^- = \left[\begin{array}{ccc} 0 & 0 & \hspace{-1.4mm}0 \\ 0 & 0 & \hspace{-1.4mm}0 \\ \hspace{-1.4mm}\frac{1}{\sqrt{2}} & \hspace{-1.4mm}-\frac{i}{\sqrt{2}} & \hspace{-1.4mm}0  \end{array} \right]\hspace{-1.4mm}, \bar{F}^- = \left[\begin{array}{ccc} \hspace{-1.4mm} 0 & 0 & \hspace{-1.4mm}\frac{i}{\sqrt{2}} \\ \hspace{-1.4mm} 0 & 0 & \hspace{-1.4mm}\frac{1}{\sqrt{2}} \\ \hspace{-1.4mm} 0 & 0 & 0  \end{array} \right].
\end{align}
Then the following linear combinations of \eqref{eq:sugener} 
\begin{equation}
\label{eq:gensu11}
H,\,\, Z,\,\,E^+,\,\,E^-,\,\,F^++\bar{F}^+,\,\, i(F^+-\bar{F}^+),\,\, F^--\bar{F}^-,\,\, i (F^- + \bar{F}^-),
\end{equation}
are suitable matrix generators of the group SU$(1,1 \vert 1)$, which is defined as the set of complex supermatrices $g$ preserving the form diag$(1,-1\vert 1)$ as
\begin{equation}
g^\dagger \left[\begin{array}{cc|c} 1 & 0 & 0 \\ 0 & -1 & 0 \\ \hline 0 & 0 & 1  \end{array} \right] g =  \left[\begin{array}{cc|c} 1 & 0 & 0 \\ 0 & -1 & 0 \\ \hline 0 & 0 & 1  \end{array} \right],
\end{equation}
where the super-adjoint ${}^{\dagger}$ is defined by the expression
\begin{equation}
\left[ \begin{array}{c|c} A & B \\ \hline C & D \end{array} \right]^\dagger \equiv \left[ \begin{array}{c|c} A^\dagger & iC^\dagger \\ \hline i B^\dagger & D^\dagger \end{array} \right].
\end{equation}

Due to the factors of $i$ in \eqref{eq:gensu11} in the fermionic generators, we conclude that this is a different real form of the same complexified algebra, unlike in the bosonic case where these are the same real form.
Note that the fermionic combinations in \eqref{eq:gensu11} are to be identified with $F_x^\pm$ and $F_y^\pm$ as defined in \eqref{eq:generatorstransf}. From this we can conclude that the Lie superalgebras of SU$(1,1\vert 1)$ (found here) and of OSp$(2\vert 2,\mathbb{R})$ (defined in \eqref{eq:fullrealalgebra}) coincide, and hence the component of their Lie supergroups connected to the identity also matches.

\section{Hamiltonian reduction and supersymmetric quantum mechanics}\label{sec:hamiltonianreduction}
In this appendix, we elaborate on the Hamiltonian reduction of section \ref{sec:hamreduction}. In particular, we check that the representation matrix elements of the principal series representation solve the Casimir eigenvalue equation. We then confirm that by choosing gravitational Brown-Henneaux boundary conditions, the Casimir acting on the group element reduces to the $\mathcal{N}=2$ minisuperspace Liouville Hamiltonian acting on the Whittaker function $\hat{H}\Psi_{k,q} (\ell,\theta)=E(k,q)\;\Psi_{k,j}(\ell,\theta)$, where the Liouville Hamiltonian is defined by \cite{Lin:2022zxd}:
\begin{align}
\label{eq:N=2Liouville}
    \Big(-\partial_\ell^2-\frac{1}{4}\partial_\theta^2+e^{-\ell} +ie^{-\ell/2-i\theta}\;\overline{\psi}_-\psi_++ie^{-\ell/2+i\theta}\;\psi_-\overline{\psi}_+\Big)\Psi_{k,q}(\ell,\theta)=E(k,q)\; \Psi_{k,q}(\ell,\theta).
\end{align}

We start by considering the expression for the opposite algebra Casimir operator in the left-regular representation, given by \eqref{eq:Casimir2}: 
\begin{align}
    \hat{\mathcal{C}}=\hat{L}_H^2-\hat{L}_Z^2+\hat{L}_{E^-}\hat{L}_{E^+}+\hat{L}_{\overline{F}^-}\hat{L}_{F^+}-\hat{L}_{F^-}\hat{L}_{\overline{F}^+}
\end{align}
and the action 
\begin{align}
    \hat{L}_{X_i}g=-X_ig,\qquad \hat{R}_{X_i}g=gX_i. \label{eq:regularaction}
\end{align}
This Casimir then acts on the representation matrix elements as: 
\begin{align}
\left(\hat{L}_H^2-\hat{L}_Z^2+\hat{L}_{E^-}\hat{L}_{E^+}+\hat{L}_{\overline{F}^-}\hat{L}_{F^+}-\hat{L}_{F^-}\hat{L}_{\overline{F}^+}\right)\braket{\nu,\psi_1,\psi_2\mid g\mid\lambda, \psi_3,\psi_4 }\nonumber\\ =\braket{\nu,\psi_1,\psi_2\mid \left(H^2-Z^2+E_+E_--F_+\overline{F}_-+\overline{F}_+F_-\right)g\mid\lambda, \psi_3,\psi_4}.
\end{align}
Note that the generators on the second line are the generators in the principal series representation. As a consequence, care has to be taken in the ordering of the fermionic generators, since both the fermionic left-regular and principal series generators are Grassmann-valued and anticommute in the consecutive left-regular action. Using the relations of the opposite algebra, we can rewrite this in terms of the opposite algebra Casimir of the principal series representation \eqref{eq:Casimir2}: 
\begin{align}
    =\braket{\nu,\psi_1,\psi_2\mid \left(H^2-Z^2+E_-E_++\overline{F}_-F_+-F_-\overline{F}_+\right)g\mid\lambda, \psi_3,\psi_4}.
\end{align} 
Since the principal series generators furnish an irreducible representation, the principal series Casimir in representation $(j,q)$ is proportional to the unit operator $\mathcal{C}_2=j^2-q^2=-k^2-q^2$, with $j=ik$:
\begin{align}
    \left(\hat{L}_H^2-\hat{L}_Z^2+\hat{L}_{E^-}\hat{L}_{E^+}+\hat{L}_{\overline{F}^-}\hat{L}_{F^+}-\hat{L}_{F^-}\hat{L}_{\overline{F}^+}\right)\braket{\nu,\psi_1,\psi_2\mid g\mid\lambda, \psi_3,\psi_4 }\nonumber\\ = (-k^2-q^2)\braket{\nu,\psi_1,\psi_2\mid g\mid\lambda, \psi_3,\psi_4 }. \label{eq:harmonicanalysis}
\end{align}
The same reasoning holds for the right-regular realization, where the opposite ordering in the Casimir is trivially preserved by the definition of the right-regular action \eqref{eq:regularaction}.

\subsection{Hamiltonian reduction of the harmonic analysis}
By construction of the mixed parabolic matrix element, the parabolic generators diagonalize to the left of the Whittaker function \eqref{eq:parabolicfactorization} \begin{align}
    &\braket{\nu, \psi_1,\psi_2\mid g\mid \lambda, \psi_3,\psi_4}=e^{-\theta_x^- \psi_1\sqrt{\nu}} e^{-\theta_y^- \psi_2 \sqrt{\nu}}e^{-i\theta_x^+\psi_3\sqrt{\lambda}}e^{-i\theta_y^+\psi_4\sqrt{\lambda}}e^{\beta\nu}e^{-\lambda \gamma} \;\Psi_{k,q}(\phi,\theta),\nonumber
\end{align} where we choose the appropriate phase factors in the left and right parabolic eigenvalues in order to have this property.

We can explicitly write the mutual Casimir evaluated in both left- and right-regular realizations \eqref{eq:leftreg},\eqref{eq:oppositealgebrarightgeneratorslightcone} as the Laplacian on the group manifold: 
\begin{align}
\hat{\mathcal{C}}_2=&\;
\frac{1}{4}\partial_\theta^2+\frac{1}{4}\partial_\phi^2+e^{-2\phi}\partial_\beta \partial_\gamma +e^{-\phi-i\theta}D_{\overline{\theta}^-}D_{\theta^+}-e^{-\phi+i\theta}D_{\theta^-}D_{\overline{\theta}^+}
\label{eq:casimirDifferentials}
\end{align}
with the superderivatives defined as: \begin{equation}\label{eq:superderivatives}
    \begin{aligned}     D_{\overline{\theta}^-}=\partial_{\overline{\theta}^-}-\frac{1}{2}\theta^-\partial_\beta, \qquad D_{\theta^+}=\partial_{\theta^+}+\frac{1}{2}\overline{\theta}^+\partial_\gamma, \\
    D_{\theta^-}=\partial_{\theta^-}-\frac{1}{2}\overline{\theta}^-\partial_\beta, \qquad D_{\overline{\theta}^+}=\partial_{\overline{\theta}^+}+\frac{1}{2}\theta^+\partial_\gamma.
    \end{aligned}
\end{equation}
These differential operators have the property to diagonalize the parabolic exponential prefactors to the right:\footnote{\label{footnote:relativesigns}The superderivatives have a relative sign difference compared to the left- and right-regular generators of the respective left- and right parabolic generators, which diagonalize the exponential prefactors to the left.} 
 \begin{equation}
    \begin{aligned}\label{eq:N=2covariantderivatives}
        &D_{\overline{\theta}^-} e^{-\theta_x^- \psi_1\sqrt{\nu}} e^{-\theta_y^- \psi_2 \sqrt{\nu}}e^{\beta \nu}=e^{-\theta_x^- \psi_1\sqrt{\nu}} e^{-\theta_y^- \psi_2 \sqrt{\nu}}e^{\beta\nu}\;\overline{\psi}_-\sqrt{\nu}, \\
    & D_{\theta^-}e^{-\theta_x^- \psi_1\sqrt{\nu}} e^{-\theta_y^- \psi_2 \sqrt{\nu}}e^{\beta\nu}= e^{-\theta_x^- \psi_1\sqrt{\nu}} e^{-\theta_y^- \psi_2 \sqrt{\nu}}e^{\beta \nu}\;(-\psi_-\sqrt{\nu}), \\
    &D_{\overline{\theta}^+}e^{-i\theta_x^+\psi_3\sqrt{\lambda}}e^{-i\theta_y^+\psi_4\sqrt{\lambda}}e^{-\lambda \gamma}=e^{-i\theta_x^+\psi_3\sqrt{\lambda}}e^{-i\theta_y^+\psi_4\sqrt{\lambda}}e^{-\lambda\gamma}\;(-i\overline{\psi}_+\sqrt{\lambda}),\\
    &D_{\theta^+}e^{-i\theta_x^+\psi_3\sqrt{\lambda}}e^{-i\theta_y^+\psi_4\sqrt{\lambda}}e^{-\lambda\gamma}= e^{-i\theta_x^+\psi_3\sqrt{\lambda}}e^{-i\theta_y^+\psi_4\sqrt{\lambda}}e^{-\lambda\gamma}\;(-i\psi_+\sqrt{\lambda}).
    \end{aligned}
\end{equation} with the coordinate transformations defined in \eqref{eq:coordiantetransf}  and the transformations between the real and lightcone fermions \eqref{eq:fermiontransf}. 

After acting on the full mixed parabolic matrix element, we can remove the parabolic exponential prefactors multiplying both sides of the Casimir eigenvalue equation on the left, leading to the reduced Casimir eigenvalue equation acting on the Whittaker function
 \begin{align}
    &\Big(-\frac{1}{4}\partial_\theta^2-\frac{1}{4}\partial_\phi^2+e^{-2\phi}\lambda\nu \nonumber+ie^{-\phi-i\theta}\sqrt{\nu\lambda}\;\overline{\psi}_-\psi_++ie^{-\phi+i\theta}\sqrt{\nu\lambda}\;\psi_-\overline{\psi}_+\Big)\; \Psi_{k,q}(\phi,\theta)\nonumber\\&=(k^2+q^2)\;\Psi_{k,q}(\phi,\theta).
\end{align}
Therefore, choosing the mixed parabolic boundary conditions on the principal series matrix element with  $\nu=\lambda=1$, the Casimir operator reduces to the $\mathcal{N}=2$ Liouville Hamiltonian \eqref{eq:N=2Liouville} acting on the Whittaker function, using the identification $\ell=2\phi$.

The solutions to this eigenvalue problem have been determined from the supersymmetric quantum mechanical algebra in \cite{Lin:2022zxd}, which by consistency coincide with the states in the $\mathcal{N}=2$ multiplet determined from the Whittaker function \eqref{eq:finalWhittakerfunction}. 

\subsection{Supersymmetric quantum mechanics from group theory}
\label{section:susyfromgrouptheory}
In the Hamiltonian reduction procedure, the superconformal $\mathfrak{osp}(2|2,\mathbb{R})$ algebra reduces to a $\mathcal{N}=(2,2)$ supersymmetry algebra with two pairs of supercharges $Q_-,\overline{Q}_-$, and $Q_+,\overline{Q}_+$ and two U$(1)$ $R$-symmetry generators. These supercharges anticommute to the Hamiltonian $\hat{H}$ when acting on the Whittaker function: 
\begin{align} \label{eq:supersymmetricQM}
    \{Q_-,\overline{Q}_-\}=\{Q_+,\overline{Q}_+\}\simeq \hat{H}. 
\end{align}
Our procedure is inspired by \cite{Bershadsky:1989mf} and \cite{Bershadsky:1989tc}, where the representation spaces of the $\mathfrak{sl}_k(2,\mathbb{R})$ and $\mathfrak{osp}_k(N|2)$ affine Lie (super)algebras reduce to the (super)conformal Virasoro algebra after imposing mixed parabolic boundary conditions on the generators. We add to that analysis explicit expressions for the $N=2$ case. \\

We start by writing the raising and lowering left and right supercharges in terms of the left- and right-regular generators respectively:
\begin{align}
    Q_-&=\frac{1}{2}\{\hat{L}_{H},\hat{L}_{F^-}\}-\frac{1}{2}\{\hat{L}_{Z},\hat{L}_{F^-}\}+\frac{1}{2}\{\hat{L}_{E^-},\hat{L}_{F^+}\},\label{eq:Qminus}\\
    \overline{Q}_-&=\frac{1}{2}\{\hat{L}_{H},\hat{L}_{\bar{F}^-}\}+\frac{1}{2}\{\hat{L}_{Z},\hat{L}_{\bar{F}^-}\}-\frac{1}{2}\{\hat{L}_{E^-},\hat{L}_{\bar{F}^+}\}, \\
    Q_+&=-\frac{1}{2}\{\hat{R}_H, \hat{R}_{F^+}\}-\frac{1}{2}\{\hat{R}_Z,\hat{R}_{F^+}\}+\frac{1}{2}\{\hat{R}_{E^+},\hat{R}_{F^-}\},\\
    \overline{Q}_+&= -\frac{1}{2}\{\hat{R}_H, \hat{R}_{\overline{F}^+}\}+\frac{1}{2}\{\hat{R}_Z, \hat{R}_{\overline{F}^+}\}-\frac{1}{2}\{\hat{R}_{E^+}, \hat{R}_{\overline{F}^-}\}.
    \label{eq:Qbarplus}
\end{align}
Inserting the explicit expressions \eqref{eq:leftreg}, \eqref{eq:oppositealgebrarightgeneratorslightcone}, we can write the result in terms of differential operators and superderivatives \eqref{eq:superderivatives} 
\begin{alignat}{2} \label{eq:leftregularQm}
    Q_-&=\left(\frac{\partial_\phi}{2}+\frac{i\partial_\theta}{2}\right)D_{\theta_-}+e^{-\phi-i\theta}D_{\theta_+}\partial_\beta, \quad
    &&\overline{Q}_-=\left(\frac{\partial_\phi}{2}-\frac{i\partial_\theta}{2}\right)D_{\overline{\theta}_-}-e^{-\phi+i\theta}D_{\overline{\theta}_+}\partial_\beta, \nonumber \\
    Q_+&=\left(-\frac{\partial_\phi}{2}+\frac{i}{2}\partial_\theta\right)D_{\theta^+}+e^{-\phi+i\theta}D_{\theta^-}\partial_\gamma,\quad
    &&\overline{Q}_+=\left(-\frac{\partial_\phi}{2}-\frac{i}{2}\partial_\theta\right)D_{\overline{\theta}^+}-e^{-\phi-i\theta}D_{\overline{\theta}^-}\partial_\gamma.
\end{alignat} 
Using the action of the superderivatives \eqref{eq:N=2covariantderivatives} and setting $\nu=\lambda=1$, these are in agreement with the quantum mechanical supercharges found in \cite{Lin:2022zxd} acting on the Whittaker function:
\begin{alignat}{2}
    Q_-&=\left(-\frac{\partial_\phi}{2}-\frac{i}{2}\partial_\theta\right)\psi_--ie^{-\phi-i\theta}\psi_+, \quad & \overline{Q}_-&=\left(\frac{\partial_\phi}{2}-\frac{i}{2}\partial_\theta\right)\overline{\psi}_-+ie^{-\phi+i\theta}\;\overline{\psi}_+,\nonumber\\
    Q_+&=\left(\frac{i}{2}\partial_\phi+\frac{\partial_\theta}{2}\right)\psi_++e^{-\phi+i\theta}\psi_-, \quad &  \overline{Q}_+&=\left(\frac{i}{2}\partial_\phi-\frac{\partial_\theta}{2}\right)\overline{\psi}_++e^{-\phi-i\theta}\;\overline{\psi}_-.\label{eq:reducedcharges}
\end{alignat}

One can easily check that the anticommutation relations between the left- and right-regular supercharges satisfy
 \begin{align}
    \{Q_-,\overline{Q}_-\}=\hat{\mathcal{C}}_2 \hat{L}_{E_-},\qquad
    \{Q_+,\overline{Q}_+\}=\hat{\mathcal{C}}_2 \; \hat{R}_{E^+},
\end{align} 
where $\hat{\mathcal{C}}_2$ is the Casimir operator. We see that the supercharges indeed square to the Casimir up to the multiplicative action of a parabolic generator. By  acting on an eigenstate of this parabolic generator with $\hat{L}_{E^-}=\hat{R}_{E^+}=-1$, this last generator is immaterial. Additionally, as checked above, the Casimir reduces to the Liouville Hamiltonian $\hat{\mathcal{C}}_2\simeq -\hat{H}$ when acting on the Whittaker function $\Psi_{j,q}(\phi,\theta)$, leading to the supersymmetric algebra \eqref{eq:supersymmetricQM}.

The $\mathcal{N}=(2,2)$ supersymmetric algebra also exhibits a pair of U$(1)$ $R$-symmetry generators $\hat{J}_\pm$, defined by the relations
\begin{alignat}{2}
   [\hat{J}_-,Q_-]&=Q_-, \qquad &[\hat{J}_-,\overline{Q}_-]&=-\overline{Q}_-, \label{eq:Ralgebra1}\\ 
   [\hat{J}_+,Q_+]&=Q_+,\qquad &[\hat{J}_+,\overline{Q}_+]&=-\overline{Q}_+.\label{eq:Ralgebra2}
\end{alignat}
To recover these $R$-symmetry generators from the $\mathfrak{osp}(2|2,\mathbb{R})$ algebra, we consider the following combinations of the regular generators, including the U$(1)$-generators $\hat{L}_Z$ and $\hat{R}_Z$: 
\begin{align}
    \hat{J}_-=-2\hat{L}_Z\hat{L}_{E_-}+\frac{1}{2}[ \hat{L}_{\overline{F}_-},\hat{L}_{F_-}], \qquad \hat{J}_+=-2\hat{R}_Z\hat{R}_{E^+}+\frac{1}{2}[\hat{R}_{\overline{F}^+},\hat{R}_{F^+}].
\end{align} 
In explicit coordinates, these become:
\begin{align}
    \hat{J}_-&=i\partial_\theta\partial_\beta+\partial_{\overline{\theta}_-}\partial_{\theta}+\frac{1}{4}\theta_-\overline{\theta}_-\partial_\beta^2+\frac{1}{2}\overline{\theta}_-\partial_{\overline{\theta}_-}\partial_\beta-\frac{1}{2}\theta_-\partial_{\theta_-}\partial_\beta,\\  \hat{J}_+&=i\partial_\theta\partial_\gamma+\partial_{\overline{\theta}^+}\partial_{\theta^+}+\frac{1}{4}\theta^+\overline{\theta}^+\partial_\gamma^2-\frac{1}{2}\overline{\theta}^+\partial_{\overline{\theta}^+}\partial_\gamma+\frac{1}{2}\theta^+\partial_{\theta^+}\partial_\gamma.
\end{align} 
We can now check that we indeed recover the supersymmetric algebra relations \eqref{eq:Ralgebra1}-\eqref{eq:Ralgebra2} in the form:
\begin{alignat}{2}
   [\hat{J}_-,Q_-]&=-Q_-\;\hat{L}_{E^-}, \qquad &[\hat{J}_-,\overline{Q}_-]&=\overline{Q}_-\;\hat{L}_{E^-}, \\ 
   [\hat{J}_+,Q_+]&=-Q_+\;\hat{L}_{E^+},\qquad &[\hat{J}_+,\overline{Q}_+]&=\overline{Q}_+.\;\hat{L}_{E^+},
\end{alignat}
by again acting on states that diagonalize the parabolic generators $\hat{L}_{E^-}=\hat{R}_{E^+}=-1$. 
In particular, we recover the quantum mechanical $R$-symmetry generators of \cite{Lin:2022zxd} acting on the Whittaker function
\begin{align}
     \hat{J}_-=i\partial_\theta-\frac{1}{2}[\overline{\psi}_-,\psi_-], \qquad \hat{J}_+=-i\partial_\theta -\frac{1}{2}[\overline{\psi}_+,\psi_+]. \label{eq:reducedRcharges}
\end{align}

\subsection{Discrete series wavefunctions}\label{section:discreteseriesharmonicanalysis}
In this subsection, we will determine the mixed parabolic discrete series matrix elements from the reduced harmonic analysis, and match the highest weight matrix element with the direct computation of the Whittaker function in section \ref{section:discreteseries}. 

The Casimir operator acts diagonally on the discrete series matrix elements as
\begin{align}
&\frac{1}{4}\partial_\theta^2+\frac{1}{4}\partial_\phi^2-e^{-2\phi}\lambda\nu -ie^{-\phi-i\theta}\sqrt{\nu\lambda}\;\overline{\psi}_-\psi_+-ie^{-\phi+i\theta}\sqrt{\nu\lambda}\;\psi_-\overline{\psi}_+,
\end{align} 
with eigenvalue $j^2-Q^2$. The solutions to this eigenvalue equation are given by the principal series matrix elements in the region $\lambda\nu>0$. In the region $\lambda\nu<0$, delta-normalizable energy solutions exist as well, yielding the discrete series matrix elements which have a smooth $\lambda\nu\rightarrow0$ highest weight limit. We choose the branch defined by $\lambda\nu=|\lambda\nu|e^{i\pi}$:
\begin{align}
&\frac{1}{4}\partial_\theta^2+\frac{1}{4}\partial_\phi^2+e^{-2\phi}|\lambda\nu| +e^{-\phi-i\theta}\sqrt{|\nu\lambda|}\;\overline{\psi}_-\psi_++e^{-\phi+i\theta}\sqrt{|\nu\lambda|}\;\psi_-\overline{\psi}_+.
\end{align}
Acting on the vacuum state in the multiplet, the operator reduces to
\begin{align}
\frac{1}{4} \partial_\phi^2+e^{-2\phi}|\lambda\nu| +\frac{1}{4}\partial_\theta^2.
\end{align} 
Separation of variables leads to a solution of the form $e^{2iQ\theta}f(\phi)$, with 
\begin{align}
\left(\frac{1}{4}\partial_\phi^2+ e^{-2\phi}|\lambda\nu|\right)f(\phi)=j^2f(\phi).
\end{align}
The solution with regular asymptotics as $x\rightarrow0$ is given by $J_{2j}(2\sqrt{|\lambda\nu|}e^{-\phi})$, yielding the bottom component of the discrete series matrix element: 
\begin{align}
    \ket{F^+}=e^{2iQ\theta}J_{2j}(2\sqrt{|\lambda\nu|}e^{-\phi}) \;\ket{0}.
\end{align}
Using the asymptotics of the Bessel functions $J_{\alpha}(x)\sim x^{|\alpha|}$ for $x\rightarrow 0$, we can write the highest weight matrix element with $|\lambda\nu|\rightarrow 0$  as: $\sim e^{2iQ\theta}e^{2j\phi}\ket{0}$, since $2j<0$.

The other states in the multiplet can be constructed by acting with the reduced supercharges \eqref{eq:reducedcharges}. In particular, the analytically continued right supercharges which anticommute to the Hamiltonian are: 
\begin{align}
    Q_+&= \psi_+\left(\frac{i}{2}\partial_\phi+\frac{\partial_\theta}{2}\right)+ie^{-\phi+i\theta}\sqrt{|\lambda\nu|}\;\psi_-, \\
    \overline{Q}_+&= \overline{\psi}_+\left(\frac{i}{2}\partial_\phi-\frac{\partial_\theta}{2}\right)+i e^{-\phi-i\theta}\sqrt{|\lambda\nu|}\;\overline{\psi}_-.
\end{align} 
Acting on the bottom state yields: 
\begin{align}
    \ket{L}&=\frac{\overline{Q}_+}{\sqrt{E}}\left(e^{2iQ\theta}J_{2j}(2\sqrt{|\lambda\nu|}e^{-\phi})\right)\ket{0}\nonumber\\
    &=\Big[\overline{\psi}_+\left((j-Q)J_{2j}(.)-\sqrt{|\lambda\nu|}e^{-\phi}J_{2j-1}(.)\right)+\overline{\psi}_-e^{-\phi-i\theta}\sqrt{|\lambda\nu|}J_{2j}(.)\Big]\frac{ie^{2iQ\theta}}{\sqrt{E}}\ket{0},
\end{align}
where we suppressed the argument $2\sqrt{|\lambda\nu|}e^{-\phi}$ of the Bessel functions in our notation. In the highest weight limit $|\nu\lambda|\rightarrow0$, this expression reduces to the expected 
\begin{align}
    \ket{L}= \frac{j-Q}{\sqrt{E}}i\;e^{2iQ\theta}e^{2\phi j}\; \overline{\psi}_+\ket{0}.
\end{align}
Analogously, the analytically continued left charges are \begin{align}
    Q_-&= \psi_-\left(-\frac{\partial_\phi}{2}-\frac{i}{2}\partial_\theta\right)+e^{-\phi-i\theta}\sqrt{|\lambda\nu|}\;\psi_+,\\
    \overline{Q}_-&=\overline{\psi}_-\left(\frac{\partial_\phi}{2}-\frac{i}{2}\partial_\theta\right)-e^{-\phi+i\theta}\sqrt{|\lambda\nu|}\;\overline{\psi}_+.
\end{align}
Acting with the raising charge on the bottom component readily yields 
\begin{align}
    \ket{H}=\Big[\overline{\psi}_-\left((Q+j)J_{2j}(.)-\sqrt{|\lambda\nu|}e^{-\phi}J_{2j-1}(.)\right)-\overline{\psi}_+e^{-\phi+i\theta}\sqrt{|\lambda\nu|}J_{2j}(.)\Big]\frac{e^{2iQ\theta}}{\sqrt{E}}\ket{0}.
\end{align}
For the highest weight state, this is 
\begin{align}
    \ket{H}=\frac{j+Q}{\sqrt{E}}\; e^{2iQ\theta}e^{2\phi j} \;\overline{\psi}_-\ket{0}.
\end{align} 
The top component coincides with the bottom component: \begin{align}
    \ket{F^-}=e^{2iQ\theta}J_{2j}(2\sqrt{|\lambda\nu|}e^{-\phi})\;\overline{\psi}_+\overline{\psi}_-\ket{0},
\end{align}
and has the same highest weight state. Note that this procedure of solving the harmonic analysis does not fix the relative phase factors between the different states. In particular, we are free to bundle the highest weight Wilson line operator insertion $\mathcal{W}$ of the multiplet representation in the columns of the $4$-dimensional matrix representations to coincide with the highest weight Whittaker function \eqref{eq:highestweightwhittakerfunction} as determined in the main text.

\section{ $\mathcal{N}=1$ JT supergravity revisited} \label{appendix:N=1}
In this appendix, we revisit the OSp$^+(1|2,\mathbb{R})$ structure governing the amplitudes of $\mathcal{N}=1$ JT supergravity of \cite{Fan:2021wsb} using the fermionic multiplet perspective as introduced in section \ref{eq:whitdefi} in this work. 

The generators of the OSp$(1|2,\mathbb{R})$ principal series representation are given by the differential operators acting on the space of square integrable functions $L^2(\mathbb{R}^{1|1})$ \cite{Fan:2021wsb}: 
\begin{align}
    E^-&=\partial_x, \qquad F^-=\frac{1}{2}(\partial_\vartheta+\vartheta \partial_x), \qquad H=x\partial_x+\frac{1}{2}\vartheta\partial_\vartheta-j,\\
    E^+&=-x^2\partial_x-x\vartheta\partial_\vartheta+2jx, \qquad F^+=-\frac{1}{2}x\partial_\vartheta-\frac{1}{2}x\vartheta\partial_x+j\vartheta.
\end{align}
These satisfy the opposite $\mathfrak{osp}(1|2,\mathbb{R})$ superalgebra relations, e.g.\begin{align}\label{eq:oppositeosp(1|2)}
    \{F^\pm,F^\pm\}=\mp \frac{E^\pm}{2}.
\end{align}

\subsection{Gravitational matrix elements}
We consider the $\mathcal{N}=1$ mixed parabolic matrix element $\braket{\nu, \psi_1\mid g\mid \lambda, \psi_2}$ in the principal series irrep, using the familiar Gauss-Euler coordinatization:
\begin{align}
    g= e^{2\theta^-F^-}e^{\beta E^-}e^{2\phi H} e^{\gamma E^+ }e^{2\theta^+F^+}.
    \label{eq:GaussEulerN1}
\end{align} 
This patch, which is smoothly connected to the identity, corresponds physically to the Ramond ($\mathbf{R}$) sector. We will make some comments on the $\mathbf{NS}$ sector below.

\subsubsection{Whittaker vectors}
We propose for the left and right parabolic eigenstates respectively: 
\begin{align}
    \braket{\nu, \psi_1\mid x,\vartheta}&=\frac{1}{\sqrt{2\pi}} e^{-\nu x}e^{-\psi_1\sqrt{\nu}\vartheta}= \frac{1}{\sqrt{2\pi}} e^{-\nu x} (1-\psi_1\sqrt{\nu} \vartheta), \\
    \braket{x,\vartheta\mid\lambda, \psi_2}&=\frac{1}{\sqrt{2\pi}}e^{-\lambda/x}x^{2j}e^{-i\psi_2\sqrt{\lambda}\vartheta/x}=\frac{1}{\sqrt{2\pi}} e^{-\lambda/x}x^{2j} \left(1-i\psi_2\sqrt{\lambda}\frac{\vartheta}{x}\right),
\end{align}
where $\psi_1$ and $\psi_2$ are two Majorana fermions satisfying the Dirac algebra\begin{align}
    \{\psi_i,\psi_j\}=2\delta_{ij},
\end{align}
while still anticommuting with all other Grassmann variables. 

We observe that these are eigenvectors of the bosonic parabolic generators \begin{align}
    \braket{\nu, \psi_1\mid x,\vartheta} E^-&=\braket{\nu, \psi_1\mid x,\vartheta}(-\overleftarrow{\partial}_x)=\nu \braket{\nu, \psi_1\mid x,\vartheta}, \\
    E^+\braket{x,\vartheta\mid \lambda, \psi_2}&=-\lambda \braket{x,\vartheta\mid\lambda, \psi_2},
\end{align} 
and are diagonalized by the fermionic generators to the left as:
\begin{align}
    \braket{\nu, \psi_1\mid x,\vartheta} F^-&= \braket{\nu, \psi_1\mid x,\vartheta} \tfrac{1}{2}(\overleftarrow{\partial}_\vartheta-\vartheta\overleftarrow{\partial}_x)=-\tfrac{1}{2} \psi_1\sqrt{\nu}\braket{\nu, \psi_1\mid x,\vartheta}, \\
    F^+ \braket{x,\vartheta\mid \lambda, \psi_2}&=-\tfrac{i}{2} \psi_2\sqrt{\lambda} \braket{x,\vartheta\mid\lambda,\psi_2}.
\end{align}
These actions are compatible with the opposite algebra relations \eqref{eq:oppositeosp(1|2)}. Note that compared to the case of $\mathcal{N}=2$, the algebra on both sides only requires one-dimensional representations of the Clifford algebra which square to $1$. In particular, there are no anticommutation relations on each side, and we are free to use $\mathbb{Z}_2$ sign factors $\epsilon$ for both fermionic eigenvalues to satisfy the algebra, following the analysis of \cite{Fan:2021wsb}. However, in the interest of streamlining to higher supersymmetric cases, we use an independent realization of the Clifford algebra on both sides with anticommuting fermionic numbers.

The mixed parabolic matrix element thus diagonalizes the exponential parabolic factors to the left of the Whittaker function $\Psi_j(\phi)\equiv  \braket{\nu, \psi_1\mid e^{2\phi H}\mid \lambda, \psi_2} $: \begin{align}
   & \braket{\nu, \psi_1\mid e^{2\theta^-F^-}e^{\beta E^-}e^{2\phi H} e^{\gamma E^+ }e^{2\theta^+F^+}\mid \lambda, \psi_2}=e^{-\theta^-\psi_1\sqrt{\nu}} e^{-i\theta^+\psi_2\sqrt{\lambda}}e^{\beta \nu} e^{-\gamma\lambda}\;\Psi_j(\phi).\label{Eq:Exponentiated}
\end{align}
\subsubsection{Whittaker function}
The Whitakker function of the semigroup  can be easily deduced from the group action \begin{align}
    (e^{2\phi H}\cdot f)(x, \vartheta)=e^{-2j\phi}f(e^{2\phi}x,e^{\phi}\vartheta),
\end{align}
and the integral representation of the modified Bessel functions \eqref{eq:integralBessel}: \begin{align}
    \Psi_j(\phi)= \frac{1}{\pi}\frac{\lambda^{j+1/2}}{\nu^{j}}e^{-\phi}\left(\psi_1K_{2j+1}(2\sqrt{\nu\lambda}e^{-\phi})+i\psi_2 K_{2j}(2\sqrt{\nu\lambda}e^{-\phi})\right).\label{eq:N=1Whittaker}
\end{align} The continuous representation label $j=ik-1/4$ is related to the momentum $k\in \mathbb{R}^+$ by unitarity \cite{Fan:2021wsb}. One arrives at the gravitational wavefunction by imposing the (renormalized) Brown-Henneaux boundary conditions $\nu=\lambda=1$ \begin{align}
    \Psi_{k}(\phi)=\raisebox{-0.3\height}{\includegraphics[height=0.8cm]{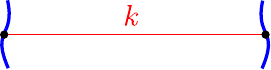}}= \frac{1}{\pi} e^{-\phi}\left(\psi_1K_{2j+1}(2e^{-\phi})+i\psi_2 K_{2j}(2e^{-\phi})\right).
\end{align}

\subsection{Hamiltonian reduction}
The two-sided gravitational wavefunctions of $\mathcal{N}=1$ JT supergravity are the quantum mechanical solutions to the Liouville minisuperspace Schr\"odinger equation with energy variable $E(k)=k^2$ \cite{Douglas:2003up}: \begin{align}\label{eq:N=1Liouville}
    \left(-\partial_\ell^2 +e^{-\ell}+\frac{i}{2}e^{-\ell/2}\psi_1\psi_2\right) \psi_k(\ell)=E(k)\; \psi_k(\ell),
\end{align}
in terms of the two Majorana fermions $\psi_{1,2}$. The two-sided gravitational solutions span the bulk Hilbert space of $\mathcal{N}=1$ JT supergravity in terms of the renormalized geodesic separation between the two boundaries $\ell$. The aim of this section is to establish the gravitational wavefunctions as Whittaker functions satisfying the Casimir eigenvalue equation.

 The left- and right-regular generators \eqref{eq:left}, acting on group elements of the principal series representations are:
\begin{align}
    \hat{L}_{F^-}&=-\frac{1}{2}\left(\partial_{\theta^-}-\theta^-\partial_{\beta}\right),\qquad
    \hat{L}_{E^-}=-\partial_{\beta},\qquad
    \hat{L}_H=-\frac{1}{2}\partial_{\phi}+\beta\partial_{\beta}+\frac{1}{2}\theta^-\partial_{\theta^-},\nonumber\\
    \hat{L}_{E^+}&=-e^{-2\phi}\partial_{\gamma}-\beta\partial_{\phi}+\beta^2\partial_{\beta}+\beta\theta^-\partial_{\theta^-}+e^{-\phi}\theta^-(\partial_{\theta^+}+\theta^+\partial_{\gamma}),\nonumber\\
    \hat{L}_{F^+}&=-\frac{1}{2}e^{-\phi}(\partial_{\theta^+}+\theta^+\partial_{\gamma})-\frac{\beta}{2}(\partial_{\theta^-}-\theta^-\partial_{\beta})-\frac{1}{2}\theta^-\partial_{\phi},\nonumber\\
    \hat{R}_{F_+}&= \frac{1}{2}\partial_{\theta_+}-\frac{\theta_+}{2}\partial_{\gamma},\qquad
    \hat{R}_{E_+}=\partial_{\gamma}\qquad
    \hat{R}_H=\frac{1}{2}\partial_\phi-\gamma\partial_{\gamma}-\frac{\theta_+}{2}\partial_{\theta_+},\nonumber\\
    \hat{R}_{E_-}&= \gamma\partial_\phi-\gamma^2\partial_{\gamma}+e^{-2\phi}\partial_{\beta}+e^{-\phi}\theta_+(\partial_{\theta_-}+\theta_-\partial_{\beta})-\theta_+\gamma\partial_{\theta_+},\nonumber\\
    \hat{R}_{F_-}&=-\frac{\theta_+}{2}\partial_\phi+\frac{e^{-\phi}}{2}(\partial_{\theta_-}+\theta_-\partial_{\beta})-\frac{\gamma}{2}(\partial_{\theta_+}-\theta_+\partial_{\gamma}).
\end{align}
The Casimir eigenvalue equation in either representations reads: \begin{align}
    &\hat{\mathcal{C}}\; \braket{\nu, \psi_1\mid g\mid \lambda, \psi_2}\nonumber\\&=\left(\frac{1}{4}\partial_\phi^2+\frac{1}{4} \partial_\phi+e^{-2\phi}\partial_{\beta}\partial_{\gamma}-\frac{1}{2}e^{-\phi}(\partial_{\theta^-}+\theta^-\partial_{\beta})(\partial_{\theta^+}+\theta^+\partial_{\gamma})\right) \braket{\nu,\psi_1\mid g\mid \lambda,\psi_2}\nonumber\\&= j(j+1/2) \;\braket{\nu, \psi_1\mid g\mid \lambda, \psi_2}.
\end{align}
The combination \begin{align}
    \Xi\equiv(\partial_{\theta_-}+\theta_-\partial_{\beta})(\partial_{\theta_+}+\theta_+\partial_{\gamma})
\end{align} is a product of two  superderivatives, which each get diagonalized by the parabolic exponential prefactors to the right: \begin{align}
    D_{\theta_-}&=(\partial_{\theta_-}+\theta_-\partial_{\beta})\;\rightarrow\; -\psi_1\sqrt{\nu}\label{eq:N=1superderivative1},\\
    D_{\theta_+}&=(\partial_{\theta_+}+\theta_+\partial_{\gamma})\;\rightarrow\; -i\psi_2\sqrt{\lambda}.\label{eq:N=1superderivative2}
\end{align}

In the end, choosing mixed parabolic boundary conditions, the full Casimir acting on the principal series matrix element reduces to a Schr\"odinger equation of the Whittaker function $\Psi_j(\phi)$ \begin{equation}\label{eq:N=1reducedcasimir}
    \left(\frac{1}{4}\partial_\phi^2+\frac{1}{4} \partial_\phi-e^{-2\phi}\nu\lambda-\frac{i}{2}e^{-\phi}\psi_1\psi_2\sqrt{\nu\lambda}\right) \; \Psi_j(\phi)=j(j+1/2) \;\Psi_j(\phi).
\end{equation}
Setting $\nu=\lambda=1$ and $\ell=2\phi$ to go to gravity and shifting
$\Psi_k(\phi)=e^{-\phi/2}\psi_k(\phi)$, this eigenvalue equation becomes the $\mathcal{N}=1$ minisuperspace Liouville equation \eqref{eq:N=1Liouville}.

\subsection{Multiplet representation}
We can represent the fermions by using the $\mathcal{N}=1$ multiplet states $\ket{0}, \overline{\psi}\ket{0}$, with the lightcone combinations
\begin{align}
    \psi=\tfrac{1}{2}(\psi_1+i\psi_2), \qquad \overline{\psi}=\tfrac{1}{2}(\psi_1-i\psi_2),
\end{align}
satisfying the Dirac algebra 
\begin{align}
    \{\psi,\overline{\psi}\}=1. 
\end{align} 
The vacuum state itself is annihilated by the lowering operator $\psi\ket{0}=0$. 
In terms of the matrix multiplet representation with ordering $\ket{0}=\begin{bmatrix}
    1\\0
\end{bmatrix},\; \overline{\psi}\ket{0}=\begin{bmatrix}
    0\\1
\end{bmatrix}$, the fermions are represented using the Dirac algebra: 
\begin{align}
    \psi=\begin{bmatrix}
        0&1\\0&0
    \end{bmatrix},\qquad \overline{\psi}=\begin{bmatrix}
        0&0\\1&0
    \end{bmatrix},
\end{align} 
leading to a two-dimensional realization of the gamma matrices 
\begin{align}
    \psi_1=\begin{bmatrix}
        0&1\\1&0
    \end{bmatrix}=\sigma_x, \qquad \psi_2=\begin{bmatrix}
        0&-i\\i&0
    \end{bmatrix}=\sigma_y.
\end{align}
The shifted Whittaker function $\psi_k(\phi)$ then explicitly contains the two linearly independent gravitational wavefunctions in its two columns:
\begin{align}
    &\psi_k(\phi)= \raisebox{-0.3\height}{\includegraphics[height=0.8cm]{gravprop5.pdf}}\nonumber\\&  =e^{-\phi/2} \begin{bmatrix}
        0& K_{2ik+1/2}(2e^{-\phi})+K_{2ik-1/2}(2e^{-\phi})\\ K_{2ik+1/2}(2e^{-\phi})-K_{2ik-1/2}(2e^{-\phi})&0
    \end{bmatrix},\label{eq:N=1matrixwhittaker}
\end{align} 
which each diagonalize the Hamiltonian \eqref{eq:N=1Liouville}, leading to the two linearly independent eigenstates: 
\begin{align}
    \ket{F^-}&=e^{-\phi/2}\left[K_{2ik+1/2}(2e^{-\phi})-K_{2ik-1/2}(2e^{-\phi})\right]\overline{\psi}\ket{0}, \\
    \ket{F^+}&=e^{-\phi/2}\left[K_{2ik+1/2}(2e^{-\phi})+K_{2ik-1/2}(2e^{-\phi})\right]\ket{0}.
\end{align}
with odd and even fermionic parity respectively.

\subsection{Supersymmetric quantum mechanics from group theory}
\label{appendix:susqmN1}
 We mirror section \ref{section:susyfromgrouptheory} 
 and generate an emergent $\mathcal{N}=(1,1)$ supersymmetric algebra from the $\mathcal{N}=1$ $\mathfrak{osp}(1 \vert 2,\mathbb{R})$ superalgebra.

The left and right supercharges can be constructed from the combinations:
\begin{align}
    Q_-&=\{\hat{L}_{E_-}, \hat{L}_{F_+}\}+\{\hat{L}_H,\hat{L}_{F_-}\},\\
    Q_+&=\{\hat{R}_{H},\hat{R}_{F^+}\}-\{\hat{R}_{F^-},\hat{R}_{E^+}\},
\end{align}
which can be worked out in terms of the $\mathcal{N}=1$ superderivatives \eqref{eq:N=1superderivative1}-\eqref{eq:N=1superderivative2}:
\begin{align}
    Q_-&=e^{-\phi}D_{\theta_+}\partial_{\beta}+\frac{1}{2}\partial_\phi D_{\theta_-}+\frac{1}{4}D_{\theta_-},\\
     Q_+&=-e^{-\phi}D_{\theta_-}\partial_{\gamma}+\frac{1}{2}D_{\theta_+}\partial_\phi+\frac{1}{4} D_{\theta_+}.
\end{align}  
Squaring these supercharge operators yields the  Casimir multiplied by a left or right parabolic generator:
\begin{align}
    Q_-^2&=\left(\mathcal{C}_2+\frac{1}{16}\right)(-\hat{L}_{E_-}),\\
     Q_+^2&=\left(\mathcal{C}_2+\frac{1}{16}\right)\hat{R}_{E_+}.
\end{align} 
The parabolic generators are by construction diagonalized by the mixed parabolic matrix element, leading to the Liouville Hamiltonian acting on the Whittaker function \eqref{eq:N=1reducedcasimir}. 

\subsection{Thermofield double state}
The physical boundary conditions imposed on the thermofield double state depend on the fermions.
The Ramond ($\mathbf{R}$)-sector corresponds to the sector connected to the identity group element, whereas the Neveu-Schwarz ($\mathbf{NS}$)-sector corresponds to the sector disconnected to the identity.
\subsubsection{Ramond sector}
The Hartle-Hawking state is obtained by propagating the unit group element: 
\begin{align}
\raisebox{-0.45\height}{\includegraphics[height=2cm]{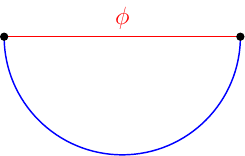}}=\braket{\phi\mid e^{-\beta H/2 }\mid \mathbb{1}}=\int_0^\infty dk \braket{\phi\mid k} \braket{k\mid \mathbb{1}} e^{-\beta E(k)/2}.
\end{align}  
The representation matrices are the Whittaker functions \eqref{eq:N=1matrixwhittaker} normalized by the Plancherel measure $\rho(k)=\cosh2\pi k$ of the OSp$^+(1|2,\mathbb{R})$ \emph{semi}group \cite{Fan:2021wsb}:
\begin{align}
    \braket{\phi\mid k}&=\sqrt{\rho(k)}\Psi_j(\phi)= \sqrt{\cosh(2\pi k)}\;e^{-\phi}\begin{bmatrix}
        0&f_2\\f_1&0
    \end{bmatrix},
\end{align}  
with 
\begin{align}
    f_1&=K_{2ik+1/2}(2e^{-\phi})-K_{2ik-1/2}(2e^{-\phi}),\\ f_2&=K_{2ik+1/2}(2e^{-\phi})+K_{2ik-1/2}(2e^{-\phi}).
\end{align}
The unit group element in a mixed parabolic basis corresponds to the coordinate region $\phi\rightarrow-\infty$. Using the leading asymptotics of the Bessel function, this becomes up to an overall infinitesimal $k$-independent constant \begin{align}
    \braket{\mathbb{1}\mid k}=\sqrt{\cosh(2\pi k)}\;\begin{bmatrix}
    0&1\\0&0
    \end{bmatrix}.
\end{align}
Using the conjugate form in the Hartle-Hawking state, we determine the single linearly independent thermofield double state:
\begin{align}
    \ket{\text{TFD}}=\int_0^\infty dk \cosh(2\pi k)e^{-\beta E(k)/2}\;e^{-\phi}[K_{2ik+1/2}(2e^{-\phi})+K_{2ik-1/2}(2e^{-\phi})]\ket{0},
\end{align} proportional to the bottom state of the multiplet with even parity.

\subsubsection{Neveu-Schwarz sector}
The gravitational wavefunctions of the $\mathbf{NS}$-sector correspond to the representation matrices of the opposite patch of the OSp$(1|2,\mathbb{R})$ group manifold
\begin{align}
     g= e^{2\theta^-F^-}e^{\beta E^-}e^{2\phi H}(-)^F e^{\gamma E^+ }e^{2\theta^+F^+},
\end{align}
obtained by inserting the ``sCasimir'' operator $(-)^F=1-2\vartheta\partial_\vartheta$. This leads to the normalized wavefunctions propagating in the bulk \begin{align}
    \braket{\phi\mid k}=\sqrt{\cosh(2\pi k)}\;e^{-\phi}\begin{bmatrix}
        0&g_2\\g_1&0
    \end{bmatrix},
\end{align}
where now
\begin{align}
    g_1&=K_{2ik+1/2}(2e^{-\phi})+K_{2ik-1/2}(2e^{-\phi}),\\ g_2&=K_{2ik+1/2}(2e^{-\phi})-K_{2ik-1/2}(2e^{-\phi}).
\end{align} 
The ``unit'' group element is regularized by the asymptotics of the Bessel functions as 
\begin{align}
    \braket{\mathbb{1}\mid k}=\sqrt{\cosh(2\pi k)}\begin{bmatrix}
        0&0\\1&0
    \end{bmatrix},
\end{align}
leading to the physical thermofield double state 
\begin{align}
     \ket{\text{TFD}}=\int_0^\infty dk\;\cosh(2\pi k)e^{-\beta E(k)/2}\;e^{-\phi}\left[K_{2ik+1/2}(2e^{-\phi})+K_{2ik-1/2}(2e^{-\phi})\right]\overline{\psi}\ket{0},
\end{align}
proportional to the top state of the multiplet with odd parity. In terms of the $\mathbb{Z}_2$ sign labels $\epsilon$ used to describe the different states in \cite{Fan:2021wsb}, the $\mathbf{R}$-sector selects the same periodicity of the fermions on both sides, whereas the $\mathbf{NS}$-sector selects the opposite signs on the two sides. 

\bibliographystyle{ourbst}
\bibliography{Grav_matrix.bib}
\end{document}